hep-th/9812110
\def\Ad            {\mathrm{Ad}}

\newcommand\aepm[1]{\alpha^\pm_{{#1}}}

\newcommand\aed[2] {\alpha_{({#1},{#2})}}
\newcommand\aedv[3] {\alpha_{({#1},{#2})}^{#3}}
\newcommand\aedp[2]{\alpha^+_{({#1},{#2})}}
\newcommand\aedm[2]{\alpha^-_{({#1},{#2})}}
\newcommand\aedx[3]{\alpha_{({#1},{#2})}^{(#3)}}
\newcommand\aedpm[2]{\alpha^\pm_{({#1},{#2})}}
\newcommand\aedmp[2]{\alpha^\mp_{({#1},{#2})}}

\newcommand\aedprod[4]{\alpha^+_{({#1},{#2})}\circ\alpha^-_{({#3},{#4})}}
\newcommand\aefp[3] {\alpha^+_{({#1},{#2},{#3})}}
\newcommand\aefm[3] {\alpha^-_{({#1},{#2},{#3})}}

\def\ala           {\alpha_\lambda}
\def\alap          {\alpha^+_\lambda}
\def\alapm         {\alpha^\pm_\lambda}
\def\alamp         {\alpha^\mp_\lambda}
\def\alam          {\alpha^-_\lambda}

\def\aLap          {\alpha^+_\Lambda}

\def\aLapm         {\alpha^\pm_\Lambda}
\def\aLam          {\alpha^-_\Lambda}
\def\aLamb         {\alpha^-_{\overline{\Lambda}}}
\def\aLaps         {\alpha^+_{\Lambda'}}
\def\aLapbs        {\alpha^+_{\overline{\Lambda'}}}

\def\aLams         {\alpha^-_{\Lambda'}}
\def\aLambs        {\alpha^-_{\overline{\Lambda'}}}
\def\aLapmbs       {\alpha^\pm_{\overline{\Lambda'}}}

\def\amup          {\alpha^+_\mu}
\def\amupm         {\alpha^\pm_\mu}
\def\amum          {\alpha^-_\mu}
\def\amump         {\alpha^\mp_\mu}

\def\amubm         {\alpha^-_\mub}
\def\anup          {\alpha^+_\nu}

\def\anum          {\alpha^-_\nu}

\def\arhom         {\alpha^-_\rho}

\newcommand\alfp[1]{\alpha^+_{\Lambda_{({#1})}}}
\newcommand\alfm[1]{\alpha^-_{\Lambda_{({#1})}}}
\newcommand\alfpm[1]{\alpha^\pm_{\Lambda_{({#1})}}}

\newcommand\ASU[2] {\cA^{({#1},{#2})}}
\newcommand\as[1]  {[\alpha_{{#1}}]}
\newcommand\asp[1] {[\alpha^+_{{#1}}]}
\newcommand\asm[1] {[\alpha^-_{{#1}}]}
\newcommand\aspm[1]{[\alpha^\pm_{{#1}}]}
\newcommand\asprod[2]{[\alpha^+_{{#1}}\circ\alpha^-_{{#2}}]}
\newcommand\asprodpm[2]{[\alpha^\pm_{{#1}}\circ\alpha^\mp_{{#2}}]}
\newcommand\asdprod[4]{[\alpha^+_{({#1},{#2})}\circ\alpha^-_{({#3},{#4})}]}
\newcommand\asx[2] {[\alpha_{{#1}}^{({#2})}]}
\newcommand\asxp[2]{[\alpha_{{#1}}^{+({#2})}]}
\newcommand\asxm[2]{[\alpha_{{#1}}^{-({#2})}]}
\newcommand\asxpm[2]{[\alpha_{{#1}}^{\pm({#2})}]}
\newcommand\asd[2] {[\alpha_{({#1},{#2})}]}
\newcommand\asdx[3]{[\alpha_{({#1},{#2})}^{({#3})}]}
\newcommand\asdp[2]{[\alpha^+_{({#1},{#2})}]}
\newcommand\asdxp[3]{[\alpha_{({#1},{#2})}^{+({#3})}]}
\newcommand\asdm[2]{[\alpha^-_{({#1},{#2})}]}
\newcommand\asdxm[3]{[\alpha_{({#1},{#2})}^{-({#3})}]}
\newcommand\asdpm[2]{[\alpha^\pm_{({#1},{#2})}]}
\newcommand\asdmp[2]{[\alpha^\mp_{({#1},{#2})}]}
\newcommand\asdxpm[3]{[\alpha_{({#1},{#2})}^{\pm({#3})}]}
\newcommand\asf[3] {[\alpha_{({#1},{#2},{#3})}]}
\newcommand\asfx[4]{[\alpha_{({#1},{#2},{#3})}^{(#4)}]}

\newcommand\asfp[3] {[\alpha^+_{({#1},{#2},{#3})}]}
\newcommand\asfm[3] {[\alpha^-_{({#1},{#2},{#3})}]}
\newcommand\asfpm[3]{[\alpha^\pm_{({#1},{#2},{#3})}]}

\newcommand\asfxpm[4]{[\alpha_{({#1},{#2},{#3})}^{\pm(#4)}]}
\newcommand\asfprod[6]{[\alpha^+_{({#1},{#2},{#3})}\circ
                    \alpha^-_{({#4},{#5},{#6})}]}

\def\asibp         {\alpha^+_\sib}
\def\asibm         {\alpha^-_\sib}
\def\asibpm        {\alpha^\pm_\sib}

\def\asidm         {\alpha^-_\sid}
\def\asidpm        {\alpha^\pm_\sid}
\def\Aut           {\mathrm{Aut}}
\def\bbC           {\mathbb{C}}

\def\bbN           {\mathbb{N}}

\def\bbZ           {\mathbb{Z}}
\def\bfe           {{\bf1}}

\def\can           {\gamma}
\def\cani          {\gamma^{-1}}
\def\canr          {\theta}
\newcommand\cdd[2] {h_{({#1},{#2})}}
\newcommand\cdf[3] {h_{({#1},{#2},{#3})}}
\def\cA            {\mathcal{A}}

\def\cD            {\mathcal{D}}
\def\cE            {\mathcal{E}}

\def\cG            {\mathcal{G}}
\def\cH            {\mathcal{H}}

\def\cM            {\mathcal{M}}
\def\cN            {\mathcal{N}}
\def\cO            {\mathcal{O}}
\def\cP            {\mathcal{P}}

\def\cT            {\mathcal{T}}
\def\cV            {\mathcal{V}}
\def\cW            {\mathcal{W}}
\def\cX            {\mathcal{X}}
\def\cY            {\mathcal{Y}}

\newcommand\chid[2]{\chi_{({#1},{#2})}}

\newcommand\co[1]  {\overline{{#1}}}

\def\DelNIo        {\Delta_\cN(\Io)}
\def\DelMIo        {\Delta_\cM(\Io)}

\def\dim           {\mathrm{dim}}
\newcommand\del[2] {\delta_{{#1},{#2}}}
\def\dLam          {{\dot{\Lambda}}}
\def\ddLam         {{\ddot{\Lambda}}}

\def\E             {\mathrm{e}}
\def\Eig           {\mathrm{Eig}}
\def\End           {\mathrm{End}}
\newcommand\eps[2] {\varepsilon({#1},{#2})}
\newcommand\epsm[2]{\varepsilon^-({#1},{#2})}
\newcommand\epsp[2]{\varepsilon^+({#1},{#2})}
\newcommand\epspm[2]{\varepsilon^\pm({#1},{#2})}

\newcommand\epsr[2]{\varepsilon_\mathrm{r}({#1},{#2})}
\newcommand\erf[1] {Eq.\ (\ref{#1})}
\def\Exp           {\mathrm{Exp}}

\def\Gtwo          {\mathrm{G}_2}
\def\Hom           {\mathrm{Hom}}
\def\I             {{\rm i}}
\def\Bild          {\mathrm{Im}\,}

\def\id            {\mathrm{id}}
\def\Io            {I_\circ}
\def\iotab         {\co\iota}
\def\Jz            {\mathcal{J}_z}

\def\la            {\langle}

\newcommand\lasd[2]{[\lambda_{({#1},{#2})}]}

\def\LG            {\mathit{LG}}

\def\LIG           {\mathit{L}_I\mathit{G}}

\def\LISUn         {\mathit{L}_I\mathit{SU}(n)}

\newcommand\ls[1]  {[\lambda_{{#1}}]}

\def\LSUn          {\mathit{LSU}(n)}

\def\LTSN          {[\Delta]_\cN(\Io)}

\def\MIo           {{M(\Io)}} 
\def\mod           {\mathrm{mod}}
\def\Mor           {\mathrm{Mor}}
\def\mub           {{\overline{\mu}}}

\newcommand\N[3]   {N_{{#1},{#2}}^{{#3}}}
\def\NIo           {{N(\Io)}}
\newcommand\Nd[6]  {N_{({#1},{#2}),({#3},{#4})}^{({#5},{#6})}}

\def\pio           {\pi_0}

\def\Qbp           {Q_{\beta,+}}
\def\Qbm           {Q_{\beta,-}}
\def\Qbpm          {Q_{\beta,\pm}}
\def\Qbmp          {Q_{\beta,\mp}}
\def\Qdp           {Q_{\delta,+}}

\def\Qdpm          {Q_{\delta,\pm}}
\def\ra            {\rangle}

\def\Sect          {\mathrm{Sect}}

\def\sib           {{\sigma_\beta}}
\def\sid           {{\sigma_\delta}}

\def\SOf           {\mathit{SO}(5)}

\def\SUd           {\mathit{SU}(3)}

\def\SUm           {\mathit{SU}(m)}
\def\SUn           {\mathit{SU}(n)}

\def\SUz           {\mathit{SU}(2)}

\def\SUf           {\mathit{SU}(4)}

\def\triv          {\mathrm{triv}}

\def\ucpm          {u_{\canr,\pm}}

\newcommand\V[3]   {V_{{#1};{#2}}^{#3}}

\def\be            {\begin{equation}}
\def\bearl         {\begin{array}{l}}
\def\bearll        {\begin{array}{ll}}
\def\bearlll       {\begin{array}{lll}}
\def\bearrl        {\begin{array}{rl}}
\def\bea           {\begin{eqnarray}}
\def\beaa          {\begin{eqnarray*}}
\def\ee            {\end{equation}}
\def\eear          {\end{array}}
\def\eea           {\end{eqnarray}}
\def\eeaa          {\end{eqnarray*}}
\newcommand\bproof {\noindent {\it Proof. }}
\newcommand\eproof {\hspace*{\fill}\nolinebreak\hspace*{\fill}$\Box$
                   \par\vspace{3mm}}

\newcommand\labl[1]{\label{#1}\ee}
\newcommand\lablth[1]{\label{#1}}
\newcommand\lablsec[1]{\label{#1}}

\def\per           {positive energy representation}
\newcommand\ddA[1] {$\mathrm{A}_{#1}$}
\newcommand\ddD[1] {$\mathrm{D}_{#1}$}
\newcommand\ddE[1] {$\mathrm{E}_{#1}$}

\newcommand\ddDgx[1]{$\mathcal{D}^{({#1})}$}
\newcommand\ddEgx[1]{$\mathcal{E}^{({#1})}$}
\def\Deven         {$\mathrm{D}_{\mathrm{even}}$}

\documentclass[11pt]{article}
\usepackage{amssymb,amsfonts,latexsym,epic,eepic}
\begin{document}

%%%%%%%%%%%%%%%%%%%%%%%%%%%%%%%%%%%%%%%%%%%%%%%%%%%%%%%%%%%%%%%%%%%%%%%

\title{Modular Invariants, Graphs and $\alpha$-Induction for
Nets of Subfactors. III}
\author{{\sc Jens B\"ockenhauer} and {\sc David E. Evans}\\ \\
School of Mathematics\\University of Wales Cardiff\\
PO Box 926, Senghennydd Road\\Cardiff CF2 4YH, Wales, U.K.}
\maketitle

\begin{abstract}
In this paper we further develop the theory of
$\alpha$-induction for nets of subfactors,
in particular in view of the system of sectors
obtained by mixing the two kinds of induction
arising from the two choices of braiding.
We construct a relative braiding between the
irreducible subsectors of the two ``chiral''
induced systems, providing a proper
braiding on their intersection. We also express
the principal and dual principal graphs of the local
subfactors in terms of the induced sector systems.
This extended theory is again applied to conformal
or orbifold embeddings of $\SUn$ WZW models.
A simple formula for the corresponding modular invariant
matrix is established in terms of the two inductions, and we
show that it holds if and only if the sets of irreducible
subsectors of the two chiral induced systems intersect
minimally on the set of marked vertices, i.e.\ on the
``physical spectrum'' of the embedding theory,
or if and only if the canonical endomorphism sector
of the conformal or orbifold inclusion subfactor
is in the full induced system. We can prove either
condition for all simple current extensions of
$\SUn$ and many conformal inclusions, covering in
particular all type \nolinebreak I modular invariants of
$\SUz$ and $\SUd$, and we conjecture that it holds also
for any other conformal inclusion of $\SUn$ as well.
As a by-product of our calculations, the dual principal graph
for the conformal inclusion $\SUd_5\subset\mathit{SU}(6)_1$
is computed for the first time.
\end{abstract}

\newpage

\tableofcontents

%%%%%%%%%%%%%%%%%%%%%%%%%%%%%%%%%%%%%%%%%%%%%%%%%%%%%%%%%%%%%%%%%%%%%%%%%%%

\newtheorem{definition}{Definition}[section]
\newtheorem{lemma}[definition]{Lemma}
\newtheorem{corollary}[definition]{Corollary}
\newtheorem{theorem}[definition]{Theorem}
\newtheorem{proposition}[definition]{Proposition}
\newtheorem{conjecture}[definition]{Conjecture}

%%%%%%%%%%%%%%%%%%%%%%%%%%%%%%%%%%%%%%%%%%%%%%%%%%%%%%%%%%%%%%%%%%%%%%%%%%%

\section{Introduction}

In a previous paper \cite{boev2}, motivated by work of Feng Xu
\cite{xu4}, we analyzed nets of subfactors $\cN\subset\cM$
associated to type \nolinebreak I (or block-diagonal) modular
invariants through a notion of induction and restriction of
sectors between the two nets of factors \cite{boev1} --- a
notion introduced by Longo and Rehren in \cite{lore}.
As the main application we considered
type \nolinebreak I modular invariants of $\SUn$.

Here we take the analysis further to understand the modular
invariant matrix $Z$ in terms of the inductions $\alpha^+$
and $\alpha^-$, which depend on the choice of braiding and
opposite braiding in the $\SUn_k$ sectors of the
smaller net $\cN$. In fact we find for our examples
(and believe it to be true in general)
\[ Z_{\Lambda,\Lambda'} = \la \aLap,\aLams \ra_M \,, \]
where $\la \aLap,\aLams \ra_M$ is the dimension of the
intertwiner space $\Hom_M(\aLap,\aLams)$ and
where $\Lambda,\Lambda'$ are weights in the Weyl alcove
$\ASU nk$, labelling the $\SUn_k$ sectors, and $M$ is a
local factor of the enveloping net $\cM$.

We recall the story so far. The fusion graph of
$[\alfp 1]$ (or $[\alfm 1]$), where $\Lambda_{(1)}$
is the (first) fundamental weight and corresponds to
the generator of the $\SUn_k$ fusion algebra, is
the graph which in the $\SUz$ case appears in the
A-D-E classification of Capelli, Itzykson and Zuber
\cite{caiz} and empirically associated to the $\SUd$
modular invariants by Di Francesco and Zuber
\cite{frzu1,frzu2}. In general, the non-zero diagonal
terms of the modular invariant matrix correspond
exactly to the eigenvalues of (the adjacency matrix of)
the fusion graph of $[\alfp 1]$ (or $[\alfm 1]$),
as long as the fusion coefficients of the sectors
of the extended theory are diagonalized by the
corresponding modular S-matrix.

Let us restrict our discussion to the
conformal inclusion case for a while. The set $\cT$ of the
original sectors of the extended net appears in the set
$\cV^+$ of irreducible subsectors of the induced system
$\{[\aLap]:\Lambda\in\ASU nk\}$, and similarly in $\cV^-$
corresponding to $\{[\aLam]:\Lambda\in\ASU nk\}$. Consequently,
the ``chiral'' sets of sectors $\cV^+$ and $\cV^-$ intersect at
least on $\cT$, the ``marked vertices''. Note that although
there is a canonical bijection between $\cV^+$ and $\cV^-$
(see Subsect.\ \ref{relpm} below), they rarely coincide
as sets of sectors. Indeed it will be shown in
Proposition \ref{4equiv} that the following conditions are
equivalent:
\begin{itemize}
\item $Z_{\Lambda,\Lambda'} = \la \aLap,\aLams \ra_M$ for all
      $\Lambda,\Lambda'\in\ASU nk$,
\item $\cT=\cV^+\cap\cV^-$.
\end{itemize}
Although it is shown that the matrix $\la \aLap,\aLams \ra_M$
is $T$-invariant (see Lemma \ref{disj} below) we have no direct
proof why it is $S$-invariant or why either of the conditions
holds in the general framework. However, the above conditions
are also shown to be equivalent to either of the following which
say that the set $\cV$ of irreducible subsectors of the full
induced system $\{[\aLap\circ\aLams] : \Lambda,\Lambda'\in\ASU nk\}$
is complete in a certain sense:
\begin{itemize}
\item Each irreducible subsector of the canonical endomorphism
      sector $[\can]$ belongs to $\cV$,
\item $\sum_{x\in\cV} d_x^2 = \sum_{\Lambda\in\ASU nk} d_\Lambda^2$.
\end{itemize}
Here the $d$'s denote the statistical dimensions of the sectors.
In concrete examples, in particular for the conformal embeddings
$\SUz_{4}\subset\SUd_1$, $\SUz_{10}\subset\SOf_1$,
$\SUz_{28}\subset(\Gtwo)_1$, $\SUd_3\subset\mathit{SO}(8)_1$,
$\SUd_5\subset\mathit{SU}(6)_1$, $\SUd_9\subset(\mathrm{E}_6)_1$,
$\SUd_{21}\subset(\mathrm{E}_7)_1$ and
$\SUf_4\subset\mathit{SO}(15)_1$, such conditions can be
shown to be satisfied, and thus $\cV^+$ and $\cV^-$ only
intersect on the marked vertices or we recover the
modular invariant matrix $Z$ from the induced
$\la \aLap,\aLams \ra_M$.

The completeness of the induced system has another
important aspect. If each irreducible subsector of
the canonical endomorphism sector $[\can]$ belongs to $\cV$,
then, besides the principal graph, also the dual principal graph
of the conformal inclusion subfactor is determined in terms
of the induced system. We use this fact to compute the two
basic graphical invariants of conformal inclusion subfactors
in examples. This includes the computation of the dual
principal graph for the conformal embedding
$\SUd_5\subset\mathit{SU}(6)_1$ which has, to the best
of our knowledge, not been obtained before.

We also extend the discussion of $\bbZ_n$ orbifold inclusions
(or ``simple current extensions'') in \cite{boev2} to the
$\bbZ_m$ case, where $m$ is
any divisor of $n$ if $n$ is not prime, and this covers
all simple current extension modular invariants of $\SUn$.
For these cases we can in fact show that
$Z_{\Lambda,\Lambda'} = \la \aLap,\aLams \ra_M$ holds
(see Theorem \ref{Zapamorb} below), and in consequence
that each irreducible subsector of
the canonical endomorphism sector $[\can]$ belongs
to the induced system and that
$\sum_{x\in\cV} d_x^2 = \sum_{\Lambda\in\ASU nk} d_\Lambda^2$.
The intersection $\cV^+\cap\cV^-$ corresponds to the ``localized
sectors'' or the ``physical spectrum'' of the extended theory
and is expected to coincide with the labelling set of the
conjectured extended S-matrix in \cite{fss2}.
In fact, we construct a non-degenerate braiding on this
intersection (see Theorem \ref{Tnd} below), and by
Rehren's methods \cite{reh0} this provides a representation
of the modular group, thus a matrix $S$. Although we have no
proof we expect that this is the S-matrix of \cite{fss2}.

Together with our conformal embedding examples we obtain
completeness of the induced system for all the
type \nolinebreak I cases of the modular invariants of $\SUz$
and $\SUd$ which were classified by Cappelli, Itzykson
and Zuber \cite{caiz} and Gannon \cite{gan2}.

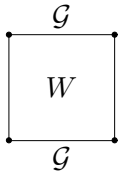
\begin{figure}[htb]
\begin{center}
\begin{picture}(40,60)
\path(0,10)(40,10)(40,50)(0,50)(0,10)
\multiput(0,10)(0,40){2}{\circle*{2}}
\multiput(40,10)(0,40){2}{\circle*{2}}
\put(20,3){\makebox(0,0){$\cG$}}
\put(20,57){\makebox(0,0){$\cG$}}
\put(20,30){\makebox(0,0){$W$}}
\end{picture}
\end{center}
\caption{A bi-unitary connection $W$}
\label{biuc}
\end{figure}

Ocneanu has classified in \cite{ocng} irreducible
bi-unitary connections on the A-D-E Dynkin diagrams.
The family of bi-unitary connections as in
Fig.\ \nolinebreak \ref{biuc},
%
%%%%%%%%%%%% Connection W %%%%%%%%%%%%%
where the horizontal graph $\cG$ is an A-D-E Dynkin
diagram and the vertical graphs are arbitrary, form
a fusion ring with generators $W$ and $\co W$. He then
obtains the graphs of Figs.\ \ref{E6pm}, \ref{E8pm},
\ref{D4pm} and \ref{Devenpm} below, the vertices
describing all such irreducible connections, and the
edges arise from the fusion graphs of the generators.
The open string bimodule construction
(see \cite{asha} for details) identifies such connections
or vertices with bimodules, arising from the
Goodman-de la Harpe-Jones \cite{ghj}
inclusion $N\subset M$. If $Z=(Z_{i,j})_{i,j=0}^k$
is the $\SUz_k$ modular invariant matrix associated
to the graph $\cG$ by \cite{caiz} then the sum
$\sum_{j,j'} Z_{j,j'}^2$ coincides with the total number
of vertices, and the irreducible $M$-$M$ bimodules
form a subset of the even vertices, which exhausts all of
them in the \ddE 6 and \ddE 8 cases. Each non-zero entry
$Z_{j,j'}$ of the modular invariant matrix is claimed to be
identified with an irreducible representation of the full
fusion ring (cf.\ Proposition \ref{ZExp} and
Conjecture \ref{regV} below).
A relative braiding between the chiral halves is also
constructed (cf.\ Proposition \ref{relbra} below) which
yields a braiding on the ``ambichiral'' intersection
(cf. Corollary \ref{Tbraid} below). The off-diagonal
terms in the modular invariant matrix
\[ Z_{j,j'} = \sum_{t\in\cT} b_{t,j} b_{\omega(t),j'}  \,,\]
is given a subfactor interpretation as
$b_{t,j}\equiv V_{j,\bfe}^t$ can be computed in terms
of ``essential paths''; here $V$ is the
$A-\cG_\mathrm{flat}$ intertwiner matrix introduced
in \cite{frzu2} (cf.\ also \cite[Sect.\ 5.4]{pezu2}),
$\cG_\mathrm{flat}$ the ``flat part'' of the graph
$\cG$ ($\cG_\mathrm{flat}=\mathrm{D}_{10}$ for $\cG=\mathrm{E}_7$,
$\cG_\mathrm{flat}=\mathrm{A}_{4\ell-1}$ for $\cG=\mathrm{D}_{2\ell+1}$,
$\ell=2,3,...$, and $\cG_\mathrm{flat}=\cG$ for the
type \nolinebreak I cases A, \Deven, \ddE 6 and \ddE 8),
$\cT$ the set of ``marked vertices'' of the
modular invariant labelled by $\cG_\mathrm{flat}$
and $\omega$ the corresponding fusion rule automorphism
of $\cT$ (which is trivial in the type \nolinebreak I case).
The relationship between our net of subfactors
approach and Ocneanu's bimodule-connection approach
will be discussed in \cite{bek}.

\section{Preliminaries}
In this section we recall several mathematical tools
we use and the general framework of \cite{boev1,boev2}.

\subsection{Sectors between different factors}
\lablsec{preizu}

For our purposes it turns out to be convenient to make use of
the formulation of sectors between different factors, we follow
here (up to minor notational changes) Izumi's presentation
\cite{izu3,izu4} based on Longo's sector theory \cite{lon2}.
Let $A$, $B$ infinite factors. We denote by
$\Mor(A,B)$ the set of unital $\ast$-homomorphisms from
$A$ to $B$. For $\rho\in\Mor(A,B)$ we define the statistical
dimension $d_\rho=[B:\rho(A)]^{1/2}$, where $[B:\rho(A)]$ is the
minimal index \cite{jon,kosa}. A morphism $\rho\in\Mor(A,B)$ is
called irreducible if the subfactor $\rho(A)\subset B$ is
irreducible, i.e.\ if the relative commutant
$\rho(A)'\cap B$ consists only of scalar multiples of
the identity in $B$. Two morphisms $\rho,\rho'\in\Mor(A,B)$
are called equivalent if there exists a unitary $u\in B$
such that $\rho'(a)=u\rho(a)u^*$ for all $a\in A$.
We denote by $\Sect(A,B)$ the quotient of $\Mor(A,B)$ by
unitary equivalence, and we call its elements $B$-$A$ sectors.
Similar to the case $A=B$, $\Sect(A,B)$ has the natural
operations, sums and products:  For $\rho_1,\rho_2\in\Mor(A,B)$
choose generators $t_1,t_2\in B$ of a Cuntz algebra $\cO_2$,
i.e.\ such that $t_i^*t_j=\del ij \bfe$ and
$t_1t_1^*+t_2t_2^*=\bfe$. Define $\rho\in\Mor(A,B)$ by
putting $\rho(a)=t_1\rho_1(a)t_1^*+t_2\rho_2(a)t_2^*$
for all $a\in A$, and then the sum of sectors is defined
as $[\rho_1]\oplus[\rho_2]=[\rho]$. The product of sectors
comes from the composition of endomorphisms,
$[\rho_1]\times[\rho_2]=[\rho_1\circ\rho_2]$.
The statistical dimension is an invariant for sectors
(i.e.\ equivalent morphisms have equal dimension) and
is additive and multiplicative with respect to these
operations. Moreover, for $[\rho]\in\Sect(A,B)$ there
is a unique conjugate sector $\co{[\rho]}\in\Sect(B,A)$
such that, if $[\rho]$ is irreducible and has finite
statistical dimension,
$\co{[\rho]}\times[\rho]$ contains the
identity sector $[\id_A]$ and $[\rho]\times\co{[\rho]}$
contains $[\id_B]$ precisely once. Then we denote a suitable
representative endomorphism of $\co{[\rho]}$
naturally by $\co\rho$, thus $[\co\rho]=\co{[\rho]}$.
For conjugates we have $d_{\co\rho}=d_\rho$. As for
bimodules one may decorate $B$-$A$ sectors $[\rho]$
with suffixes, ${}_B[\rho]_A$, and then we can
multiply ${}_B[\rho]_A \times {}_A[\sigma]_B$ but
not, for instance, ${}_B[\rho]_A$ with itself.
For $\rho,\tau\in\Mor(A,B)$ we denote
\[ \Hom_{A,B}(\rho,\tau) = \{ t\in B :
t\, \rho(a) = \tau (a) \, t \,,\,\,\, a\in A \} \]
and
\[ \la \rho, \tau \ra_{A,B} = \dim\,\Hom_{A,B}(\rho,\tau) \,.\]
If $[\rho]=[\rho_1]\oplus[\rho_2]$ then
\[ \la \rho, \tau \ra_{A,B} = \la \rho_1, \tau \ra_{A,B}
+ \la \rho_2, \tau \ra_{A,B} \,. \]
If $A=B$ we just write $\Hom_A(\rho,\tau)$ and
$\la \rho, \tau \ra_A$ for
$\rho,\tau\in\Mor(A,A)\equiv\End(A)$,
as usual. If $C$ is another factor,
$\rho\in\Mor(A,B)$, $\sigma\in\Mor(A,C)$,
$\tau\in\Mor(B,C)$ are morphisms with finite statistical
dimension and $\co\tau\in\Mor(C,B)$,
$\co\rho\in\Mor(B,A)$, representative conjugates of
$\rho$ and $\tau$, respectively, then we have
Frobenius reciprocity \cite{izu3,lon4},
\[ \la \tau\circ\rho, \sigma \ra_{A,C} =
\la \rho, \co\tau \circ \sigma \ra_{A,B} =
\la \tau, \sigma \circ \co\rho \ra_{B,C} \,. \]

Now let $N\subset M$ be an infinite subfactor of finite
index. Let $\can\in\End(M)$ be a canonical endomorphism from
$M$ into $N$ and $\canr=\can|_N\in\End(N)$. By $\iota\in\Mor(N,M)$
we denote the injection map, $\iota(n)=n\in M$, $n\in N$.
Then a conjugate $\iotab\in\Mor(M,N)$ is given by
$\iotab(m)=\can(m)\in N$, $m\in M$. (These formulae
could in fact be used to define the canonical and
dual canonical endomorphism.) Note that
$\can=\iota\circ\iotab$ and $\canr=\iotab\circ\iota$.
Denote by $\cP_0\subset\Sect(N)$, $\cP_1\subset\Sect(M,N)$,
$\cD_0\subset\Sect(M)$ and $\cD_1\subset\Sect(N,M)$ the set
of all irreducible subsectors of $[\canr^\ell]$,
$[\canr^\ell\circ\iotab]$, $[\can^\ell]$ and
$[\can^\ell\circ\iota]$, $\ell=0,1,2,3...$, respectively.
Note that there is a bijection from $\cP_1$ to
$\cD_1$ arising from sector conjugation.
The principal graph of the inclusion $N\subset M$ is
obtained as follows. The even vertices are labelled by
the elements of $\cP_0$, the odd vertices by those of
$\cP_1$, and we connect any even vertex labelled by
$[\lambda]\in\cP_0$ with any odd vertex labelled by
$[\rho]\in\cP_1$ by $\la\lambda\circ\iotab,\rho\ra_{N,M}$
edges. Similarly we obtain the dual principal graph.
We label the even vertices by $\cD_0$ and the odd
vertices by $\cD_1$, and then connect even vertices
labelled by $[\beta]\in\cD_0$ with odd vertices labelled
by $[\tau]\in\cD_1$ by $\la\beta\circ\iota,\tau\ra_{M,N}$
edges.

\subsection{Braidings}

Let $A$ be an infinite factor and
$\Delta\subset\End(A)$ a subset such
that $\Ad(u)\in\Delta$ for any unitary $u\in A$ and
$\lambda\circ\mu\in\Delta$ whenever $\lambda,\mu\in\Delta$,
moreover, if $t_1,t_2\in A$ are Cuntz algebra ($\cO_2$)
generators,
i.e.\ $t_i^*t_j=\del ij \bfe$ and $t_1t_1^*+t_2t_2^*=\bfe$,
and $\lambda,\lambda_1,\lambda_2\in\End(A)$ such that
$\lambda(a)=t_1\lambda_1(a)t_1^*+t_2\lambda_2(a)t_2^*$
for all $a\in A$, then $\lambda\in\Delta$ whenever
$\lambda_1,\lambda_2\in\Delta$ and conversely
$\lambda_1,\lambda_2\in\Delta$ whenever $\lambda\in\Delta$.
In other words, $\Delta$ is a set of
representative endomorphisms of some set of sectors which is
closed under products and sums and decomposition.
We say that $\Delta$ is braided if for any pair
$\lambda,\mu\in\Delta$ there is a unitary operator
$\eps\lambda\mu\in\Hom_A(\lambda,\mu)$, called
braiding operator, subject to initial conditions
\be
\eps {\id}\mu=\eps\lambda{\id}=\bfe \,,
\labl{ini}
composition rules ($\nu\in\Delta$)
\be
\eps {\lambda\circ\mu}\nu = \eps \lambda\nu \,
\lambda (\eps \mu\nu) \,, \qquad
\eps \lambda{\mu\circ\nu} = \mu (\eps \lambda\nu) \,
\eps \lambda\mu \,,
\labl{comp}
and whenever $t\in \Hom_A (\lambda,\mu)$
we have the naturality equations ($\rho\in\Delta$)
\be
\rho(t) \, \eps \lambda \rho = \eps \mu \rho \, t \,,
\qquad t \, \eps \rho \lambda = \eps \rho \mu \, \rho (t) \,.
\labl{nat}
Note that from Eqs.\ (\ref{comp}) and (\ref{nat}) one
obtains easily the braiding fusion equations, that is if 
$s\in \Hom_A (\lambda,\mu\circ\nu)$ we have
\be
\begin{array}{rl}
\rho(s) \, \eps \lambda \rho 
&=\,\, \eps \mu \rho \, \mu (\eps \nu \rho) \, s \,, \\[.4em]
s \, \eps \rho \lambda
&=\,\, \mu (\eps \rho \nu) \, \eps \rho \mu \, \rho (s) \,.
\eear
\labl{BFE}
Also note that for a unitary $u\in A$ the braiding
operators transform as
\be
\eps {\Ad(u)\circ\lambda}\mu = \mu(u)\, \eps \lambda\mu \, u^* \,,
\quad \eps \lambda{\Ad(u)\circ\mu} = u \, \eps \lambda\mu \,
\lambda(u)^* \,,
\labl{compu}
by \erf{nat}. Similarly, if we have Cuntz algebra ($\cO_n$)
generators $t_i$ and endomorphisms $\lambda,\lambda_i\in\Delta$,
$i=1,2,...,n$, such that
$\lambda(a)=\sum_{i=1}^n t_i\lambda_i(a)t_i^*$ for all $a\in A$,
then we find by \erf{nat},
\be
\eps \lambda\rho = \sum_{i=1}^n \rho(t_i) \, \eps {\lambda_i}\rho
\, t_i^* \,, \qquad \eps \rho\lambda = \sum_{i=1}^n t_i \,
\eps \rho{\lambda_i} \rho(t_i^*) \,.
\labl{esum}
Moreover, putting $\epsm \lambda\mu = (\epsp \mu\lambda )^*$,
$\eps \mu\lambda\equiv\epsp \mu\lambda$
gives another ``opposite'' braiding, i.e. satisfying
the same relations.

Now let $\cX\subset\Sect(A)$ be a sector basis.
(A sector basis is a finite set of irreducible
sectors of finite dimensions containing the trivial
sector and being closed under conjugation and irreducible
decomposition of sector products.) We obtain a set
$\Delta\equiv\Delta(\cX)\subset\End(A)$ from $\cX$
by taking all representative endomorphisms of all
sector products and sums.
We say $\cX$ is braided if $\Delta$ is braided. Note that,
if we take a choice of representatives for all elements
of $\cX$ and there is a collection of unitaries satisfying
the braiding fusion relation \erf{BFE} for these
representatives, then we obtain a
braiding of $\Delta$ by using Eqs.\ (\ref{comp}),
(\ref{compu}) and (\ref{esum}) as definitions. In particular,
if $\rho\in\End(A)$ is a representative for $[\rho]\in\cX$
(hence irreducible)
then $\eps\id\rho$ (and $\eps\rho\id$) is a phase, and from
\erf{BFE} with $\lambda=\mu=\nu=\id$, $s=\bfe$,
it follows that it is idempotent, hence the initial
condition. 

For a sector basis $\cX\in\Sect(A)$ we may make a choice
of representative endomorphisms, as usual denoted by
$\lambda$ for $[\lambda]\in\cX$. We call a braiding on
$\cX$ non-degenerate if for some $[\lambda]\in\cX$
trivial monodromy, $\eps\lambda\mu \eps\mu\lambda = \bfe$
for all $[\mu]\in\cX$ implies that $[\lambda]$ is the
trivial sector. Note that by \erf{compu} this
definition does not depend on the choice of representatives.

\subsection{Nets of subfactors and $\alpha$-induction}

Here we briefly review our basic notation and some results
of our previous papers \cite{boev1,boev2}.
There we considered certain nets of subfactors
$\cN\subset\cM$ on the punctured circle, i.e.\ we were dealing with
a family of subfactors $N(I)\subset M(I)$ on a Hilbert space $\cH$,
indexed by the set $\Jz$ of open intervals $I$ on the unit circle
$S^1$ that do neither contain nor touch a distinguished point
``at infinity'' $z\in S^1$. The defining representation of $\cN$
possesses a subrepresentation $\pio$ on a distinguished
subspace $\cH_0$ giving rise to another net
$\cA=\{A(I)=\pio(N(I))\,,\,\,I\in\Jz\}$. We assumed this net
to be strongly additive and to satisfy Haag duality, and also
locality of the net $\cM$. Fixing an interval $\Io\in\Jz$ we
used the crucial observation in \cite{lore} that there is an
endomorphism $\can$ of the $C^*$-algebras $\cM$ into $\cN$ (the
$C^*$-algebras associated to the nets are denoted by the same
symbols as the nets itself, as usual) such that it restricts to
a canonical endomorphism of $M(I)$ into $N(I)$ whenever
$I\supset\Io$. By $\canr$ we
denote its restriction to $\cN$. We denote by $\DelNIo$ the set
of transportable endomorphisms localized in $\Io$. It is a
result of \cite{lore} that $\canr\in\DelNIo$.
Elements of $\DelNIo$ leave $\NIo$ invariant and can
therefore also be considered as elements of $\End(\NIo)$.
The elements of $\DelNIo$ are braided endomorphisms,
and the braiding operators
$\eps \lambda\mu \equiv \epsp \lambda\mu$,
$\epsm \lambda\mu = (\epsp \mu\lambda )^*$,
$\epspm\lambda\mu\in\Hom_\NIo(\lambda,\mu)$,
for $\lambda,\mu\in\DelNIo$ are given by the
DHR statistics operators \cite{dhr3,haag}.
The $\pm$-sign here is due to the two possibilities of the
statistics operators coming from the non-trivial space-time
topology of the punctured circle (see \cite{frs1,frs2}).
Therefore the two statistics operators, corresponding to
braiding and opposite braiding, are in general different
but they may coincide for some $\lambda$ and $\mu$.
We call $\alpha$-induction the two maps
$\DelNIo\rightarrow\End(\cM)$, $\lambda\mapsto\alapm$,
where
\[ \alapm = \cani \circ \Ad (\epspm \lambda\canr)
\circ \lambda \circ \can \,. \]
As endomorphisms in $\DelNIo$ leave $\NIo$ invariant it makes
sense to define the quotient $\LTSN$ by inner equivalence in $\NIo$.
Similarly, the endomorphisms $\alapm$ leave $\MIo$ invariant,
hence we can consider them also as elements of $\End(\MIo)$
and form their inner equivalence classes $[\alapm]$ in $\MIo$.
We derived that in terms of these sectors,
$\alpha$-induction preserves sums and products,
and we proved
\[ \la\alapm,\amupm\ra_\MIo = \la \canr\circ\lambda,\mu\ra_\NIo\,,
\qquad \lambda,\mu\in\DelNIo\,, \]
which is useful to determine the structure of the induced sectors.
We also have a map $\End(\cM)\rightarrow\End(\cN)$,
$\beta\mapsto\sib$, where $\sib=\can\circ\beta|_\cN$,
called $\sigma$-restric\-tion.
If $\beta$ is a localized (in $\Io$) and transportable endomorphism
of $\cM$ which leaves $\MIo$ invariant (the invariance follows
automatically from the localization in $\Io$ if the net $\cM$
is Haag dual --- as is the case in all our
applications) then $\sib\in\DelNIo$, and then we
have also $\alpha\sigma$-reciprocity,
\[ \la \ala,\beta \ra_\MIo = \la \lambda,\sib \ra_\NIo \,.\]

In \cite{boev2} we have already applied this theory to several
conformal and to the $\bbZ_n$ orbifold inclusions of $\SUn$.
If $\SUn_k\subset G_1$ is a conformal inclusion at level $k$
with $G$ some connected compact simple Lie group then the
associated net of subfactors is given in terms of the local
inclusions defined by local loop groups,
\[ N(I)=\pi^0(\LISUn)'' \subset \pi^0(\LIG)'' = M(I)\,, \]
where $\pi^0$ is the level $1$ vacuum representation of $\LG$.
In the orbifold case the net of subfactors is obtained
by a certain crossed product construction from the net
of factors $A(I) = \pio(\LISUn)''$, relative to a simple current;
here $\pio$ denotes the level $k$ vacuum representation of $\LSUn$.
The orbifold inclusions appear only for certain values of the level,
and this turns out to be related to the locality condition of the
extended net. In any case we apply induction to the set of
sectors $[\lambda_\Lambda]$ which correspond to the positive
energy representation $\pi_\Lambda$ of $\LSUn$,
$\Lambda\in\ASU nk$, and obey the Verlinde fusion rules
by the results of Wassermann \cite{wass3}.

\section{Mixing Two Inductions}

From now on let us assume that we are dealing with a given
quantum field theoretic net of subfactors $\cN\subset\cM$
over the index set $\Jz$ as in \cite{boev1}, i.e.\ we assume
that $\cN$ is strongly additive and Haag dual in the vacuum
representation and we assume $\cM$ to be local. We also
assume the index to be finite. We fix an arbitrary interval
$\Io\in\Jz$ and take the endomorphism $\can\in\End(\cM)$ of
\cite{lore} which restricts to a canonical endomorphism from
$\MIo$ into $\NIo$, and we denote $\canr=\can|_\cN$.
To simplify notation, we will abbreviate $N=\NIo$ and $M=\MIo$
for the rest of this paper.

\subsection{Subsectors of $[\alap]$ and $[\alam]$}

We now consider $\alpha$-induction defined by means of the
two different braidings simultaneously.

\begin{lemma}
Let $\lambda\in\DelNIo$ and $\beta\in\End(M)$ such that
$[\beta]$ is a subsector of $[\alamp]$. If there is a $\mu\in\DelNIo$
such that $[\beta]$ is a subsector of $[\amupm]$ as well,
then $[\beta]$ is also subsector of $[\alapm]$.
\lablth{lamupm}
\end{lemma}

\bproof
By assumption there is an isometry $t\in\Hom_M(\beta,\alamp)$.
Now if $[\beta]$ is also a subsector of $[\amupm]$ for some
$\mu\in\DelNIo$ then we have an isometry
$s\in\Hom_M(\beta,\amupm)$, hence
\[ ts^* \, \amupm(m) = t\, \beta(m) \, s^* =
\alamp (m) \, ts^* \,, \qquad m\in M.\]
Restriction yields $ts^*\mu(n)=\lambda(n)ts^*$ for all $n\in N$,
hence, by (the plus- and minus-versions of) \cite[Lemma 3.5]{boev1},
\[ ts^* \, \amupm (m) = \alapm (m) \, ts^* = t \, \beta(m) \, s^*\,,
\qquad m\in M \,.\]
Right multiplication by $s$ yields
$t\in\Hom_M(\beta,\alapm)$.
\eproof

\begin{lemma}
Let $\beta_i\in\End(M)$ such that $[\beta_i]$ is a
subsector of $[\alpha^\pm_{\lambda_i}]$,
$\lambda_i\in\DelNIo$, $i=1,2$. If $u\in M$ fulfills
$u\beta_1(n)=\beta_2(n)u$, $n\in N$, then
$u\in\Hom_M(\beta_1,\beta_2)$.
\lablth{key'}
\end{lemma}

\bproof
By assumption there are isometries
$t_i\in\Hom_M(\beta_i,\alpha^\pm_{\lambda_i})$.
If $u\in M$ fulfills
$u\beta_1(n)=\beta_2(n)u$, then
$t_2ut_1^* \lambda_1(n)=\lambda_2(n)t_2ut_1^* $, $n\in N$,
hence
$t_2ut_1^*\in\Hom_M(\alpha^\pm_{\lambda_1},\alpha^\pm_{\lambda_2})$
by \cite[Lemma 3.5]{boev1}, and thus
\[ u\,\beta_1(m)=t_2^*t_2ut_1^*  \, \alpha^\pm_{\lambda_1}(m) \, t_1
=t_2^* \, \alpha^\pm_{\lambda_2}(m) \, t_2ut_1^*t_1
= \beta_2 (m) \, u \,, \]
we are done.
\eproof

Next we present a slight generalization of our ``main formula'',
\cite[Thm.\ 3.9]{boev1}. Recall that $v\in M$ and $w\in N$
are the intertwining isometries from the identity of
$\cM$ and $\cN$ to $\can$ and $\canr$, respectively, and
satisfying $w^*v=[M:N]^{-1/2}\bfe$. Also recall that we
have pointwise equality $M=Nv$ \cite{lore}.

\begin{proposition}
Let $\beta\in\End(M)$ such that $[\beta]$ is a subsector of
$[\amupm]$ for some $\mu\in\DelNIo$.
Then we have
\be
\la \alapm, \beta \ra_M = \la \lambda, \sib \ra_N
\ee
for all $\lambda\in\DelNIo$.
\lablth{genmain}
\end{proposition}

\bproof
First we show ``$\le$'': Assume $s\in\Hom_M(\alapm,\beta)$.
Then, by restriction,
$s\lambda(n)=\beta(n)s$ for all $n\in N$, hence
$\can(s)\in\Hom_N(\canr\circ\lambda,\sib)$, hence
$\can(s)w\in\Hom_N(\lambda,\sib)$. As the map
$s\mapsto\can(s)w$ is injective
\cite[Lemma 3.8]{boev1}, this proves ``$\le$''.

Next we show ``$\ge$'': Let $r\in\Hom_N(\lambda,\sib)$.
Put $s=v^*r$. Then
\[ s\,\lambda(n) = v^*r \,\lambda(n) = v^* \cdot \can\circ\beta(n)
\cdot r = \beta(n)\,v^*r=\beta(n)\,s\,, \qquad n\in N\,.\]
By Lemma \ref{key'} it follows $s\in\Hom_M(\alapm,\beta)$.
Now the map $r\mapsto s=v^*r$ is injective, proving ``$\ge$''.
\eproof

Let $\lambda\in\DelNIo$ and let
\[ \alap(m) = \sum_a t_a \, \beta^+_a (m) \, t_a^* \,,
\qquad m\in M \,, \]
be an irreducible decomposition with some set
$\{ t_a \}_a$ of Cuntz algebra generators.
Here we allow multiplicities, so some of the
$\beta_a^+$'s may be equivalent. Now
\[ t_a t_a^* \in \alap(M)' \cap M = \lambda(N)'
\cap M = \alam(M)' \cap M \,, \]
again by \cite[Lemma 3.5]{boev1}.
Therefore putting
\[ \beta^-_a(m) = t_a^* \, \alam(m) \, t_a \,, \qquad m\in M\,,\]
defines endomorphisms of $M$ and we have
\[ \alam(m) = \sum_a t_a \, \beta^-_a(m) \, t_a^* \,, \qquad m\in M.\]
Clearly, $\beta^+_a(n) = t_a^* \lambda(n) t_a = \beta^-_a(n)$
for all $n\in N$. We now easily obtain from Lemma \ref{key'}
the following

\begin{corollary}
The $\beta^-_a$'s are irreducible as well. Moreover,
if $[\beta^+_a]=[\beta^-_b]$ then
$[\beta^+_a]=[\beta^-_a]=[\beta^+_b]=[\beta^-_b]$.
We have $[\beta^+_a]=[\beta^+_b]$ if and only if
$[\beta^-_a]=[\beta^-_b]$.
\lablth{relpm}
\end{corollary}

We also find

\begin{lemma}
We have $d_{\beta^+_a}=d_{\beta^-_a}$.
\end{lemma}

\bproof
Since $\sigma_{\beta^\pm_a}$ is the restriction of
$\can\circ\beta^\pm_a$
to $N$ we have $d_{\sigma_{\beta^\pm_a}}=d_\can d_{\beta^\pm_a}$,
cf.\ \cite[Subsect.\ 3.3]{boev1},
but also $\sigma_{\beta^+_a}=\sigma_{\beta^-_a}$, implying
the statement.
\eproof

\subsection{Comparing $[\alap]$ and $[\amum]$}

For $\mu\in\DelNIo$ define
\[ \triv(\mu,\canr) = \{ t\in M:\,\,\eps\mu\canr
\eps\canr\mu \, \can(t) = \can(t) \} \,.\]
Recall from \cite[Subsect.\ 3.1]{boev1} that we have
$\amupm(v)=\epspm\mu\canr ^*v$.

\begin{lemma}
For $\lambda,\mu\in\DelNIo$ we have
\be
\Hom_M(\alap,\amum)=\Hom_M(\alap,\amup) \cap
\triv(\mu,\canr) \,.
\ee
\lablth{pmpp}
\end{lemma}

\bproof
Ad ``$\subset$'': Let $t\in\Hom_M(\alap,\amum)$.
Restriction and \cite[Lemma 3.5]{boev1} clearly implies
$t\in\Hom_M(\alap,\amup)$. Moreover,
$t\in\Hom_M(\alap,\amum)$ implies also
\[ t \, \alap(v) = \amum(v)\,t = \epsm\mu\canr ^*v \, t =
\eps\canr\mu \, \can(t)v \,.\]
Whereas $t\in\Hom_M(\alap,\amup)$ yields
\[ t \alap(v) = \amup(v)\,t =
\eps \mu\canr ^* \, \can(t)v \,,\]
thus
\[ \eps\mu\canr\eps\canr\mu \can(t)v = \can(t)v \,. \]
By \cite[Lemma 3.8]{boev1} this implies
\[ \eps\mu\canr\eps\canr\mu \can(t) = \can(t) \,, \]
proving ``$\subset$''.

Ad ``$\supset$'': Let
$t\in\Hom_M(\alap,\amup) \cap \triv(\mu,\canr)$. As
$\alap$ and $\amupm$ restrict to $\lambda$ and $\mu$ on $N$,
respectively, it suffices to show $t\alap(v)=\amum(v)t$.
From $\eps\mu\canr\eps\canr\mu\can(t)=\can(t)$ we obtain
$\eps\mu\canr^*\can(t)=\epsm\mu\canr ^* \can(t)$,
hence
\[ t\alap(v)=\amup(v)t=\epsp\mu\canr ^*\can(t)v
=\epsm\mu\canr ^*vt = \amum(v)t \,, \]
proving the lemma.
\eproof

Trivially, we obtain

\begin{corollary}
We have $\la\alap,\amum\ra_M\le\la\alap,\amup\ra_M$ for all
$\lambda,\mu\in\DelNIo$.
\lablth{estpm}
\end{corollary}

For a reducible $\mu\in \DelNIo$ take an irreducible decomposition
\[ \mu(n) = \sum_{i=1}^s t_i \, \mu_i(n) \, t_i^* \,,
\qquad n\in N \,, \]
with $\mu_i\in\DelNIo$ and a set $\{t_i: i=1,2,...,s\}$
of Cuntz algebra generators in $N$. We allow that
some of the irreducible $\mu_i$'s may be equivalent.

\begin{lemma}
For $\lambda\in\DelNIo$ and $\mu$ as above we have
$\eps\lambda\mu\eps\mu\lambda=\bfe$ if and only if
$\eps\lambda{\mu_i}\eps{\mu_i}\lambda=\bfe$ for all
$i=1,2,\ldots,s$.
\lablth{decmon}
\end{lemma}

\bproof
Since $t_i\in\Hom(\mu_i,\mu)$ we have
the naturality equations
\[ \lambda (t_i) \, \eps{\mu_i}\lambda
= \eps\mu\lambda \, t_i \,, \qquad
t_i \, \eps\lambda{\mu_i}
= \eps\lambda\mu \, \lambda (t_i) \,.\]
Therefore
\[ \eps\lambda\mu\eps\mu\lambda
= \sum_{i=1}^s \eps\lambda\mu\eps\mu\lambda \, t_it_i^*
= \sum_{i=1}^s t_i \, \eps\lambda{\mu_i}
\eps{\mu_i}\lambda \, t_i^* \,,\]
hence $\eps\lambda{\mu_i}\eps{\mu_i}\lambda=
t_i^* \eps\lambda\mu\eps\mu\lambda t_i$.
\eproof

Now let $[\canr]=\bigoplus_{i=1}^s [\canr_i]$ be
an irreducible decomposition where
$\canr_i\in\DelNIo$, $i=1,2,...,s$.
We recall that the monodromy is
diagonalized as follows \cite[Lemma 3.3]{frs2}
(see also \cite{evka}, Sect.\ 8.2,
in particular Figs.\ 8.30 and 8.31):
\be
\eps\mu\nu\eps\nu\mu \, t =
\frac{\kappa_\lambda}{\kappa_\nu \kappa_\mu} \, t \,,
\qquad  t\in\Hom_N(\lambda,\nu\circ\mu)\,,
\labl{mondiag}
for irreducible $\lambda,\mu,\nu\in\DelNIo$,
where the $\kappa$'s are the statistical phases which are
invariants of the sectors.
Now for any $\lambda\in\DelNIo$ write the statistical phase
as $\kappa_\lambda=\E^{2\pi \I h_\lambda}$ with some
$h_\lambda \ge 0$. In our applications, $h_\lambda$ will be
the conformal dimension of the sector $[\lambda]$, and for the
susbsectors of $[\canr]$ we will also have
$h_{\canr_i}=0\,\,\mod\,\,\bbZ$, i.e.\ $\kappa_{\canr_i}=1$
for all $i=1,2,...,s$.
We then obtain easily from \cite[Prop.\ 3.23]{boev1}
the following

\begin{corollary}
Let $\lambda\in\DelNIo$ be irreducible. If there is an
$i=1,2,\ldots, s$ and a $\mu\in\DelNIo$ such that
$\kappa_{\canr_i}=1$, 
$\N {[\canr_i]}{[\lambda]}{[\mu]}\equiv
\la\mu,\canr_i\circ\lambda\ra_N\neq 0$ and
$h_\mu \neq h_\lambda \,\,\mod\,\,\bbZ$, then
$[\alap]\neq[\alam]$.
\lablth{channel}
\end{corollary}

Similarly we have

\begin{lemma}
Assume $\kappa_{\canr_i}=1$ for all $i=1,2,\ldots, s$
and let $\lambda,\mu\in\DelNIo$ be irreducible. Then
$\la\alap,\amum\ra_M=0$ if
$h_\mu \neq h_\lambda \,\,\mod\,\,\bbZ$.
\lablth{disj}
\end{lemma}

\bproof
Assume that $\la\alap,\amum\ra_M\neq 0$, i.e.\ there is a non-zero
intertwiner $t\in\Hom_M(\alap,\amum)$. It follows that
$t\lambda(n)=\mu(n)t$ for all $n\in N$,
hence $\can(t)\in\Hom_N(\canr\circ\lambda,\canr\circ\mu)$,
and we have $\eps\mu\canr\eps\canr\mu\can(t)=\can(t)$ by
Lemma \ref{pmpp}. It follows that
\[ \eps\mu\canr\eps\canr\mu \, \can(t)w=\can(t)w \,, \]
hence with isometries $w_i\in\Hom_N(\canr_i,\canr)$
such that $\canr(n)=\bigoplus_{i=1}^s w_i \canr_i(n) w_i^*$
for $n\in N$ (so that one may choose $w_1=w$) we obtain
\[ \eps\mu{\canr_i}\eps{\canr_i}\mu \, w_i^* \can(t) w =
 w_i^* \can(t) w \]
for all $i=1,2,...,s$. Now
$w_i^* \can(t) w\in\Hom_N(\lambda,\canr_i\circ\mu)$,
hence this is
\[ \frac{\kappa_\lambda}{\kappa_\mu} \,  w_i^* \can(t) w =
w_i^* \can(t) w \,.\]
Multiplying by $w_i$ from the left and summing over $i$
yields
\[ \frac{\kappa_\lambda}{\kappa_\mu} \, \can(t) w = \can(t) w \,,\]
and $\can(t)w\neq 0$ since $t\neq 0$ by \cite[Lemma 3.8]{boev1},
hence $\kappa_\lambda=\kappa_\mu$.
\eproof

\subsection{A relative braiding}

Representative endomorphisms of subsectors of
$[\alap]$ (or $[\amum]$) will not possess a braiding
since they do not even commute as sectors in general.
However, we have seen in \cite[Prop.\ 3.26]{boev1}
that if $\beta,\delta\in\End(M)$ are such that $[\beta]$
is a subsector of $[\alap]$ and $[\delta]$ is a subsector
of $[\amum]$ for some $\lambda,\mu\in\DelNIo$, then $[\beta]$
and $[\delta]$ commute, $[\beta\circ\delta]=[\delta\circ\beta]$,
and that a relevant unitary which we will now denote by
$\epsr\beta\delta$ is given by
\be
\epsr\beta\delta = s^* \amum(t^*) \, \eps \lambda\mu \, \alap(s)t
\in \Hom_M(\beta\circ\delta,\delta\circ\beta)
\labl{releps}
with isometries $t\in\Hom_M(\beta,\alap)$ and
$s\in\Hom_M(\delta,\amum)$.

\begin{lemma}
The operator $\epsr\beta\delta$ of \erf{releps} does not
depend on $\lambda,\mu\in\DelNIo$ and not on the isometries
$s,t$, in the sense that, if
there are isometries $x\in\Hom_M(\beta,\anup)$ and
$y\in\Hom_M(\delta,\arhom)$ with some $\nu,\rho\in\DelNIo$,
then
\be
\epsr\beta\delta = y^* \arhom(x^*) \, \eps \nu\rho \, \anup(y)x \,.
\ee
\lablth{invlm}
\end{lemma}

\bproof
If $x,y$ are as above then clearly $xt^*\in\Hom_M(\alap,\anup)$
and $sy^*\in\Hom_M(\arhom,\amum)$. Hence
$\amum(xt^*)\eps\lambda\mu=\eps\nu\mu xt^*$ and also
$sy^*\eps\nu\rho = \eps\nu\mu \anup (sy^*)$ by
\cite[Lemma 3.25]{boev1}, therefore
\[ \bearll
y^* \arhom(x^*) \, \eps \nu\rho \, \anup(y)x
&= s^*sy^* \arhom(x^*) \, \eps \nu\rho \, \anup(y)xt^*t \\[.4em]
&= s^* \amum(x^*) sy^* \, \eps \nu\rho \, \anup(y)xt^*t \\[.4em]
&= s^* \amum(x^*) \, \eps \nu\mu \, \anup(s)xt^*t \\[.4em]
&= s^* \amum(x^*) \, \eps \nu\mu \, xt^*\alap(s)t \\[.4em]
&= s^* \amum(t^*) \, \eps \lambda\mu \, \alap(s)t \,,
\eear \]
yielding the invariance.
\eproof

We will now strengthen the relative commutativity statement
of \cite[Prop.\ 3.26]{boev1} to the following

\begin{proposition}
The system of unitaries of \erf{releps} provides a relative
braiding between representative endomorphisms of subsectors
of $[\alap]$ and $[\amum]$ in the sense that, if
$\beta,\delta,\omega,\xi\in\End(M)$ are such that
$[\beta],[\delta],[\omega],[\xi]$ are subsectors of
$[\alap],[\amum],[\anup],[\arhom]$, respectively,
$\lambda,\mu,\nu,\rho\in\DelNIo$, then we have initial
conditions
\be
\epsr{\id_M}\delta=\epsr\beta{\id_M}=\bfe \,,
\labl{inirel}
composition rules
\be
\epsr {\beta\circ\omega}\delta = \epsr \beta\delta \,
\beta (\epsr \omega\delta) \,, \quad
\epsr \beta{\delta\circ\xi} = \delta (\epsr \beta\xi) \,
\epsr \beta\delta \,,
\labl{comprel}
and whenever $q_+\in \Hom_M(\beta,\omega)$
and $q_-\in \Hom_M(\delta,\xi)$ then
\be
\delta(q_+) \, \epsr \beta\delta = \epsr \omega\delta \, q_+ \,, \qquad
q_- \, \epsr \beta\delta = \epsr \beta\xi \, \beta (q_-) \,.
\labl{natrel}
\lablth{relbra}
\end{proposition}

\bproof
For $\beta=\id_M$ ($\delta=\id_M$) we are free to choose
$\lambda=\id_N$ ($\mu=\id_N$) and $t=\bfe$ ($s=\bfe$)
by Lemma \ref{invlm}, and then \erf{inirel} is obvious.
To show \erf{comprel} we first note that, if
$t\in\Hom_M(\beta,\alap)$ and $x\in\Hom_M(\omega,\anup)$
are isometries then
$\alap(x)t\in\Hom_M(\beta\circ\omega,\alpha^+_{\lambda\circ\nu})$
is an isometry. With an isometry $s\in\Hom_M(\delta,\amum)$ we
can therefore write
\[ \bearl
\epsr {\beta\circ\omega}\delta
= s^* \amum (t^*\alap(x^*)) \, \eps {\lambda\circ\nu}\mu \,
\alpha^+_{\lambda\circ\nu}(s) \alap(x)t \\[.4em]
= s^* \amum (t^*) \cdot \amum \circ \alap(x^*) \cdot
\eps \lambda \mu \, \lambda (\eps \nu\mu) \cdot
\alap\circ\anup(s) \cdot \alap(x)t \\[.4em]
= s^* \amum (t^*) \, \eps \lambda \mu \cdot
\alap \circ \amum (x^*) \cdot \lambda (\eps \nu\mu) \cdot
\alap\circ\anup(ss^*s) \cdot \alap(x)t \\[.4em]
= s^* \amum (t^*) \, \eps \lambda \mu \cdot
\alap \circ \amum (x^*) \cdot \alap (ss^*) \,
\lambda (\eps \nu\mu) \cdot
\alap\circ\anup(s) \cdot \alap(x)t \\[.4em]
= s^* \amum (t^*) \, \eps \lambda \mu \, \alap (ss^*)
\cdot \alap \circ \amum (x^*)
\cdot \lambda (\eps \nu\mu) \cdot
\alap\circ\anup(s) \cdot \alap(x)t \\[.4em]
= s^* \amum (t^*) \, \eps \lambda \mu \, \alap (s) t
\cdot \beta (s^* \amum (x^*) \, \eps \nu\mu \, \anup (s)x) \\[.4em]
= \epsr \beta\delta \, \beta (\epsr \omega\delta)    \,,
\eear \]
where we used \cite[Lemmata 3.24 and 3.25]{boev1}.
The proof for the second relation in \erf{comprel} is analogous.
Now let $q_+\in\Hom_M(\beta,\omega)$. Note that then
$xq_+t^*\in\Hom_M(\alap,\anup)$. Hence
\[ \bearll
\delta(q_+) \, \epsr \beta\delta 
&= s^*\amum(q_+)ss^*\amum(t^*) \, \eps \lambda\mu \,
\alap (s)t \\[.4em]
&= s^*\amum(q_+t^*) \, \eps \lambda\mu \, \alap (s)t \\[.4em]
&= s^*\amum(x^*x q_+t^*) \, \eps \lambda\mu \, \alap (s)t \\[.4em]
&= s^*\amum(x^*) \, \eps \nu\mu \, xq_+t^* \alap (s)t \\[.4em]
&= s^*\amum(x^*) \, \eps \nu\mu \, \anup (s)xq_+\\[.4em]
&= \epsr \omega\delta \, q_+ \,,
\eear \]
where we used \cite[Lemma 3.25]{boev1} again.
The proof for the second relation in \erf{natrel} is analogous.
\eproof

Now consider sectors $[\beta]$ that can be obtained by both
inductions, i.e.\ $[\beta]$ is a subsector of $[\alap]$
and $[\alam]$ for some $\lambda\in\DelNIo$,
cf.\ Lemma \ref{lamupm}. (And for a representative
$\beta\in\End(M)$ we can in fact use the same intertwining
isometry.) We easily obtain the following

\begin{corollary}
For the collection of endomorphisms $\beta,\delta\in\End(M)$
of that kind that $[\beta]$ is a subsector of both,
$[\alap]$ and $[\alam]$, and similarly $[\delta]$ is a
subsector of $[\amup]$ and $[\amum]$ for some (varying)
$\lambda,\mu\in\DelNIo$, the unitaries $\epsr \beta\delta$
and $\epsr \delta\beta ^*$ define a braiding.
\lablth{Tbraid}
\end{corollary}

Later we will use the following

\begin{lemma}
Let $\beta\in\End(M)$ such that $[\beta]$ is
a subsector of both, $[\alap]$ and $[\alam]$,
for some irreducible $\lambda\in\DelNIo$.
Let further $\mu\in\DelNIo$ such that
$[\amup]=[\amum]$, and let $\delta_i\in\End(M)$
such that $[\amupm]=\bigoplus_{i=1}^q[\delta_i]$.
Then if $\epsr \beta{\delta_i}=\epsr {\delta_i}\beta ^*$
for all $i=1,2,\ldots,q$, then
$\eps\lambda\mu \eps\mu\lambda=\bfe$.
\lablth{monlm=1}
\end{lemma}

\bproof
First note that if $t\in\Hom_M(\beta,\alapm)$
is an isometry then
$\sib(n)=\can(t)^*\cdot\canr\circ\lambda(n)\cdot\can(t)$
for all $n\in N$, and by extending this formula to
$n\in\cN$ we can consider $\sib\in\DelNIo$. Note that
for our isometry $v\in\Hom_M(\id_M,\can)$ we also
have $v\in\Hom_M(\beta,\asibpm)$ by, for instance,
Lemma \ref{key'}. Let $s_i\in\Hom_M(\delta_i,\amupm)$,
$i=1,2,...,q$, be isometries generating a Cuntz algebra
(recall $\amup=\amum$ by \cite[Prop.\ 3.23]{boev1}).
Then
$\epsr \beta{\delta_i}=\epsr {\delta_i}\beta ^*$
yields
\[ s_i^* \amum (v^*) \, \eps \sib\mu \,
\asibp(s_i) vv^* \asibm (s_i^*) \, \eps \mu\sib \,
\amup(v)s_i = \bfe \,. \]
Since we can switch the $\pm$-signs as $\amup=\amum$
and $\asibp(s_i)v=v\beta(s_i)=\asibm(s_i)v$ we obtain
by left multiplication by $\amupm(v)s_i$ and by use
of \cite[Lemma 3.25]{boev1}
\[ \eps \sib\mu \eps \mu\sib \, \amup(v)s_i = \amup(v)s_i\,,\]
and we obtain 
$\eps \sib\mu \eps \mu\sib \amup(v) = \amup(v)$
by right multiplication by $s_i^*$ and summation over $i$.
Now recall $\amup(v)=\eps\mu\canr ^* v$, and therefore
we obtain $\eps \sib\mu \eps \mu\sib = \bfe$
by \cite[Lemma 3.8]{boev1}. Now
$\la\lambda,\sib\ra_N=\la\alapm,\beta\ra_M\neq 0$
by Prop.\  \ref{genmain},
hence $[\lambda]$ is a subsector of $[\sib]$,
and hence $\eps \lambda\mu \eps \mu\lambda = \bfe$
by Lemma \ref{decmon}.
\eproof

For the rest of this subsection we assume that the
enveloping net $\cM$ is Haag dual. Let $\beta\in\DelMIo$
where $\DelMIo$ denotes the set of localized transportable
endomorphisms, localized in $\Io$.
Then $\sib\in\DelNIo$, in particular if
$\Qbpm\in\cM$ and $\ucpm\in\cN$ are unitaries such that
$\beta_\pm=\Ad(\Qbpm)\circ\beta\in\Delta_\cM(I_\pm)$
and $\canr_\pm=\Ad(\ucpm)\circ\canr\in\Delta_\cN(I_\pm)$
with intervals $I_+,I_-\in\Jz$ lying in the right respectively
left complement of $\Io$, one checks easily
(cf.\ \cite[Lemma 3.18]{boev1})
\[ \sigma_{\beta,\pm} = \Ad (\ucpm \can(\Qbpm)) \circ\sib
\in \Delta_\cN (I_\pm) \,, \]
so that (cf.\ \cite[Lemma 3.19]{boev1})
\[ \epspm \sib\canr = \can^2 (\Qbmp)^* \epspm \canr\canr
\can (\Qbmp) \,. \]
Now $[\beta]$ is a subsector of both, $[\asibp]$ and $[\asibm]$,
in particular we have $v\in\Hom_M(\beta,\asibpm)$.
Note that we therefore find
\[ \can(\Qbp^*) vv \Qbp = \can(\Qbm^*) vv \Qbm
= v\beta(v) = \asibp (v)v = \asibm (v)v \,, \]
since $v\beta(v)=v\Qbpm^* \beta_\pm (v)\Qbpm=
v\Qbpm^* v\Qbpm = \can(\Qbpm^*) vv \Qbpm$. 
Now let also $\delta\in\DelMIo$ and choose unitaries
$\Qdpm\in\cM$ such that
$\Ad(\Qdpm)\circ\delta\in\Delta_\cM(I_\pm)$. Putting
$u_{\sid,\pm}=\ucpm \can(\Qdpm)$ we can write
\[ \bearll
\epspm \sib\sid &= u_{\sid,\pm}^* \sib (u_{\sid,\pm})
=  \can (\Qdpm^*) \, \ucpm^* \sib(\ucpm) \cdot
\can\circ\beta\circ\can(\Qdpm) \\[.4em]
&= \can (\Qdpm^*) \, \epspm \sib\canr \cdot
\can\circ\beta\circ\can(\Qdpm) \,.
\eear \]
Endomorphisms in $\DelMIo$ are clearly braided and the
statistics operators are given by
\[ \epspm \beta\delta = \Qdpm^* \beta (\Qdpm) \,. \]
We now have the following

\begin{proposition}
Assume that $\cM$ is Haag dual. For $\beta,\delta\in\DelMIo$
we have
\be
\epsr \beta\delta = \epsp \beta\delta \,, \qquad
\epsr \delta\beta ^* = \epsm \beta\delta \,.
\ee
\lablth{oreps}
\end{proposition}

\bproof
Since $v\in\Hom_M(\beta,\asibpm)$ and $v\in\Hom_M(\delta,\asidpm)$,
we can write
\[ \bearll
\epsr \beta\delta &= v^* \asidm (v^*) \, \eps \sib\sid \,
\asibp (v)v \\[.4em]
&= \Qdp^* v^* v^* \can(\Qdp) \, \eps \sib\sid \,
v\beta(v) \\[.4em]
&= \Qdp^* v^* v^* \, \eps \sib\canr \cdot
\can\circ\beta\circ\can(\Qdp) \cdot v\beta(v) \\[.4em]
&= \Qdp^* v^* v^* \, \eps \sib\canr \, v\beta(v) \,
\beta (\Qdp) \\[.4em]
&= \Qdp^* v^* v^* \can^2(\Qbm^*) \eps \canr\canr
\can(\Qbm) \, v\beta(v) \, \beta (\Qdp) \\[.4em]
&= \Qdp^* \Qbm^* v^* v^* \eps \canr\canr \, v
\Qbm \beta(v) \, \beta (\Qdp) \\[.4em]
&= \Qdp^* \Qbm^* v^* v^* \eps \canr\canr \, vv \,
\Qbm \, \beta (\Qdp) \\[.4em]
&= \Qdp^* \beta (\Qdp) \equiv \epsp \beta\delta \,,
\eear \]
where we used the locality relation $\eps\canr\canr v^2=v^2$
from  \cite{lore} (or \cite[Lemma 3.4]{boev1}).
The second relation follows from
$\epsm \beta\delta = \epsp \delta\beta ^*$.
\eproof

\subsection{Subsectors of $[\can]$ and $[\alap\circ\amum]$}

Let $\iota\in\Mor(N,M)$ be the injection map from $N$ into
$M$ and recall that $\iotab\in\Mor(M,N)$ given by
$\iotab(m)=\can(m)\in N$ for $m\in M$ is a conjugate,
see Subsect.\ \ref{preizu}. We first note a simple fact.

\begin{lemma}
We have $\la\can,\can\ra_M=\la\canr,\canr\ra_N$.
\lablth{gg=tt}
\end{lemma}

\bproof
This is just because we have $[\can]=[\iota\circ\iotab]$ as a
sector of $M$ and $[\canr]=[\iotab\circ\iota]$ as a sector of
$N$. Hence
\[ \la\can,\can\ra_M = \la\iota\circ\iotab,
\iota\circ\iotab\ra_M=
\la\iota,\iota\circ\iotab\circ\iota\ra_{N,M}
=\la\iotab\circ\iota,\iotab\circ\iota\ra_N
=\la\canr,\canr\ra_N  \]
by Frobenius reciprocity.
\eproof

The extension property of $\alpha$-induction,
$\alapm(n)=\lambda(n)$ for all $n\in N$, $\lambda\in\DelNIo$,
can be written as $\alapm\circ\iota=\iota\circ\lambda$
as morphisms in $\Mor(N,M)$.
Now recall
$\la \alapm , \amupm \ra_M = \la\canr\circ\lambda,\mu\ra_N$
by \cite[Thm.\ 3.9]{boev1} .

\begin{lemma}
For any $\lambda,\mu\in\DelNIo$ we have
\be
\la \alapm,\iota\circ\mu\circ\iotab \ra_M =
\la\canr\circ\lambda,\mu\ra_N \equiv
\la \alapm, \amupm \ra_M\,.
\ee
\lablth{alaimuib}
\end{lemma}

\bproof
We compute
\[ \la \alapm,\iota\circ\mu\circ\iotab \ra_M
= \la \alapm\circ\iota , \iota\circ\mu \ra_{N,M}
= \la \iota\circ\lambda , \iota\circ\mu \ra_{N,M}
= \la \iotab\circ\iota\circ\lambda , \mu \ra_N \]
by Frobenius reciprocity.
\eproof

Taking $\mu$ to be trivial we immediately obtain the following

\begin{corollary}
Only the identity sector $[\id_M]$ can be a common subsector
of $[\alapm]$ and $[\can]$ for $\lambda\in\DelNIo$. 
\lablth{idonly}
\end{corollary}

Now assume $d_\mu<\infty$ and let
$\mub\in\DelNIo$ be a conjugate of $\mu\in\DelNIo$.
Then $[\amubm]=[\co{\amum}]$ by \cite[Lemma 3.14]{boev1},
and therefore we find
\[ \la\alap\circ\amubm,\can\ra_M 
= \la\alap,\amum\circ\can \ra_M
= \la\alap,\iota\circ\mu\circ\iotab \ra_M
= \la\alap,\amup\ra_M \,. \]
Thus we have the following

\begin{corollary}
For $\lambda,\mu\in\DelNIo$, $d_\mu<\infty$, and
$\mub\in\DelNIo$ a conjugate of $\mu$ we have
$\la\alap\circ\amubm,\can\ra_M = \la\alap,\amup\ra_M$.
In particular, if $[\amupm]$ is irreducible then
$[\amup\circ\amubm]$ has one irreducible subsector in
common with $[\can]$ which cannot be the identity if
$[\amup]\neq[\amum]$.
\lablth{congam}
\end{corollary}

Recall that $[\amup]\neq[\amum]$ if and only if
the monodromy $\eps\mu\canr\eps\canr\mu$ is non-trivial
\cite[Prop.\ 3.23]{boev1}.

We now further investigate subsectors of mixed products
$[\alap\circ\amum]$.
Recall that subsectors of $[\amup]$ commute with $[\alap]$,
$\lambda,\mu\in\DelNIo$, by \cite[Prop.\ 3.16]{boev1}.
We will now generalize this result.

\begin{lemma}
Let $\beta\in\End(M)$ such that $[\beta]$ is a subsector of
$[\amup\circ\anum]$ for some $\mu,\nu\in\DelNIo$. Then
$[\alapm\circ\beta]=[\beta\circ\alapm]$ for any
$\lambda\in\DelNIo$.
\lablth{abelpm}
\end{lemma}

\bproof
For any $\lambda,\mu\in\DelNIo$ we have by
(the plus- and minus-version of) \cite[Cor.\ 3.11]{boev1}
\[ \epspm\lambda\mu \cdot \alapm\circ\amupm (m) =
\amupm\circ\alapm (m) \cdot \epspm\lambda\mu  \]
and similarly by \cite[Lemma 3.24]{boev1}
\[ \epspm\lambda\mu \cdot \alapm\circ\amump (m) =
\amump\circ\alapm (m) \cdot \epspm\lambda\nu \,.\]
Recall
$\epspm \lambda{\mu\circ\nu} = \mu(\epspm\lambda\nu)\epspm\lambda\mu$
for $\lambda,\mu,\nu\in\DelNIo$. Therefore
\[ \bearll
\epspm \lambda{\mu\circ\nu} \cdot \alapm\circ\amup\circ\anum (m) \!\!\!
&= \mu(\epspm\lambda\nu) \cdot \amup\circ\alapm\circ\anum(m) \cdot
\epspm \lambda\nu \\[.4em]
&= \amup\circ\anum\circ\alapm (m) \cdot \epspm \lambda{\mu\circ\nu}
\eear \]
for all $m\in M$. By assumption, there is an isometry
$t\in\Hom_M(\beta,\amup\circ\anum)$. Hence
\[ \bearl
t^* \epspm\lambda{\mu\circ\nu} \alapm(t) \cdot
\alapm\circ\beta (m) =\\[.4em]
\qquad\qquad\qquad \bearl=
t^* \epspm\lambda{\mu\circ\nu} \cdot
\alapm\circ\amup\circ\anum(m) \cdot \alapm(t) \\[.4em]
= t^* \cdot \amup\circ\anum\circ\alapm(m) \cdot
\epspm\lambda{\mu\circ\nu} \alapm(t) \\[.4em]
= \beta\circ\alapm(m) \cdot t^* \epspm\lambda{\mu\circ\nu} \alapm(t)
\eear \eear\]
for all $m\in M$. It remains to be shown that
$u=t^* \epspm\lambda{\mu\circ\nu} \alapm(t)$ is unitary. Note that
$tt^*\cdot\mu\circ\nu(n)=\mu\circ\nu(n) \cdot tt^*$ for all
$n\in N$, hence
\[ tt^* \epspm\lambda{\mu\circ\nu} = \epspm\lambda{\mu\circ\nu}
\alapm (tt^*) \]
by \cite[Lemma 3.25]{boev1}. With this relation one checks
easily that $u^*u=uu^*=\bfe$.
\eproof

\section{\sloppy Induction-Restriction Graphs and
$\gamma$-Mul\-ti\-pli\-ca\-tion Graphs}

In this section we will relate $\alpha$-induction to
basic invariants of the subfactor $N\subset M$ (and
hence to each local subfactor $N(I)\subset M(I)$,
$I\in\Jz$, since the choice of the interval $\Io$
was arbitrary), namely the principal graph and the
dual principal graph. In our applications these results
can be used to determine the graphs for several examples.

\subsection{$\alpha$-induction and (dual) principal graphs}

Choose a sector basis
$\cW\subset\LTSN$ which contains (at least) all the
irreducible subsectors of $[\canr]$. Since a sector basis
is by definition finite and closed under products
(after irreducible decomposition) such a choice is
possible if and only if the subfactor $N\subset M$
has finite depth. A representative
endomorphism of $\Lambda\in\cW$ is denoted by
$\lambda_\Lambda$, $[\lambda_\Lambda]\equiv\Lambda$.
We define the chiral induced sector bases
$\cV^+,\cV^-\subset\Sect(M)$ to be the
sector bases given by all irreducible subsectors of
$[\aLap]$, $[\aLam]$, $\Lambda\in\cW$, respectively,
$\aLapm\equiv\alpha^\pm_{\lambda_\Lambda}$.
We denote representative endomorphisms of $a\in\cV^+$
by $\beta^+_a$, $[\beta^+_a]\equiv a$.
If for $a\in\cV^+$ we have $\beta^+_a(m)=t^*\aLap(m)t$,
$m\in M$, for some $\Lambda\in\cW$ and some isometry
$t\in M$, we denote by $\beta^-_a\in\End(M)$ the endomorphism
given by $\beta^-_a(m)=t^*\aLam(m)t$, $m\in M$. Note
that then $\cV^-=\{[\beta^-_a]\,,\,\,a\in\cV^+\}$ by
Corollary \ref{relpm}. Also note that
$\beta^+_a(n)=\beta^-_a(n)$ for all $n\in N$, thus
$\beta^+_a\circ\iota=\beta^-_a\circ\iota$ for the
injection map $\iota$.
Now let $\co\cY\subset\Sect(N,M)$ be the set of all irreducible
subsectors of $M$-$N$ sectors $[\iota\circ\lambda_\Lambda]$,
$\Lambda\in\cW$.

\begin{lemma}
We have $\co\cY=\{[\beta^\pm_a\circ\iota]\,,\,\,\,a\in\cV^+\}$.
\lablth{YisoVpm}
\end{lemma}

\bproof
For a representative endomorphism $\beta^\pm_a\in\End(M)$ for
$a\in\cV^+$ we have, by definition, some $\Lambda\in\cW$
and some isometry $t\in\Hom_M(\beta^\pm_a,\aLapm)$.
Put $\tau_a=\beta^\pm_a\circ\iota\in\Mor(N,M)$. Note that the
definition does not depend on the $\pm$-sign. Then
$t\tau_a(n)=t\beta_a(n)=\lambda_\Lambda(n)t$ for all $n\in N$,
hence $[\tau_a]$ is a subsector of $[\iota\circ\lambda_\Lambda]$.
Moreover, as
$\Hom_{N,M}(\tau_a,\tau_b)=\Hom_M(\beta^\pm_a,\beta^\pm_b)$
for $a,b\in\cV^+$ by Lemma \ref{key'}, the $[\tau_a]$'s
are irreducible and $[\tau_a]=[\tau_b]$ if and only if
$[\beta^\pm_a]=[\beta^\pm_b]$, i.e.\ $a=b$.

Conversely, let $[\tau]\in\co\cY$, i.e.\ $[\tau]$ is an irreducible
$M$-$N$ sector and there is some $\Lambda\in\cW$ and an
isometry $t\in\Hom_{N,M}(\tau,\iota\circ\lambda_\Lambda)$.
Hence $tt^*\in\lambda_\Lambda(N)'\cap M=\aLapm(M)'\cap M$,
and therefore putting $\beta^\pm_\tau(m)=t^*\aLapm(m)t$, $m\in M$,
defines $\beta^\pm_\tau\in\End(M)$ fulfilling
$\beta^\pm_\tau\circ\iota=\tau$. As
$\beta^\pm_\tau(M)'\cap M=\beta^\pm_\tau(N)'\cap M
=\tau(N)'\cap M=\bbC\bfe$ by Lemma \ref{key'} we find
that $\beta^\pm_\tau$ is irreducible, thus
$[\beta^\pm_\tau]\in\cV^\pm$. Similarly,
$[\beta^\pm_\tau]=[\beta^\pm_{\tau'}]$ if and only if
$[\tau]=[\tau']$ for $[\tau']\in\co\cY$.
\eproof

From now on, we use the notation
$[\tau_a]=[\co{\beta^\pm_a}\circ\iota]\in\co\cY$ for $a\in\cV^+$.
This makes sense since $\cV^+$ (and $\cV^-$)
is closed under conjugation.
We also denote $[\rho_a]=[\co{\tau_a}]=[\iotab\circ\beta^\pm_a]$,
$a\in\cV^+$, and define the set
$\cY=\{[\rho_a]\,,\,\,a\in\cV^+\}$.
Next we define $\tilde{\cV}\subset\Sect(M)$ to be the
set of all irreducible subsectors of some
$[\beta^\pm_a\circ\can]$, $a\in\cV^+$,
and also this definition is obviously independent
of the $\pm$-sign. We denote representative endomorphisms of
$x\in\tilde{\cV}$ by $\beta_x$, $[\beta_x]\equiv x$.
Clearly $[\id_M]\in\tilde{\cV}$.

We can now draw a bipartite graph as follows. We label the even
vertices by the elements of $\cW$ and the odd vertices by
the elements of $\cY$. We connect any even vertex
labelled by $\Lambda\in\cW$ with any odd vertex labelled
by $[\rho_a]$, $a\in\cV^+$, by
$\la\lambda,\sigma_{\beta^\pm_a}\ra_N$
edges. Due to Prop.\ \ref{genmain} we call the
(possibly disconnected) graph obtained this way
the {\em induction-restriction graph}.
We can draw another bipartite graph as follows. We label the
even vertices by the elements of $\tilde{\cV}$ and the odd vertices
by the elements of $\co\cY$. We connect any even vertex
labelled by $x\in\tilde{\cV}$ with any odd vertex labelled
by $[\tau_a]$, $a\in\cV^+$, by
$\la\beta_x,\co{\beta^\pm_a}\circ\can\ra_M$
edges. We call the (possibly disconnected) graph obtained this way
the {\em $\gamma$-multiplication graph}.

\begin{theorem}
The principal graph of the inclusion $N\subset M$ is given by the
connected component of $[\id_N]\in\cW$ of the induction-restriction
graph. The dual principal graph is given by the connected component
of $[\id_M]\in\tilde{\cV}$ of the $\gamma$-multiplication graph.
\lablth{pdpg}
\end{theorem}

\bproof
Note that $\cP_0$, defined in Subsect.\ \ref{preizu}, is
contained in $\cW$ since it is closed under reduction
of products and contains the irreducible subsectors of $[\canr]$.
As $\co\cY$ is the set of irreducible subsectors of
$[\iota\circ\lambda_\Lambda]$, it follows that $\cY$
is the set of irreducible subsectors of
$[\co{\lambda_\Lambda}\circ\iotab]$, $\Lambda\in\cW$.
Since $\cW$ is closed under conjugation, it follows in
particular that $\cP_1\subset\cY$. Recall that
the elements of $\cY$ are of the form
$[\rho_a]=[\iotab\circ\beta^\pm_a]$.
Now for $\Lambda\in\cW$ and $a\in\cV^+$ we have
\[ \la\lambda_\Lambda\circ\iotab,\rho_a\ra_{M,N}
= \la\lambda_\Lambda\circ\iotab,\iotab\circ\beta^\pm_a\ra_{M,N}
= \la\lambda_\Lambda,\iotab\circ\beta^\pm_a\circ\iota\ra_N
= \la\lambda_\Lambda,\sigma_{\beta^\pm_a}\ra_N \]
by Frobenius reciprocity, therefore the induction-restriction
graph has the principal graph as a subgraph. This must
be the connected component of $[\id_N]\in\cP_0$.

Similarly, as $\cP_1\subset\cY$, we have $\cD_1\subset\co\cY$.
Since any subsector in $\cD_0$ can be obtained by
decomposing sectors $[\tau\circ\iotab]$, $[\tau]\in\cD_1$,
we find that also $\cD_0\subset\tilde{\cV}$. Now for
$x\in\tilde{\cV}$ and $a\in\cV^+$ we have
\[ \la\beta_x\circ\iota,\tau_a\ra_{N,M}=\la\beta_x\circ\iota,
\co{\beta^\pm_a}\circ\iota\ra_{N,M}=
\la\beta_x,\co{\beta^\pm_a}\circ\can\ra_M \,. \]
Hence the connected component of $[\id_M]\in\tilde{\cV}$
is the dual principal graph.
\eproof

Now let $\cV^\pm_0\subset\cV^\pm$ be the subset of those
sectors $[\beta^\pm_a]$ such that
$[\beta^+_a\circ\iota]=[\beta^-_a\circ\iota]$ appears
(as a label of some odd vertex) in the dual principal graph.
As $[\can]$ possesses the identity sector as a subsector,
$[\beta^+_a\circ\can]=[\beta^-_a\circ\can]$ contains
$[\beta^+_a]$ and, if different, also $[\beta^-_a]$
as a subsector. Recall that a sector algebra associated
to a sector basis is the vector space with the sector basis
as a basis endowed with the sector operations as algebraic
structure. From Theorem \ref{pdpg} we now obtain immediately
the following

\begin{corollary}
Elements of $\cV^+_0\cup\cV^-_0$ appear as labels of the
even vertices of the dual principal graph. Therefore, as
sector algebras, the algebra of $M$-$M$ sectors of the dual
principal graph possesses two subalgebras corresponding to the
sector bases $\cV^+_0$ and $\cV^-_0$ (which may be identical).
\lablth{subdual}
\end{corollary}

\subsection{Global indices}

It is known that the $N$-$N$, $N$-$M$, $M$-$N$ and $M$-$M$
bimodules arising from a subfactor $N\subset M$ and labelling
the vertices of the principal and dual principal graph have the
same global indices. Here we mean by global index the sum over
the squares of the Perron-Frobenius weights which correspond
to the statistical dimensions in the sector context.
We will now show that an analogous
statement holds for the sectors labelling the vertices of
the (possibly larger) induction-restriction and
$\gamma$-multiplication graphs. We denote
$d_\Lambda\equiv d_{\lambda_\Lambda}$, $\Lambda\in\cW$,
$d_a\equiv d_{\beta^\pm_a}$, $a\in\cV^+$,
$d_x\equiv d_{\beta_x}$, $x\in\tilde{\cV}$, and
define global indices
\[ [[\cW]]=\sum_{\Lambda\in\cW} d_\Lambda^2 \,,\qquad
[[\cV^\pm]]=\sum_{a\in\cV^+} d_a^2 \,,\qquad
[[\tilde{\cV}]]=\sum_{x\in\tilde{\cV}} d_x^2 \,.\]
Recall $d_\can=[M:N]$.

\begin{lemma}
We have $[[\cW]]=d_\gamma [[\cV^\pm]]$.
\lablth{W=dV}
\end{lemma}

\bproof
We define a rectangular matrix $P$ by
\[ P_{a,\Lambda} = \la \sigma_{\beta^\pm_a},\lambda_\Lambda 
\ra_N \,,\qquad a\in\cV^+\,,\quad \Lambda\in\cW \,. \]
We then have
$[\sigma_{\beta^\pm_a}]=\bigoplus_{\Lambda\in\cW} P_{a,\Lambda}
[\lambda_\Lambda]$, hence 
$d_\can d_a =\sum_{\Lambda\in\cW} P_{a,\Lambda} d_\Lambda$.
As then $P_{a,\Lambda}=\la \beta^\pm_a\circ\iota,
\iota\circ\lambda_\Lambda \ra_{N,M}$ by Frobenius reciprocity,
and all irreducible subsectors of $[\iota\circ\lambda_\Lambda]$
are of the form $[\beta^\pm_a\circ\iota]$, $a\in\cV^+$,
we have similarly
$[\iota\circ\lambda_\Lambda]=\bigoplus_{a\in\cV^+} P_{a,\Lambda}
[\beta^\pm_a\circ\iota]$, hence 
$d_\Lambda =\sum_{a\in\cV^+} P_{a,\Lambda} d_a$.
Therefore
\[ [[\cW]] = \sum_{\Lambda\in\cW} d_\Lambda^2 =
\sum_{\Lambda\in\cW} \sum_{a\in\cV^+} d_\Lambda P_{a,\Lambda} d_a
= \sum_{a\in\cV^+} d_\can d_a^2 = d_\can [[\cV^\pm]] \,, \]
and so we are done.
\eproof

\begin{lemma}
We have $[[\cW]]=[[\tilde{\cV}]]$.
\lablth{W=tV}
\end{lemma}

\bproof
We define a rectangular matrix $D$ by
\[ D_{a,x} = \la \beta^\pm_a\circ\can , \beta_x
\ra_N \,,\qquad a\in\cV^+\,,\quad x\in\tilde{\cV} \,. \]
Since then
$[\beta^\pm_a\circ\can]=\bigoplus_{x\in\tilde{\cV}} D_{a,x}
[\beta_x]$ we find $d_\can d_a =\sum_{x\in\tilde{\cV}} D_{a,x} d_x$.
As $[\beta_x]\in\tilde{\cV}$ is (by definition)
a subsector of $\beta^\pm_b\circ\can$ for some $b\in\cV^+$,
we find that $[\beta_x\circ\iota]$ is a subsector of
\[ \bearll [\beta^\pm_b\circ\can\circ\iota]=
[\beta^\pm_b\circ\iota\circ\canr] &=
\bigoplus_{c\in\cV^+} \la\canr,\sigma_{\beta^\pm_c}\ra_N
[\beta^\pm_b\circ\beta^\pm_c\circ\iota] \\[.4em]
&=\bigoplus_{a,c\in\cV^+} \la\canr,\sigma_{\beta^\pm_c}\ra_N
\la\beta^\pm_b\circ\beta^\pm_c,\beta^\pm_a\ra_M
[\beta^\pm_a\circ\iota] \,,
\eear \]
hence $[\beta_x\circ\iota]$ decomposes only into
elements of $\co\cY$. Therefore
$[\beta_x\circ\iota]=\bigoplus_{a\in\cV^+} D_{a,x}
[\beta^\pm_a\circ\iota]$, hence 
$d_x =\sum_{a\in\cV^+} D_{a,x} d_a$.
Therefore
\[ [[\tilde{\cV}]] = \sum_{x\in\tilde{\cV}} d_x^2 =
\sum_{x\in\tilde{\cV}} \sum_{a\in\cV^+} d_x D_{a,x} d_a
= \sum_{a\in\cV^+} d_\can d_a^2 = d_\can [[\cV^\pm]] \,, \]
and the statement now follows from Lemma \ref{W=dV}.
\eproof

Let $\cV$ be the set of irreducible subsectors of
$[\aLap\circ\aLams]$, $\Lambda,\Lambda'\in\cW$. As the maps
$[\lambda_\Lambda]\mapsto[\aLapm]$ are multiplicative,
conjugation preserving and $[\aLap]$ and $[\aLams]$ commute,
$\cV$ must be in fact a sector basis and we call it
the full induced sector basis.

\begin{lemma}
We have $\cV\subset\tilde{\cV}$.
\lablth{VintV}
\end{lemma}

\bproof
Let $[\beta]$ be an irreducible subsector
of $[\aLap\circ\aLams]$ for
some $\Lambda,\Lambda'\in\cW$. Then there are sectors
$a,b\in\cV^+$ such that
$[\beta]$ is a subsector of $[\beta^+_a\circ\beta^-_b]$,
hence of
\[ [\beta^+_a\circ\beta^-_b\circ\can] = 
[\beta^+_a\circ\beta^+_b\circ\can] = \bigoplus_{c\in\cV^+}
\la \beta^+_a\circ\beta^+_b, \beta^+_c \ra_M
[\beta^+_c\circ\can] \,, \]
and therefore $[\beta]$ must be a subsector of
$[\beta^+_c\circ\can]$ for some $c\in\cV^+$.
\eproof

Since $\cV\supset\cV^\pm$ is a sector basis,
we have equality $\cV=\tilde{\cV}$ if and only if
each irreducible subsector of $[\can]$ is in $\cV$. This
is not the case in general but we will find this situation
in our conformal field theory examples. The point is that
then the $\alpha$-induction machinery provides useful
methods to compute, besides the principal graphs, the
dual principal graphs of conformal inclusion subfactors.

\section{Two Inductions and Modular Invariants}

Before turning to the concrete examples, we will now
discuss the application of $\alpha$-induction to certain
conformal or orbifold embeddings involving $\SUn$, and we
relate the $\pm$-inductions to the entries of the
corresponding modular invariant mass matrix $Z$.

\subsection{Some equivalent conditions}

Let us now consider the more specific situation as already
treated in \cite{boev2}, namely that the net of subfactors
$\cN\subset\cM$ arises from a conformal or orbifold inclusion
of $\SUn$. We extend the orbifold analysis from the $\bbZ_n$
case of \cite{boev2} to the $\bbZ_m$ case, where $m$ is any
divisor of $n$ since there are also associated
type \nolinebreak I modular invariants;
that these inclusions also lead to suitable
nets of subfactors will be shown in Subsect.\ \ref{morbn}.
For later reference, we now also include the
conformal inclusions $\SUn_k\otimes\SUm_\ell\subset G_1$,
with $G$ some simple Lie group,
in our discussion. Let $Z_{\Lambda,\Lambda'}$
denote the entries of the mass matrix of the corresponding 
modular invariant. Here $\Lambda$ denotes weights in the
Weyl alcove $\ASU nk$ in the former case, and in the
latter case it labels pairs of
weights, denoted $\Lambda=(\dLam,\ddLam)$, with
$\dLam\in\ASU nk$ and $\ddLam\in\ASU m\ell$. Therefore
we are dealing with a fusion algebra $(\cW,W)$ in $\DelNIo$,
where $\cW=\{[\lambda_\Lambda]\}$.
We sometimes identify $\cW$ with its labelling set
$\ASU nk$ or $\ASU nk \times \ASU m\ell$.
As usual, we write
$\aLapm\equiv\alpha^\pm_{\lambda_\Lambda}$.
We obtain two chiral induced sector bases $\cV^\pm$
given by all irreducible subsectors of the $[\aLapm]$'s.
Further, we obtain the full induced sector basis
$\cV$ by taking all the irreducible subsectors
of $[\aLap\circ\aLams]$. Clearly, $\cV^\pm\subset\cV$.
We denote representative endomorphisms for $x\in\cV$ by
$\beta_x$ so that we may identify $[\beta_x]\equiv x$.

Let us now define a matrix $\tilde{Z}$ by
\[ \tilde{Z}_{\Lambda,\Lambda'} = \la \aLap , \aLams \ra_M \,,
\qquad \Lambda,\Lambda'\in\cW \,. \]
We remark that Lemma \ref{disj} states $T$-invariance of this matrix.
Let $M_y$ be the sector product matrices $M_y$ of $(\cV,V)$, with
\[ (M_y)_{x,z}\equiv M_{x,y}^z = \la \beta_x \circ \beta_y,
\beta_z \ra_M \,, \qquad x,y,z\in\cV\,. \]
We define a collection of matrices $R^{\Lambda,\Lambda'}$,
$\Lambda,\Lambda'\in\cW$ by
\[ R^{\Lambda,\Lambda'}_{x,y} = \la \beta_x \circ \aLap \circ \aLams,
\beta_y \ra_M \,, \qquad x,y\in\cV \,. \]
First note that
$\tilde{Z}_{\Lambda,\Lambda'}=R^{\Lambda,\overline{\Lambda'}}_{0,0}$.
As $[\aLap]$ and $[\aLams]$ commute with $[\beta_x]$,
the matrices $R^{\Lambda,\Lambda'}$ commute with $M_x$,
$\Lambda,\Lambda'\in\cW$, $x\in\cV$. 
It follows from the homomorphism property of $\alpha$-induction
that
\[ R^{\Lambda,\Lambda'} \, R^{\Omega,\Omega'} =
\sum_{\Phi,\Phi'\in\cW} \N \Lambda\Omega\Phi
\N {\Lambda'}{\Omega'}{\Phi'} \cdot R^{\Phi,\Phi'} \,,\qquad
\Lambda,\Lambda',\Omega,\Omega' \in \cW \,, \]
where the $N$'s are the fusion coefficients in $W$.
Thus these matrices constitute a representation of the
fusion algebra $W\otimes W$ and hence must decompose into
its characters $\gamma_{\Phi_1}\otimes\gamma_{\Phi_2}$, where
$\gamma_{\Phi_\epsilon}(\Lambda)=
S_{\Lambda,\Phi_\epsilon}/S_{0,\Phi_\epsilon}$,
$\Lambda,\Phi_\epsilon\in\cW$, $\epsilon=1,2$.
Here $S$ is the S-matrix of the $\SUn_k$
(or the $\SUn_k\otimes\SUm_\ell$) theory,
implementing the modular transformations of the conformal
characters and diagonalizing the sector fusion rules
at the same time by Wassermann's result \cite{wass3}.
Similar to the procedure
in \cite[Subsect.\ 4.2]{boev2} we conclude that
there is an orthonormal basis $\{\xi^i:i=0,1,....,D-1\}$,
where $D=|\cV|$, with components $\xi^i_x\in\bbC$,
indexed by $x\in\cV$, such that
\[ R^{\Lambda,\Lambda'}_{x,y} = \sum_{i=0}^{D-1}
\frac{S_{\Lambda,\Phi_1(i)}}{S_{0,\Phi_1(i)}} \cdot
\frac{S_{\Lambda',\Phi_2(i)}}{S_{0,\Phi_2(i)}} \cdot
\xi^i_x (\xi^i_y)^* \]
with a map
$\Phi:i\mapsto (\Phi_1(i),\Phi_2(i)) \in \cW\times\cW$.
We have $S^*ZS=S$ by modular invariance,
hence in particular $(S^*ZS)_{0,0}=1$.
By $d_\Lambda$ and $d_x$ we denote the
statistical (or ``quantum'') dimension of
$\lambda_\Lambda$, $\Lambda\in\cW$, and $\beta_x$,
$x\in\cV$, respectively. We have in particular
$d_\Lambda=S_{\Lambda,0}/S_{0,0}$.

Let us now concentrate on the conformal inclusion case.
As in \cite{boev1}, we denote by $(T,\cT)$ the fusion algebra
corresponding to the (level one) \per s of the ambient theory.
We know from $\alpha\sigma$-reciprocity that
\[  \cT \subset \cV^+ \cap \cV^- \,. \]
Recall that
$Z_{\Lambda,\Lambda'}=\sum_{t\in\cT} b_{t,\Lambda} b_{t,\Lambda'}$,
where $b_{t,\Lambda}=\la\lambda_\Lambda,\sigma_{\beta_t}\ra_N$
are the restriction coefficients.

\begin{proposition}
For conformal inclusions we have
$Z_{\Lambda,\Lambda'}\le \la \aLap,\aLams \ra_M$,
$\Lambda,\Lambda'\in\cW$ and
$\sum_{x\in\cV} d_x^2 \le \sum_{\Lambda\in\cW} d_\Lambda^2$.
Moreover, the following conditions are equivalent:
\begin{enumerate}
\item $\cT=\cV^+\cap\cV^-$,
\item $Z_{\Lambda,\Lambda'}=\la \aLap,\aLams \ra_M$,
      $\Lambda,\Lambda'\in\cW$,
\item $\sum_{x\in\cV} d_x^2 = \sum_{\Lambda\in\cW} d_\Lambda^2$,
\item each irreducible subsector of $[\can]$ is in $\cV$.
\end{enumerate}
\lablth{4equiv}
\end{proposition}

\bproof
Let us first show the inequality for the matrix elements
of the $Z$'s and equivalence $1\Leftrightarrow 2$:
We have by $\alpha\sigma$-reciprocity,
\[ \bearll
Z_{\Lambda,\Lambda'} &=\sum_{t\in\cT} b_{t,\Lambda} b_{t,\Lambda'} \\[.4em]
&= \sum_{t\in\cT} \la\lambda_\Lambda,\sigma_{\beta_t}\ra_N
\la \sigma_{\beta_t}, \lambda_{\Lambda'} \ra_N \\[.4em]
&= \sum_{t\in\cT} \la \aLap,\beta_t \ra_M \la \beta_t, \aLams \ra_M \\[.4em]
&\le \sum_{x\in\cV^+\cap\cV^-} \la \aLap,\beta_x \ra_M
\la \beta_x, \aLams \ra_M \\[.4em]
&=\sum_{x\in\cV} \la \aLap,\beta_x \ra_M
\la \beta_x, \aLams \ra_M \\[.4em]
&= \la \aLap,\aLams \ra_M \,,
\eear \]
and it is clear that we have equality for all
$\Lambda,\Lambda'\in\cW$ if and only if
$\cT=\cV^+\cap\cV^-$.

Next we show the inequality for the dimensions and
equivalence $2\Leftrightarrow 3$: We compute
\[ \bearll
(S^*\tilde{Z}S)_{0,0} &= \displaystyle\sum_{\Lambda,\Lambda'\in\cW}
(S^*)_{0,\Lambda} \tilde{Z}_{\Lambda,\Lambda'} S_{\Lambda',0} \\[1.0em]
&=  \displaystyle\sum_{\Lambda,\Lambda'\in\cW}
\displaystyle\sum_{i=0}^{D-1}
\displaystyle\frac{(S^*)_{0,\Lambda} S_{\Lambda,\Phi_1(i)}
S_{\overline{\Lambda'},\Phi_2(i)} S_{\Lambda',0}}
{S_{0,\Phi_1(i)}S_{0,\Phi_2(i)}} \,\, |\xi^i_0|^2 \\[1.0em]
&= \displaystyle\sum_{i=0}^{D-1} \del {\Phi_1(i)}0 \del {\Phi_2(i)}0
\displaystyle\frac{|\xi^i_0|^2}{S_{0,0}^2} \,,
\eear \]
where we used
$\tilde{Z}_{\Lambda,\Lambda'}=R^{\Lambda,\overline{\Lambda'}}_{0,0}$
and
$S_{\overline{\Lambda'},\Phi_2(i)}=S_{\Phi_2(i),\overline{\Lambda'}}
=(S^*)_{\Phi_2(i),\Lambda'}$.
The sector product matrices obey
$\sum_{z\in\cV}M_{x,y}^z d_z = d_xd_y$, hence
$\xi^0_x=\|\xi^0\|^{-1} d_x$, $x\in\cV$, realizes a
(normalized) eigenvector for each $M_y$
with eigenvalue $d_y$ and since
$R^{\Lambda,\Lambda'}=\sum_y \la\aLap\circ\aLams, \beta_y\ra_M M_y$,
$\xi^0$ is also an eigenvector
for each matrix $R^{\Lambda,\Lambda'}$.
The corresponding eigenvalues are given by
\[ \sum_{y\in\cV} \la\aLap\circ\aLams, \beta_y\ra_M \, d_y = 
d_{\aLap} d_{\aLams} = d_\Lambda d_{\Lambda'}
= S_{\Lambda,0}S_{\Lambda',0}/S_{0,0}^2 \,,\]
thus we have
$S_{\Lambda,0}S_{\Lambda',0}/S_{0,0}^2=
S_{\Lambda,\Phi_1(0)}S_{\Lambda',\Phi_2(0)}/
S_{0,\Phi_1(0)}S_{0,\Phi_2(0)}$ for all
$\Lambda,\Lambda'\in\cW$, implying
$\Phi_1(0)=\Phi_2(0)=0$. Clearly, $\xi^0$
is also an eigenvector of the sum matrix
$Q=\sum_{\Lambda,\Lambda'} R^{\Lambda,\Lambda'}$ which is
irreducible, hence it is in fact a Perron-Frobenius
eigenvector. Irreduciblility of $Q$
(for the definition of irreducible
matrices see e.g.\ \cite{ghj}) is seen as follows. For given
$x,y\in\cV$ choose some $z\equiv[\beta_z]\in\cV$ in the
irreducible decomposition of $[\co{\beta_x}\circ\beta_y]$,
then $\la\beta_x\circ\beta_z,\beta_y\ra_M\neq 0$. Since
$z$ is realized as an irreducible subsector of
$[\aLap\circ\aLams]$ for some $\Lambda,\Lambda'\in\cW$
it follows that the corresponding matrix element
of $R^{\Lambda,\Lambda'}$ is
non-zero, $R^{\Lambda,\Lambda'}_{x,y}\neq 0$.
Hence any matrix element $Q_{x,y}$ of the sum matrix $Q$
is strictly positive, implying irreducibility.
Therefore its Perron-Frobenius eigenvector is
unique and its eigenvalue is non-degenerate,
hence there cannot be an $i\neq 0$ such that
$\Phi_1(i)=\Phi_2(i)=0$. This means
$\del {\Phi_1(i)}0 \del {\Phi_2(i)}0 = \del i0$, hence we obtain
\[ (S^* \tilde{Z} S)_{0,0} = \frac{|\xi^0_0|^2}{S_{0,0}^2} \,. \]
On the other hand we compute
\[ (S^*\tilde{Z}S)_{0,0} = \sum_{\Lambda,\Lambda'\in\cW}
(S^*)_{0,\Lambda} \tilde{Z}_{\Lambda,\Lambda'} S_{\Lambda',0}
\ge \sum_{\Lambda,\Lambda'\in\cW}
(S^*)_{0,\Lambda} Z_{\Lambda,\Lambda'} S_{\Lambda',0}
= Z_{0,0} =1 \,, \]
since $(S^*)_{0,\Lambda}=S_{\Lambda,0}>0$ for all $\Lambda\in\cW$,
and therefore we have also equality if and only if
$Z_{\Lambda,\Lambda'}=\tilde{Z}_{\Lambda,\Lambda'}$ for
all $\Lambda,\Lambda'\in\cW$. Hence we have obtained
$|\xi^0_0|^2/S_{0,0}^2 \ge 1$ with equality if and only
if $Z=\tilde{Z}$. Now, by normalization,
$|\xi^0_0|^2=(\sum_{x\in\cV} d_x^2)^{-1}$
and $S_{0,0}^2 = (\sum_{\Lambda\in\cW} d_\Lambda^2)^{-1}$,
hence
\[ \sum_{x\in\cV} d_x^2 \le \sum_{\Lambda\in\cW} d_\Lambda^2 \,,\]
and we have equality if and only if $Z=\tilde{Z}$.

Finally we show the equivalence $3\Leftrightarrow 4$:
The inequality for sums over the squared dimensions
(``global indices'') is also a corollary of Lemmata
\ref{W=tV} and \ref{VintV}, and clearly we have
exact equality if and only if $\cV=\tilde{\cV}$,
and this is clearly equivalent to having each irreducible
subsector of $[\can]$ in $\cV$, the proof is complete.
\eproof

\subsection{Modular invariants and exponents of graphs revisited}

Similar to the analysis in \cite[Subsect.\ 4.2]{boev2}
we now investigate the relation between non-vanishing
entries in the mass matrix of the modular invariant
and exponents of fusion graphs obtained by $\alpha$-induction.
We denote
\[ \Eig(\Lambda,\Lambda') = \mathrm{span} \{\xi^i :
i\in\Phi^{-1}(\Lambda,\Lambda') \} \,,\qquad
\Lambda,\Lambda' \in\cW \,. \]
Also we put
\[ \| \xi_0 \|_{\Lambda,\Lambda'} = \sqrt{
\sum_{i\in\Phi^{-1}(\Lambda,\Lambda')} |\xi^i_0|^2} \,. \]

\begin{lemma}
If $\Eig(\Lambda,\Lambda')\neq 0$ for some
$\Lambda,\Lambda'\in\cW$ then
$\|\xi_0\|_{\Lambda,\Lambda'} \neq 0$.
\lablth{nonv}
\end{lemma}

\bproof
As the matrices $R^{\Lambda,\Lambda'}$ commute with
the matrices $M_x$ we find for $i=0,1,...,D-1$,
\[ R^{\Lambda,\Lambda'} M_x \xi^i =
M_x R^{\Lambda,\Lambda'} \xi^i =
\gamma_{\Phi_1(i)}(\Lambda)\gamma_{\Phi_2(i)}(\Lambda') M_x \xi^i \,,
\quad \Lambda,\Lambda'\in\cW, \quad x\in\cV\,, \]
i.e.\ $M_x \xi^i \in\Eig(\Phi_1(i),\Phi_2(i))$. In other words,
the matrices $M_x$ are block-diagonal in the basis $\xi^i$.
It follows that there are matrices, namely the ``blocks''
$B_{\Lambda,\Lambda'}(x)$, $B_{\Lambda,\Lambda'}(x)_{i,j}\in\bbC$,
$i,j\in\Phi^{-1}(\Lambda,\Lambda')$ such that
$M_x \xi^i = \sum_{j\in\Phi^{-1}(\Lambda,\Lambda')}
B_{\Lambda,\Lambda'}(x)_{i,j}\, \xi^j$,
hence in particular for the $0$-components
\[ (M_x \xi^i)_0 = \sum_{j\in\Phi^{-1}(\Lambda,\Lambda')}
B_{\Lambda,\Lambda'}(x)_{i,j} \, \xi^j_0 \,. \]
Since $(M_x \xi^i)_0=\sum_{y\in\cV} M_{0,x}^y \xi^i_y = \xi^i_x$ we
have for any $i\in\Phi^{-1}(\Lambda,\Lambda')$ and any $x\in\cV$,
\[ \xi^i_x = \sum_{j\in\Phi^{-1}(\Lambda,\Lambda')}
B_{\Lambda,\Lambda'}(x)_{i,j} \, \xi^j_0 \,. \]
It follows if $\xi^j_0=0$ for all
$j\in\Phi^{-1}(\Lambda,\Lambda')$ then $\xi^i_x=0$
for all $i\in\Phi^{-1}(\Lambda,\Lambda')$ and $x\in\cV$,
i.e.\ $\Eig(\Lambda,\Lambda')=0$.
\eproof

Let us denote $\Exp=\Bild \Phi$, the set of exponents. Clearly
$(\Lambda,\Lambda')\in\Exp$ if and only if
$\Eig(\Lambda,\Lambda')\neq0$.

\begin{proposition}
Provided $Z=\tilde{Z}$ we have $Z_{\Lambda,\Lambda'}\neq 0$
if and only if $(\Lambda,\Lambda')\in\Exp$.
\lablth{ZExp}
\end{proposition}

\bproof
If $Z=\tilde{Z}$ then
\[ \bearll
Z_{\Lambda,\Lambda'} &= (S^*ZS)_{\Lambda,\Lambda'}
=(S^*\tilde{Z}S)_{\Lambda,\Lambda'} 
= \displaystyle\sum_{\Omega,\Omega'\in\cW}
(S^*)_{\Lambda,\Omega} \tilde{Z}_{\Omega,\Omega'}
S_{\Omega',\Lambda'} \\[.6em]
&=  \displaystyle\sum_{\Omega,\Omega'\in\cW}
\displaystyle\sum_{i=0}^{D-1}
\displaystyle\frac{(S^*)_{\Lambda,\Omega} S_{\Omega,\Phi_1(i)}
S_{\overline{\Omega'},\Phi_2(i)} S_{\Omega',\Lambda'}}
{S_{0,\Phi_1(i)}S_{0,\Phi_2(i)}} \,\, |\xi^i_0|^2 \\[1.0em]
&= \displaystyle\sum_{i=0}^{D-1} \del {\Phi_1(i)}\Lambda
\del {\Phi_2(i)}{\Lambda'} \displaystyle\frac{|\xi^i_0|^2}
{S_{0,\Lambda} S_{0,\Lambda'}} 
= \displaystyle\frac{\|\xi_0\|_{\Lambda,\Lambda'}^2}
{S_{0,\Lambda} S_{0,\Lambda'}} \,,
\eear \]
the statement follows now by Lemma \ref{nonv}.
\eproof

Recall that in \cite{boev2} we considered a set of exponents,
which we will now denote by $\Exp^+$, labelling the joint
spectrum of matrices $V_\Lambda$, where
$\V\Lambda ab = \la\beta_a\circ\aLap,\beta_b\ra_M$,
$a,b\in\cV^+$ and $\Lambda\in\cW$. We proved that
$Z_{\Lambda,\Lambda}\neq 0$ if and only if $\Lambda\in\Exp^+$,
provided that the extended S-matrix diagonalizes the
(endomorphism) fusion rules of the marked vertices $\cT$.
This condition is not particularly
difficult to prove for the conformal inclusions since
the set $\cT$ is given in terms of the level 1 \per s
of the ambient WZW theory. However, for the orbifold inclusions
this seems to be hardly possible without computer aid since
the formulae for the extended S-matrices are complicated.
On the other hand we prove $Z=\tilde{Z}$ for all orbifold
inclusions. Therefore it is useful to check the relations
between $\Exp$ and $\Exp^+$.

\begin{lemma}
If $(\Omega,\Omega')\in\Exp$ then $\Omega\in\Exp^+$.
Conversely, if $\Omega\in\Exp^+$ then there is some
$\Omega'\in\cW$ such that $(\Omega,\Omega')\in\Exp$.
\lablth{EE+}
\end{lemma}

\bproof
Since the subset $\cV^+\subset\cV$ is itself a sector basis, the
matrices $R^{\Lambda,0}$, corresponding to $[\aLap]$,
decompose block-diagonally with respect to the labels
in $\cV^+$ and $\cV\backslash\cV^+$. Thus we can write
$R^{\Lambda,0}=V_\Lambda\oplus \tilde{V}_\Lambda$.
Assume $(\Omega,\Omega')\in\Exp$. Let $\xi^i$ be a corresponding
simultaneous eigenvector of the $R^{\Lambda,\Lambda'}$'s,
i.e.\ $\Phi(i)=(\Omega,\Omega')$ and we have in particular
$R^{\Lambda,0}\xi^i=\gamma_\Omega(\Lambda)\xi^i$. We can write
$\xi^i=\psi^i\oplus\tilde{\psi}^i$, then this reads in particular
$V_\Lambda\psi^i=\gamma_\Omega(\Lambda)\psi^i$. Therefore
$\Omega\in\Exp^+$ if $\psi^i\neq 0$. However, by the same
argument as in \cite[Cor.\ 4.6]{boev2}, the eigenvectors
$\xi^i$ can be chosen such that $\xi^i_0>0$,
thus $\psi^i\neq 0$.

Conversely, assume $\Omega\in\Exp^+$. This means
$\gamma_\Omega(\Lambda)$ belongs to the spectrum of
$V_\Lambda$, therefore it belongs to the spectrum of
$R^{\Lambda,0}=V_\Lambda\oplus \tilde{V}_\Lambda$.
As the eigenvalues of the $R^{\Lambda,\Lambda'}$'s
are all of the form
$\gamma_\Omega(\Lambda)\gamma_{\Omega'}(\Lambda')$
and the characters $\gamma_\Omega$ are linearly independent,
there must be some $\Omega'\in\cW$ such that
$\gamma_\Omega(\Lambda)\gamma_{\Omega'}(\Lambda')$
gives in fact the eigenvalues of the $R^{\Lambda,\Lambda'}$'s,
i.e.\ $(\Omega,\Omega')\in\Exp$.
\eproof

Now we can prove the following

\begin{proposition}
Provided $Z=\tilde{Z}$ we have $Z_{\Lambda,\Lambda}\neq 0$
if and only if $\Lambda\in\Exp^+$.
\end{proposition}

\bproof
If $Z_{\Lambda,\Lambda}\neq 0$ then $(\Lambda,\Lambda)\in\Exp$
by Proposition \ref{ZExp}. Hence $\Lambda\in\Exp^+$ by
Lemma \ref{EE+}. Conversely, if $\Lambda\in\Exp^+$ then 
by Lemma \ref{EE+} there is a $\Lambda'\in\cW$ such that
$(\Lambda,\Lambda')\in\Exp$, hence $Z_{\Lambda,\Lambda'}\neq 0$
by Proposition \ref{ZExp}. As $Z_{\Lambda,\Lambda'}\neq 0$
implies $Z_{\Lambda,\Lambda}\neq 0$ for block-diagonal
modular invariants (cf.\ also Lemma \ref{lamupm}),
the statement follows.
\eproof

Analogous statements hold for $\Exp^-$ corresponding to $\cV^-$.
In particular $Z=\tilde{Z}$ implies $\Exp^+=\Exp^-$.

\section{Applications to Embeddings of $\SUn$: Examples}

We will now apply our results to conformal and
orbifold inclusions of $\SUn$. By the spin and
statistics theorem \cite{gulo2} we have for the
statistics phase
$\kappa_{\lambda_\Lambda}=\E^{2\pi\I h_\Lambda}$,
where $h_\Lambda$ is the conformal dimension,
$\Lambda\in\ASU nk$. Due to $T$-invariance of
the modular mass matrix $Z$ we have
$h_\Lambda=0\,\,\mod\,\,\bbZ$ whenever $[\lambda_\Lambda]$
is a subsector of $[\canr]$. With this, Corollary \ref{channel}
and Lemma \ref{disj} become useful criteria to compare
$[\aLap]$ and $[\aLam]$, and this will lead us to
the validity of
$Z_{\Lambda,\Lambda'}=\la\aLap,\aLams\ra_M$,
$\Lambda,\Lambda'\in\ASU nk$, for all our examples.

\subsection{Conformal embeddings of $\SUz$ revisited}

\ddE 6 revisited: $\SUz_{10}\subset\mathit{SO}(5)_1$.
We have derived the algebraic structure of $\cV^+$
in \cite{boev2}, and clearly the same results are obtained
for $\cV^-$, hence we have
\[ \cV^\pm = \{ \as 0 , \aspm 1 , \aspm 2 ,
\asx 31 , \aspm 9 , \as {10} \} \,,\]
where we omit here (and similar in the examples discussed below)
the $\pm$-index for the marked vertices
$\as 0$, $\asx 31$ and $\as {10}$ as we know that always
$\cT\subset\cV^+\cap\cV^-$.

\begin{lemma}
For the \ddE 6 example we have $[\can]=[\id_M]\oplus\asprod 11$.
\lablth{E6can}
\end{lemma}

\bproof
Recall that $[\canr]=\ls 0 \oplus \ls 6$. We have the
fusion rule $\N 628 =1$ but $h_8-h_2=5/3-1/6=3/2\notin\bbZ$.
Hence it follows from Corollary \ref{channel} that
$\asp 2 \neq \asm 2$. Since $\N jj2=1$ for all $j\neq 0,10$,
it follows that $\asp j \neq \asm j$ for all
$j\neq 0,10$, because equality of $\asp j$ and $\asm j$ clearly
implies equality of their squares, and if $\asm 2$ appears in
the decomposition of the square of $\asp j$, then $\asm 2$ equals
some subsector of some $\asp {j'}$, implying equality of
$\asm 2$ and $\asp 2$ by Lemma \ref{lamupm}, a contradiction.
In particular we find $\asp 1 \neq \asm 1$. Now
\[ \la \alpha^+_1 \circ  \alpha^-_1, \alpha^+_1 \circ  \alpha^-_1
\ra_M = \la \alpha^+_1 \circ  \alpha^+_1, \alpha^-_1 \circ \alpha^-_1
\ra_M = 1 + \la \alpha^+_2 , \alpha^-_2 \ra_M = 1 \,, \]
thus $\alpha^+_1\circ\alpha^-_1$ is irreducible. Moreover,
by Corollary \ref{congam} we find
that $\asprod 11$ is a subsector of $[\can]$, different
from the identity since $\asp 1 \neq \asm 1$.
Since $\la\can,\can\ra_M=\la\canr,\canr\ra_N=2$ by
Lemma \ref{gg=tt}, the statement follows.
\eproof

One can also check that
\[ d_\can = 1 + d_1^2 = 1 + d_6 = d_\canr = 3 + \sqrt 3 \,. \]
In fact, with similar arguments as used in the proof of
Lemma \ref{E6can}, it is not difficult to solve the system
completely, i.e.\ to determine the algebraic structure of $\cV$.
We find
\[ \cV  = \{ \as 0 , \asp 1 , \asm 1 ,  \asp 2 , \asm 2 ,
\asx 31 , \asp 9 , \asm 9 , \as {10} , [\delta] ,
[\zeta], [\delta'] \} \,, \]
where $[\delta]=\asprod 11$, $[\zeta]=\asprod 12 = \asprod 21$
and $[\delta']=\asprod 91 = \asprod 19$. 
The fusion graphs of $\asp 1$ (straight lines) and
$\asm 1$ (dashed lines) are given in Figure \ref{E6pm}.
We have encircled the even vertices by small circles,
the marked vertices by larger circles.
%
%%%%%%%%%%%% E_6 %%%%%%%%%%%%%
\begin{figure}[tb]
\unitlength 0.6mm
\begin{center}
\begin{picture}(100,140)
%%%%%
\thinlines 
\multiput(10,50)(0,40){2}{\circle*{2}}
\multiput(10,70)(80,0){2}{\circle{4}}
\multiput(90,50)(0,40){2}{\circle*{2}}
\multiput(50,30)(0,80){2}{\circle*{2}}
\multiput(10,70)(80,0){2}{\circle*{2}}
\multiput(50,10)(0,40){4}{\circle*{2}}
\multiput(50,10)(0,40){4}{\circle{4}}
\multiput(50,90)(0,20){3}{\circle{6}}
\Thicklines 
\path(50,130)(10,90)(10,50)(50,90) \path(50,110)(10,70) 
\dottedline{2}(50,130)(90,90)(90,50)(50,90)
\dottedline{2}(50,110)(90,70)
\path(90,90)(50.5,50)(50.5,10)(90,50) \path(90,70)(50.5,30) 
\dottedline{2}(10,90)(49.5,50)(49.5,10)(10,50)
\dottedline{2}(10,70)(49.5,30) 
\put(1,90){\makebox(0,0){$\asp 1$}}
\put(1,70){\makebox(0,0){$\asp 2$}}
\put(1,50){\makebox(0,0){$\asp 9$}}
\put(50,137){\makebox(0,0){$\as 0$}}
\put(50,118){\makebox(0,0){$\asx 31$}}
\put(50,97){\makebox(0,0){$\as {10}$}}
\put(50,57){\makebox(0,0){$[\delta]$}}
\put(56,27){\makebox(0,0){$[\zeta]$}}
\put(50,4){\makebox(0,0){$[\delta']$}}
\put(99,90){\makebox(0,0){$\asm 1$}}
\put(99,70){\makebox(0,0){$\asm 2$}}
\put(99,50){\makebox(0,0){$\asm 9$}}
\end{picture}
\caption{\ddE 6: Fusion graph of $\asp 1$ and $\asm 1$}
\label{E6pm}
\end{center}
\end{figure}
It is easy to write down the principal graph of the subfactor
$N\subset M$ by taking the connected component of $\ls 0\equiv[\id_N]$
of the induction-restriction graph, as already drawn in
\cite{boev2}. The correctly labelled graph is given in Figure
\ref{P-E6}. 
\begin{figure}[tb]
\unitlength 0.3mm
\thicklines
\begin{center}
\begin{picture}(300,100)
\multiput(50,80)(40,0){6}{\circle*{3}}
\multiput(70,20)(80,0){3}{\circle*{3}}
\put(50,80){\line(1,-3){20}}
\put(90,80){\line(-1,-3){20}}
\put(90,80){\line(1,-1){60}}
\put(130,80){\line(1,-3){20}}
\put(170,80){\line(-1,-3){20}}
\put(210,80){\line(-1,-1){60}}
\put(210,80){\line(1,-3){20}}
\put(250,80){\line(-1,-3){20}}
\put(50,92){\makebox(0,0){$\ls 0$}}
\put(90,92){\makebox(0,0){$\ls 6$}}
\put(130,92){\makebox(0,0){$\ls 2$}}
\put(170,92){\makebox(0,0){$\ls 8$}}
\put(210,92){\makebox(0,0){$\ls 4$}}
\put(250,92){\makebox(0,0){$\ls {10}$}}
\put(70,8){\makebox(0,0){$[\iotab]$}}
\put(150,8){\makebox(0,0){$[\iotab\circ\aepm 2]$}}
\put(230,8){\makebox(0,0){$[\iotab\circ\alpha_{10}]$}}
\end{picture}
\end{center}
\caption{\ddE 6: Principal graph for the conformal inclusion
$\SUz_{10}\subset\SOf_1$}
\label{P-E6}
\end{figure}
Having determined the subsectors of $[\can]$,
we can now similarly determine the dual principal graph
by taking the connected component of $\as 0\equiv[\id_M]$ of the
$\can$-multiplication graph, presented in Figure \ref{DP-E6}.
\begin{figure}[tb]
\unitlength 0.3mm
\thicklines
\begin{center}
\begin{picture}(300,100)
\multiput(50,80)(40,0){6}{\circle*{3}}
\multiput(70,20)(80,0){3}{\circle*{3}}
\put(50,80){\line(1,-3){20}}
\put(90,80){\line(-1,-3){20}}
\put(90,80){\line(1,-1){60}}
\put(130,80){\line(1,-3){20}}
\put(170,80){\line(-1,-3){20}}
\put(210,80){\line(-1,-1){60}}
\put(210,80){\line(1,-3){20}}
\put(250,80){\line(-1,-3){20}}
\put(50,92){\makebox(0,0){$\as 0$}}
\put(90,92){\makebox(0,0){$[\delta]$}}
\put(130,92){\makebox(0,0){$\asp 2$}}
\put(170,92){\makebox(0,0){$\asm 2$}}
\put(210,92){\makebox(0,0){$[\delta']$}}
\put(250,92){\makebox(0,0){$\as {10}$}}
\put(70,8){\makebox(0,0){$[\iota]$}}
\put(150,8){\makebox(0,0){$[\aepm 2\circ\iota]$}}
\put(230,8){\makebox(0,0){$[\alpha_{10}\circ\iota]$}}
\end{picture}
\end{center}
\caption{\ddE 6: Dual principal graph for the conformal inclusion
$\SUz_{10}\subset\SOf_1$}
\label{DP-E6}
\end{figure}
It is straightforward to check that the $M$-$M$ sectors,
labelling the even vertices in Figure \ref{DP-E6}, obey in
fact the fusion rules determined by Kawahigashi \cite{kaw1}
as the correct fusion table of the five possibilities
given in \cite{bis1}. Another result of \cite{kaw1},
namely that this fusion algebra contains a subalgebra
corresponding to the even vertices of \ddE 6 turns
up quite naturally here as $\as 0$, $\aspm 2$ and
$\as {10}$ appear as even vertices of the dual
principal graph due to the general fact stated in
Corollary \ref{subdual}.

\ddE 8  revisited: $\SUz_{28}\subset(\Gtwo)_1$.
Recall from \cite{boev2} that
\[ \cV^\pm = \{ \as 0 , \aspm 1 , \aspm 2 , \aspm 3 ,\aspm 4 ,
\asxpm 51 , \asxpm 52 , \asx 61 \} \,.\]

\begin{lemma}
For the \ddE 8 example we have
$[\can]=[\id_M]\oplus[\delta]\oplus[\omega]\oplus[\eta]$,
where $[\delta]=\asprod 11$ and $[\omega]=\asprod 22$
irreducible and $\asprod 33=[\eta]\oplus[\eta']$ with
$[\eta],[\eta']$ irreducible.
\lablth{E8can}
\end{lemma}

\bproof
Recall $[\canr]=\ls 0 \oplus \ls {10} \oplus \ls {18} \oplus \ls {28}$.
We have the
fusion rule $\N {10}28 =1$, but $h_8-h_2=2/3-1/15=3/5\notin\bbZ$.
Hence $\asp 2 \neq \asm 2$ by Corollary \ref{channel}.
Since $\N jj2=1$ for all $j\neq 0,28$, it follows immediately
that $\asp j \neq \asm j$ for all $j\neq 0,28$. Note that
$\asp 5=\asp {23}$. Since $h_{23}-h_5=115/24 - 7/24=9/2$ it
follows $\la \alpha^+_5,\alpha^-_5\ra_M=0$ by Lemma \ref{disj}.
Thus the subsectors of $\asp 5$ and $\asm 5$ are all
disjoint. We have shown $\la\alpha^+_j,\alpha^-_j\ra_M=0$ for
$j=1,2,3,4,5$, and since $\asp 6 \neq \asm 6$ but $\asx 61$
is a marked vertex we have $\la\alpha^+_6,\alpha^-_6\ra_M=1$. With
these relations one checks easily that $[\delta]=\asprod 11$,
$[\omega]=\asprod 22$ and $\asprod 33$ are disjoint, e.g.
\[ \la \alpha^+_2 \circ \alpha^-_2, \alpha^+_3 \circ \alpha^-_3
\ra_M = \la \alpha^+_2 \circ \alpha^+_3, \alpha^-_2 \circ \alpha^-_3
\ra_M = \sum_{j,j'=1,3,5} \la \alpha^+_j, \alpha^-_j \ra_M =0 \,,\]
and similarly that $[\delta]$ and $[\omega]$ are irreducible,
whereas
\[ \la \alpha^+_3 \circ \alpha^-_3, \alpha^+_3 \circ \alpha^-_3
\ra_M = \la \alpha^+_3 \circ \alpha^+_3, \alpha^-_3 \circ \alpha^-_3
\ra_M = \la \alpha_0,\alpha_0 \ra_M + \la \alpha^+_6,
\alpha^-_6 \ra_M =2 \,.\]
Hence $\asprod 33 = [\eta]\oplus[\eta']$ with $[\eta],[\eta']$
irreducible. Since (Corollary \ref{congam})
\[ \la \alpha^+_j\circ \alpha^-_j , \can \ra_M =
\la \alpha^+_j , \alpha^+_j \ra = 1 \,, \qquad j=1,2,3, \]
it follows that $[\delta],[\omega]$ are subsectors of $[\can]$
and $\asprod 33$ has one common subsector with $[\can]$, say
$[\eta]$. As $\la\can,\can\ra_M=\la\canr,\canr\ra_N=4$, the
statement follows.
\eproof

With a little more computation, the full induced sector basis
$\cV$ and its algebraic structure (the associated
``sector algebra'', see \cite{boev1} for definitions)
can be determined. One finds that $\cV$ has 32
elements. The fusion graphs of $\asp 1$ (straight lines) and
$\asm 1$ (dashed lines) are given in Figure \ref{E8pm}.
%
%%%%%%%%%% E_8 %%%%%%%%%%%%%
\begin{figure}[tb]
\unitlength 1.4mm
\begin{center}
\begin{picture}(80,85)
%%%%%
\thinlines 
\multiput(10,40)(0,10){2}{\circle*{1}}
\multiput(20,30)(0,10){4}{\circle*{1}}
\multiput(30,20)(0,10){6}{\circle*{1}}
\multiput(40,10)(0,10){8}{\circle*{1}}
\multiput(50,20)(0,10){6}{\circle*{1}}
\multiput(60,30)(0,10){4}{\circle*{1}}
\multiput(70,40)(0,10){2}{\circle*{1}}
\multiput(20,30)(0,10){4}{\circle{2}}
\multiput(40,10)(0,10){8}{\circle{2}}
\multiput(60,30)(0,10){4}{\circle{2}}
\multiput(40,70)(0,10){2}{\circle{3}}
\Thicklines 
\path(40,80)(20,60)(10,40)(40,70) \path(10,50)(20,50)
\dottedline{0.8}(40,80)(60,60)(70,40)(40,70)
\dottedline{0.8}(70,50)(60,50)
\path(50,70)(30,50)(20,30)(50,60) \path(20,40)(30,40)
\path(60,60)(40,40)(30,20)(60,50) \path(30,30)(40,30)
\path(70,50)(50,30)(40,10)(70,40) \path(40,20)(50,20)
\dottedline{0.8}(30,70)(50,50)(60,30)(30,60)
\dottedline{0.8}(20,60)(40,40)(50,20)(20,50)
\dottedline{0.8}(10,50)(30,30)(40,10)(10,40)
\dottedline{0.8}(60,40)(50,40)
\dottedline{0.8}(50,30)(40,30)
\dottedline{0.8}(40,20)(30,20)
\put(5,50){\makebox(0,0){\small{$\asxp 52$}}}
\put(6,40){\makebox(0,0){\small{$\asp 3$}}}
\put(15,60){\makebox(0,0){\small{$\asp 2$}}}
\put(20,54){\makebox(0,0){\small{$\asp 4$}}}
\put(21,43){\makebox(0,0){\small{$[\kappa^+]$}}}
\put(15,30){\makebox(0,0){\small{$[\zeta^+]$}}}
\put(25,70){\makebox(0,0){\small{$\asp 1$}}}
\put(29,64){\makebox(0,0){\small{$\asxp 51$}}}
\put(30,54){\makebox(0,0){\small{$[\xi^+]$}}}
\put(35,40){\makebox(0,0){\small{$[\psi^+]$}}}
\put(31,33){\makebox(0,0){\small{$[\varsigma^+]$}}}
\put(25,20){\makebox(0,0){\small{$[\tau^+]$}}}
\put(40,84){\makebox(0,0){\small{$\as 0$}}}
\put(40,74){\makebox(0,0){\small{$\asx 61$}}}
\put(40,64){\makebox(0,0){\small{$[\delta]$}}}
\put(40,54){\makebox(0,0){\small{$[\chi]$}}}
\put(40,44){\makebox(0,0){\small{$[\omega]$}}}
\put(40,26){\makebox(0,0){\small{$[\varpi]$}}}
\put(40,17){\makebox(0,0){\small{$[\eta]$}}}
\put(40,6){\makebox(0,0){\small{$[\eta']$}}}
\put(55,70){\makebox(0,0){\small{$\asm 1$}}}
\put(51,64){\makebox(0,0){\small{$\asxm 51$}}}
\put(50,54){\makebox(0,0){\small{$[\xi^-]$}}}
\put(45,40){\makebox(0,0){\small{$[\psi^-]$}}}
\put(49,33){\makebox(0,0){\small{$[\varsigma^-]$}}}
\put(55,20){\makebox(0,0){\small{$[\tau^-]$}}}
\put(65,60){\makebox(0,0){\small{$\asm 2$}}}
\put(60,54){\makebox(0,0){\small{$\asm 4$}}}
\put(59,43){\makebox(0,0){\small{$[\kappa^-]$}}}
\put(65,30){\makebox(0,0){\small{$[\zeta^-]$}}}
\put(75,50){\makebox(0,0){\small{$\asxm 52$}}}
\put(74,40){\makebox(0,0){\small{$\asm 3$}}}
\end{picture}
\caption{\ddE 8: Fusion graph of $\asp 1$ and $\asm 1$}
\label{E8pm}
\end{center}
\end{figure}
Here we denote
\[ \begin{array}{llll}
[\xi^\pm] &= \asprodpm 21 \,, \qquad & [\zeta^\pm]
&= \asprodpm 31 \,, \\[.4em]
[\psi^\pm] &= \asprodpm 41 \,, & [\kappa^\pm]
&= [\alpha^{\pm(2)}_5\circ\alpha^\mp_1] \,,\\[.4em]
[\tau^\pm] &= \asprodpm 32  \,, 
&[\varsigma^\pm] &= [\alpha^{\pm(2)}_5\circ\alpha^\mp_2] \,, \\[.4em]
[\chi] &= [\alpha^{+(1)}_5 \circ \alpha^-_1] \,, \qquad
&[\varpi] &= \asprod 42 \,.
\eear \]
By the induction-restriction mechanism it is easy to write
down the principal graph of $N\subset M$, presented in
Figure \ref{P-E8}.
\begin{figure}[tb]
\unitlength 0.4mm
\thicklines
\begin{center}
\begin{picture}(300,120)
\multiput(10,100)(20,0){15}{\circle*{3}}
\multiput(45,20)(70,0){4}{\circle*{3}}
\path(10,100)(45,20)(30,100)
\path(50,100)(45,20)(70,100)(115,20)(50,100)(185,20)(70,100)
\path(90,100)(115,20)(110,100)
\path(130,100)(115,20)(150,100)(185,20)(130,100)
\path(170,100)(185,20)(190,100)
\path(209,100)(184,20)
\path(211,100)(186,20)
\path(230,100)(115,20)(250,100)(185,20)(230,100)(255,20)(250,100)
\path(270,100)(185,20)(290,100)(255,20)(270,100)
\put(10,110){\makebox(0,0){\small{$\ls 0$}}}
\put(30,110){\makebox(0,0){\small{$\ls {28}$}}}
\put(50,110){\makebox(0,0){\small{$\ls {10}$}}}
\put(70,110){\makebox(0,0){\small{$\ls {18}$}}}
\put(90,110){\makebox(0,0){\small{$\ls 2$}}}
\put(110,110){\makebox(0,0){\small{$\ls {26}$}}}
\put(130,110){\makebox(0,0){\small{$\ls 8$}}}
\put(150,110){\makebox(0,0){\small{$\ls {20}$}}}
\put(170,110){\makebox(0,0){\small{$\ls 4$}}}
\put(190,110){\makebox(0,0){\small{$\ls {24}$}}}
\put(210,110){\makebox(0,0){\small{$\ls {14}$}}}
\put(230,110){\makebox(0,0){\small{$\ls {12}$}}}
\put(250,110){\makebox(0,0){\small{$\ls {16}$}}}
\put(270,110){\makebox(0,0){\small{$\ls 6$}}}
\put(290,110){\makebox(0,0){\small{$\ls {22}$}}}
\put(45,10){\makebox(0,0){\small{$[\iotab]$}}}
\put(115,10){\makebox(0,0){\small{$[\iotab\circ\aepm 2]$}}}
\put(185,10){\makebox(0,0){\small{$[\iotab\circ\aepm 4]$}}}
\put(255,10){\makebox(0,0){\small{$[\iotab\circ\alpha^{(1)}_6]$}}}
\end{picture}
\end{center}
\caption{\ddE 8: Principal graph for the conformal inclusion
$\SUz_{28}\subset(\Gtwo)_1$}
\label{P-E8}
\end{figure}
With a bit more calculation we can also determine the
$\can$-multiplication and therefore write down the
dual principal graph, presented in Figure \ref{DP-E8}.
\begin{figure}[tb]
\unitlength 0.39mm
\thicklines
\begin{center}
\begin{picture}(320,120)
\multiput(10,100)(20,0){16}{\circle*{3}}
\multiput(40,20)(80,0){4}{\circle*{3}}
\path(10,100)(40,20)(30,100)(120,20)(50,100)(40,20)(70,100)
(200,20)(50,100)
\path(120,20)(130,100)(200,20)(150,100)(120,20)
\path(90,100)(120,20)(110,100)
\path(170,100)(120,20)(230,100)(280,20)(170,100)(200,20)
(290,100)(280,20)(310,100)
\path(190,100)(200,20)(210,100)
\path(250,100)(200,20)(270,100)
\path(229,100)(199,20)
\path(231,100)(201,20)
\put(10,110){\makebox(0,0){\small{$\as 0$}}}
\put(30,110){\makebox(0,0){\small{$[\delta]$}}}
\put(50,110){\makebox(0,0){\small{$[\omega]$}}}
\put(70,110){\makebox(0,0){\small{$[\eta]$}}}
\put(90,110){\makebox(0,0){\small{$\asp 2$}}}
\put(110,110){\makebox(0,0){\small{$\asm 2$}}}
\put(130,110){\makebox(0,0){\small{$[\zeta^+]$}}}
\put(150,110){\makebox(0,0){\small{$[\zeta^-]$}}}
\put(170,110){\makebox(0,0){\small{$[\eta']$}}}
\put(190,110){\makebox(0,0){\small{$\asp 4$}}}
\put(210,110){\makebox(0,0){\small{$\asm 4$}}}
\put(230,110){\makebox(0,0){\small{$[\varpi]$}}}
\put(250,110){\makebox(0,0){\small{$[\kappa^+]$}}}
\put(270,110){\makebox(0,0){\small{$[\kappa^-]$}}}
\put(290,110){\makebox(0,0){\small{$[\chi]$}}}
\put(310,110){\makebox(0,0){\small{$[\alpha^{(1)}_6]$}}}
\put(40,10){\makebox(0,0){\small{$[\iota]$}}}
\put(120,10){\makebox(0,0){\small{$[\aepm 2\circ\iota]$}}}
\put(200,10){\makebox(0,0){\small{$[\aepm 4\circ\iota]$}}}
\put(280,10){\makebox(0,0){\small{$[\alpha^{(1)}_6\circ\iota]$}}}
\end{picture}
\end{center}
\caption{\ddE 8: Dual principal graph for the conformal inclusion
$\SUz_{28}\subset(\Gtwo)_1$}
\label{DP-E8}
\end{figure}
Note that the fusion algebra of the $M$-$M$ sectors
labelling the even vertices of the dual principal graph
possesses again a subalgebra (in fact two copies) which
corresponds to the even vertices of the \ddE 8 graph,
due to Corollary \ref{subdual}.

\ddD 4 revisited: $\SUz_4\subset\SUd_1$.
Recall from \cite{boev2} that
\[ \cV^\pm = \{ \as 0 , \aspm 1 , \asx 21 , \asx 22 \} \,,\]
and the marked vertices $\as 0$, $\asx 21$ and $\asx 22$
obey $\bbZ_3$ fusion rules.

\begin{lemma}
For the \ddD 4 example we have
$[\can]=[\id_M]\oplus[\epsilon]$, where $[\epsilon]$ is
an irreducible subsector of $\asprod 11$.
\lablth{D4can}
\end{lemma}

\bproof
We have $\aspm 3=\aspm 1$ but $h_3-h_1=5/8-1/8=1/2\notin\bbZ$,
hence $\asp 1\neq\asm 1$. Since
\[ \la \alpha^+_1\circ\alpha^-_1,\alpha^+_1\circ\alpha^-_1\ra_M
= \la \alpha_0,\alpha_0 \ra_M + \la \alpha_2,\alpha_2 \ra_M =3 \]
and $d_1=\sqrt 3$, we find that $\asprod 11$ decomposes into
three different subsectors,
\[ \asprod 11 = [\epsilon] \oplus [\eta] \oplus [\eta']\,,\]
with statistical dimensions $d_\epsilon=d_{\eta}=d_{\eta'}=1$.
Since
$\la\alpha^+_1\circ\alpha^-_1,\can\ra_M=\la\alpha^+_1,\alpha^+_1\ra_M=1$,
$\asprod 11$ and $[\can]$ have one subsector in common, say $[\epsilon]$.
Note that $[M:N]=d_\can=2$, hence we are dealing with the
unique $\bbZ_2$ subfactor and therefore we can choose
$\epsilon\in\Aut(M)$ such that $\epsilon^2=\id_M$ and
$N=M^\epsilon$ and $[\can]=[\id_M]\oplus[\epsilon]$.
\eproof

By Proposition \ref{4equiv} we find
$\sum_{x\in\cV} d_x^2=\sum_{j=0}^4 d_j^2=12$, hence we have already
found all the sectors of $\cV$,
\[ \cV = \{ \as 0, \asp 1, \asm 1, \asx 21, \asx 22, [\epsilon],
[\eta], [\eta'] \} \,. \]
We remark that we will show in a forthcoming joint work with
Y.\ Kawahigashi \cite{bek} that the corresponding sector algebra
is non-commutative. The simultaneous fusion graph of $\asp 1$
and $\asm 1$ is given in Figure \ref{D4pm}.
%
%%%%%%%%%%%% D_4 %%%%%%%%%%%%%
\begin{figure}[tb]
\unitlength 1.0mm
\begin{center}
\begin{picture}(60,80)
%%%%%
\thinlines 
\multiput(10,40)(40,0){2}{\circle*{1}}
\multiput(30,10)(0,10){5}{\circle*{1}}
\multiput(30,10)(0,10){5}{\circle{2}}
\multiput(30,30)(0,10){2}{\circle{3}}
\put(30,70){\circle*{1}}
\put(30,70){\circle{2}}
\put(30,70){\circle{3}}
\Thicklines 
\path(30,70)(10,40)(30,40)  \path(30,30)(10,40)
\dottedline{1.2}(30,70)(50,40)(30,40)
\dottedline{1.2}(50,40)(30,30)
\path(30,50)(50,40)(30,20)  \path(30,10)(50,40)
\dottedline{1.2}(30,50)(10,40)(30,20)
\dottedline{1.2}(10,40)(30,10)
\put(30,75){\makebox(0,0){$\as 0$}}
\put(5,40){\makebox(0,0){$\asp 1$}}
\put(55,40){\makebox(0,0){$\asm 1$}}
\put(30,54){\makebox(0,0){$[\epsilon]$}}
\put(30,45){\makebox(0,0){$\asx 21$}}
\put(30,35){\makebox(0,0){$\asx 22$}}
\put(30,16){\makebox(0,0){$[\eta]$}}
\put(30,6){\makebox(0,0){$[\eta']$}}
\end{picture}
\caption{\ddD 4: Fusion graph of $\asp 1$ and $\asm 1$}
\label{D4pm}
\end{center}
\end{figure}

\subsection{$\bbZ_m$ orbifold inclusions of $\SUn$}
\lablsec{morbn}

In \cite{boev2} we discussed the $\bbZ_n$ orbifold modular
invariants of $\SUn$ which correspond to a simple current
extension by a simple current of order $n$. However, for any
decomposition $n=mq$ with $m,q\in\bbN$ there appears a series of
block-diagonal orbifold modular invariants corresponding to a
simple current of order $m$, see \cite{scya1,scya2}. We would
like to extend our analysis to this more general situation.
Note that the case $q=1$ corresponds to the familiar $\bbZ_n$
orbifold situation whereas the other extreme case $m=1$
corresponds to the completely diagonal invariant, i.e.\ there
is no extension at all, but if $n$ is not prime then there are
intermediate cases. The $\bbZ_m$ invariants appear whenever
$2n$ divides $k'q^2$ or, equivalently, when $2m$ divides $k'q$,
where $k'=k+n$ if the level $k$ and $n$ are both odd and
$k'=k$ otherwise. One can check that this
is equivalent to the condition that $kq\in 2m\bbZ$ if $n$
is even and $kq\in m\bbZ$ if $n$ is odd. The corresponding
modular invariant mass matrix reads \cite{scya1,scya2}
\be
Z_{\Lambda,\Lambda'} = \delta^m(t(\Lambda)) \sum_{j=0}^{m-1}
\delta(\Lambda',\sigma^{jq}(\Lambda)) \,.
\labl{scez}
Here, as usual, $\sigma$ is the $\bbZ_n$ rotation of
$\ASU nk$,
\[ \sigma(\Lambda) = (k-m_1-\ldots - m_{n-1}) \Lambda_{(1)}
+ m_1 \Lambda_{(2)} + m_2 \Lambda_{(3)} + \ldots +
m_{n-2} \Lambda_{(n-1)} \,,\]
for $\Lambda=\sum_{i=1}^{n-1} m_i\Lambda_{(i)}\in\ASU nk$
with fundamental weights $\Lambda_{(i)}$,
$t(\Lambda)=\sum_{i=1}^{n-1} i\,m_i$, and $\delta^y(x)$ equals
$1$ or $0$ dependent whether or not $x/y$ is an integer,
respectively. In terms of sectors,
the $\bbZ_n$ rotation $\sigma$ corresponds to
$[\lambda_{k\Lambda_{(1)}}]$ and the $\bbZ_m$ rotation
$\sigma^q$ to $[\lambda_{k\Lambda_{(q)}}]$ which realize
the rotations $\sigma$ respectively $\sigma^q$ as fusion
rules. In fact as the vacuum block of \erf{scez} is easily
read off as $|\chi_0+\sum_{j=1}^{m-1}\chi_{k\Lambda_{(jq)}}|^2$
one notices that \erf{scez} corresponds to an extension by the
simple current $[\lambda_{k\Lambda_{(q)}}]$.
We will denote by $\sigma_q$ the representative automorphism
corresponding to the $\bbZ_m$ rotation,
$\sigma_q=\lambda_{k\Lambda_{(q)}}$. We can assume
$\sigma_q$ to be $m$-periodic, $\sigma_q^m=\id$, by
\cite[Lemma 4.4]{reh1}, since the statistics phase
fulfills $\kappa_{\sigma_q}=1$ as
$\kappa_{\sigma_q}=\E^{2\pi\I h_{\sigma_q}}$,
$h_{\sigma_q}\equiv h_{k\Lambda_{(q)}}
=kq(n-q)/2n=kq(m-1)/2m\in\bbZ$
exactly at the relevant levels.
(Note that we can no longer use our simple argument relying
on the fixed point as in \cite[Lemma 3.1]{boev2} since in this
more general case there is not necessarily a fixed point.)
Therefore we can construct the extension net of subfactors
exactly as in \cite[Subsect.\ 3.1]{boev2}, replacing $n$ by $m$.
Similarly we find that $\cM$ is local and even Haag dual since
$\eps {\sigma_q}{\sigma_q}=\kappa_{\sigma_q}\bfe=\bfe$.
By construction we have
\be
[\canr] = \bigoplus_{j=0}^{m-1} \,\, [\sigma_q^j] \,.
\labl{orbcan}
We have shown

\begin{proposition}
At levels $k$ satisfying $kq\in 2m\bbZ$ if $n$ is even and
$kq\in m\bbZ$ if $n$ is odd the simple current extension
by the simple current $\sigma_q$ is realized as a quantum
field theoretical net of subfactors $\cN\subset\cM$, where
$\cM$ is Haag dual and as a sector the dual canonical
endomorphism decomposes as in \erf{orbcan}.
\end{proposition}

Therefore we can apply $\alpha$-induction. Now
$[\lambda_{k\Lambda_{(\ell)}}] \times [\lambda_\Lambda] 
= [\lambda_{\sigma^\ell(\Lambda)}]$ is irreducible
for any $\Lambda\in\ASU nk$ and any $\ell=0,1,2,...,n-1$,
hence it follows from \erf{mondiag},
\[ \eps{\lambda_\Lambda}{\lambda_{k\Lambda_{(\ell)}}}
\eps{\lambda_{k\Lambda_{(\ell)}}}{\lambda_\Lambda}
= \E^{2\pi\I(h_{\sigma^\ell(\Lambda)}-
h_{k\Lambda_{(\ell)}}-h_\Lambda)}
\bfe \,, \]
where we also used
$\kappa_{\lambda_\Lambda}=\E^{2\pi\I h_\Lambda}$,
$\Lambda\in\ASU nk$.

\begin{lemma}
For any $\ell=1,2,\ldots,n-1$ and any
$\Lambda=\sum_{i=1}^{n-1}m_i\Lambda_{(i)}\in\ASU nk$,
we have
\be
h_{\sigma^\ell(\Lambda)}-h_\Lambda = \frac \ell n \left(
\frac{(n-\ell)k}2 - t(\Lambda) \right) +
\sum_{i=1}^{\ell-1} (\ell-i)m_{n-i} \,.
\labl{genkota}
\end{lemma}

\bproof
By induction. For $\ell=1$ the formula reduces to
\cite[Lemma 2.7]{kota},
\[ h_{\sigma(\Lambda)} - h_\Lambda = \frac 1n \left(
\frac{(n-1)k}2 - t(\Lambda) \right) \,. \]
Recall
$\sigma(\Lambda)=\sum_{i=1}^{n-1} m_i^\sigma \Lambda_{(i)}$
with $m_1^\sigma=k-m_1-...-m_{n-1}$ and
$m_i^\sigma=m_{i-1}$ for $i\ge 1$. Hence we have
$t(\sigma(\Lambda))=t(\Lambda)+k-nm_{n-1}$. The induction
from $\ell-1$ to $\ell$ is now as follows. First
\[ \bearll
h_{\sigma^\ell(\Lambda)} - h_{\sigma(\Lambda)} &=
h_{\sigma^{\ell-1}(\sigma(\Lambda))} - h_{\sigma(\Lambda)} \\[.4em]
&= \frac{\ell-1}n \left( \frac{(n-\ell+1)k}2 - t(\sigma(\Lambda)) \right)
- \sum_{i=1}^{\ell-2} (\ell-1-i) m_{n-i}^\sigma \\[.4em]
&= \frac{\ell-1}n \left( \frac{(n-\ell+1)k}2 - t(\Lambda) 
-k+nm_{n-1}\right) \\[.4em]
& \qquad\qquad\qquad\qquad -
\sum_{i=1}^{\ell-2} (\ell-1-i) m_{n-1-i} \\[.4em]
&= \frac{\ell-1}n \left( \frac{(n-\ell-1)k}2 - t(\Lambda) \right)
- \sum_{i=1}^{\ell-1} (\ell-i) m_{n-i} \,.
\eear \]
Hence
\[ \bearll
h_{\sigma^\ell(\Lambda)} - h_\Lambda &=
h_{\sigma^\ell(\Lambda)} - h_{\sigma(\Lambda)} +
h_{\sigma(\Lambda)} - h_\Lambda \\[.4em]
&= \frac{\ell-1}n \left( \frac{(n-\ell-1)k}2 - t(\Lambda) \right)
+ \frac 1n \left( \frac{(n-1)k}2 - t(\Lambda) \right) \\[.4em]
& \qquad\qquad\qquad\qquad -
\sum_{i=1}^{\ell-1} (\ell-i) m_{n-i} \\[.4em]
&= \frac \ell n \left( \frac{(n-\ell)k}2 - t(\Lambda) \right)
- \sum_{i=1}^{\ell-1} (\ell-i) m_{n-i} \,,
\eear \]
and the induction is complete.
\eproof

As $h_{k\Lambda_{(\ell)}}=k\ell(n-\ell)/2n$
we obtain immediately the following

\begin{corollary}
For $\Lambda\in\ASU nk$ and $\ell=0,1,2,...,n$ we have
$h_{\sigma^\ell(\Lambda)}-h_{k\Lambda_{(\ell)}}-h_\Lambda=
-t(\Lambda)\ell/n \,\mod\,\bbZ$ and hence
\be
\eps{\lambda_\Lambda}{\lambda_{k\Lambda_{(\ell)}}}
\eps{\lambda_{k\Lambda_{(\ell)}}}{\lambda_\Lambda}
= \E^{-2\pi\I t(\Lambda)\ell/n}
\bfe \,.
\ee
\lablth{siglmono}
\end{corollary}

Using this for $\ell=jq$, $j=0,1,2,...,m-1$,
and Lemma \ref{decmon} we finally find

\begin{corollary}
We have trivial monodromy,
$\eps{\lambda_\Lambda}\canr \eps\canr{\lambda_\Lambda} = \bfe$,
if and only if $t(\Lambda)=0\,\mod\,m$, $\Lambda\in\ASU nk$.
\lablth{monotriv}
\end{corollary}

Now we can investigate the $\alpha$-induced endomorphisms.

\begin{lemma}
For a $\bbZ_m$ orbifold inclusion of $\SUn$ we have
$[\aLap]=[\aLam]$ if $t(\Lambda)=0\,\mod\,m$ and
$\la\aLap,\aLam\ra_M=0$ if $t(\Lambda)\neq 0\,\mod\,m$,
$\Lambda\in\ASU nk$.
\lablth{eqordis}
\end{lemma}

\bproof
The first statement follows from Corollary \ref{monotriv}
and \cite[Prop.\ 3.23]{boev1}. Now note that the
decomposition $[\canr]=\bigoplus_{j=0}^{m-1} [\sigma_q^j]$
implies $[\aLapm]=[\alpha^\pm_{\sigma^q(\Lambda)}]$ since then
\[ \la\aLapm,\aLapm\ra_M =
\la\aLapm,\alpha^\pm_{\sigma^q(\Lambda)}\ra_M = 
\la\alpha^\pm_{\sigma^q(\Lambda)},\alpha^\pm_{\sigma^q(\Lambda)}\ra_M
= \la\canr\circ\lambda_\Lambda,\lambda_\Lambda\ra_N \,. \] 
Now $h_{k\Lambda_{(q)}}=kq(m-1)/2m\in\bbZ$ for the levels $k$
where the $\bbZ_m$ orbifold inclusions appear, therefore
$h_{\sigma^q(\Lambda)}-h_\Lambda\notin\bbZ$ if and only if
$t(\Lambda)\neq 0\,\mod\,m$.
It follows that then $\la\aLap,\aLam\ra_M=0$ by
Lemma \ref{disj}.
\eproof

Now we are ready to prove the main result of this subsection.

\begin{theorem}
For all $\bbZ_m$ orbifold inclusions of $\SUn$, where $n=mq$,
$m,q\in\bbN$, appearing at levels $k$ such that
$kq\in 2m\bbN$ if $n$ is even and $kq\in m\bbN$ if $n$ is odd,
we have:
\begin{enumerate}
\item $Z_{\Lambda,\Lambda'} = \la\aLap,\aLams\ra_M$ for all
$\Lambda,\Lambda'\in\ASU nk$,
\item $\sum_{x\in\cV} d_x^2 = \sum_{\Lambda\in\ASU nk} d_\Lambda^2$,
\item each irreducible subsector of $[\can]$ is in $\cV$.
\end{enumerate}
\lablth{Zapamorb}
\end{theorem}

\bproof
As by Lemma \ref{lamupm} $\la\aLap,\aLam\ra_M=0$,
$\Lambda\in\ASU nk$, implies
$\la\aLap,\alpha^-_{\Lambda'}\ra_M=0$ for all other
$\Lambda'\in\ASU nk$, we can write
\[ \la\aLap,\alpha^-_{\Lambda'}\ra_M = \delta^m(t(\Lambda))
\, \la\aLap,\alpha^+_{\Lambda'}\ra_M \]
by Lemma \ref{eqordis}. Now
\[ \la\aLap,\alpha^+_{\Lambda'}\ra_M = \la \canr \circ
\lambda_\Lambda,\lambda_{\Lambda'} \ra_N =
\sum_{j=0}^{m-1} \la \sigma_q^j \circ
\lambda_\Lambda,\lambda_{\Lambda'} \ra_N =
\sum_{j=0}^{m-1} \delta(\Lambda',\sigma^{jq}(\Lambda)) \,,\]
hence
\[ \la\aLap,\alpha^-_{\Lambda'}\ra_M = \delta^m(t(\Lambda))
\sum_{j=0}^{m-1} \delta(\Lambda',\sigma^{jq}(\Lambda)) 
= Z_{\Lambda,\Lambda'} \,,\]
proving the first statement. The second statement is derived
from the first in exactly the same way as in the proof of
Proposition \ref{4equiv} for the conformal inclusion case.
Finally the third statement follows from the second by
Lemmata \ref{W=tV} and \ref{VintV}.
\eproof

It is instructive to find the subsectors of $[\can]$ in
$\cV$ more constructively.
Since $\la\can,\can\ra_M=\la\canr,\canr\ra_N=m$, $[\can]$ contains
at most $m$ different irreducible subsectors. In fact, analogously to
\cite[Cor.\ 3.4]{boev2} we find that
$M\cong N\rtimes_{\sigma_q}\bbZ_m$,
hence it follows (see e.g.\ \cite{lon4})
\[ [\can] = \bigoplus_{j=0}^{m-1} \,\, [\epsilon^j] \,, \]
where $\epsilon\in\Aut(M)$ is $m$-periodic and $N$ is the fixed
point algebra of $M$ under the action of $\epsilon$, $N=M^\epsilon$.
Now for $i,j=1,2,...,m-1$ we have, by Lemma \ref{eqordis},
\[ \la \alfp i \circ \alfm {n-i}, \alfp j \circ \alfm {n-j} \ra_M
= \la \alfp i \circ \alfp {n-j}, \alfm i \circ \alfm {n-j} \ra_M
= 0  \]
if $i\neq j$, since then $\alfpm i \circ \alfpm {n-j}$
decomposes into $\alpha^\pm_\Lambda$'s with
$t(\Lambda)\neq 0\,\mod\,m$. Similarly
\[ \la \alfp j \circ \alfm {n-j}, \id_M \ra_M
= \la \alfp j , \alfm j \ra_M = 0 \,,
\qquad  j =1,2,\ldots,m-1 \,. \]
We conclude that $[\id_M]$ and $[\alfp j \circ \alfm {n-j}]$,
$j=1,2,...,m-1$, are all disjoint, and they all have (at least)
one subsector in common with $[\can]$ by Corollary \ref{congam}.
Therefore all the subsectors $[\epsilon^j]$ of $[\can]$ are
subsectors.

It is straightforward to determine $\cV$ for the orbifold
inclusions of $\SUz$. The simultaneous fusion graphs of
$\asp 1$ and $\asm 1$ for \ddD 6 and \ddD 8 are given in
Figure \ref{Devenpm}.
%
%%%%%%%%%%%% D_{even} %%%%%%%%%%%%%
\begin{figure}[tb]
\unitlength 0.74mm
\begin{center}
\begin{picture}(140,160)
%%%%
\thinlines 
\multiput(10,60)(0,40){2}{\circle*{1.5}}
\multiput(50,60)(0,40){2}{\circle*{1.5}}
\multiput(30,30)(0,10){5}{\circle*{1.5}}
\multiput(30,90)(0,20){3}{\circle*{1.5}}
\multiput(30,30)(0,10){5}{\circle{3}}
\multiput(30,90)(0,20){3}{\circle{3}}
\multiput(30,50)(0,10){2}{\circle{4.5}}
\multiput(30,90)(0,40){2}{\circle{4.5}}
\Thicklines 
\path(30,130)(10,100)(30,90)(10,60)(30,60)
\path(10,60)(30,50)
\dottedline{1.6}(30,130)(50,100)(30,90)(50,60)(30,60)
\dottedline{1.6}(50,60)(30,50)
\path(30,110)(50,100)(30,70)(50,60)(30,40)
\path(50,60)(30,30)
\dottedline{1.6}(30,110)(10,100)(30,70)(10,60)(30,40)
\dottedline{1.6}(10,60)(30,30)
\put(4,100){\makebox(0,0){\small{$\asp 1$}}}
\put(4,60){\makebox(0,0){\small{$\asp 3$}}}
\put(56,100){\makebox(0,0){\small{$\asm 1$}}}
\put(56,60){\makebox(0,0){\small{$\asm 3$}}}
\put(30,136){\makebox(0,0){\small{$\as 0$}}}
\put(30,115){\makebox(0,0){\small{$[\epsilon]$}}}
\put(30,95){\makebox(0,0){\small{$\as 2$}}}
\put(30,75){\makebox(0,0){\small{$[\beta_2]$}}}
\put(30,65){\makebox(0,0){\small{$\asx 41$}}}
\put(30,55){\makebox(0,0){\small{$\asx 42$}}}
\put(30,45){\makebox(0,0){\small{$[\eta]$}}}
\put(30,25){\makebox(0,0){\small{$[\eta']$}}}
\put(30,0){\makebox(0,0){\ddD 6}}
%%%%%
\thinlines 
\multiput(90,40)(0,40){3}{\circle*{1.5}}
\multiput(130,40)(0,40){3}{\circle*{1.5}}
\multiput(110,10)(0,10){5}{\circle*{1.5}}
\multiput(110,70)(0,20){5}{\circle*{1.5}}
\multiput(110,10)(0,10){5}{\circle{3}}
\multiput(110,70)(0,20){5}{\circle{3}}
\multiput(110,30)(0,10){2}{\circle{4.5}}
\multiput(110,70)(0,40){3}{\circle{4.5}}
\Thicklines 
\path(110,150)(90,120)(110,110)(90,80)(110,70)
(90,40)(110,40)
\path(90,40)(110,30)
\dottedline{1.6}(110,150)(130,120)(110,110)(130,80)(110,70)
(130,40)(110,40)
\dottedline{1.6}(130,40)(110,30)
\path(110,130)(130,120)(110,90)(130,80)(110,50)
(130,40)(110,20)
\path(130,40)(110,10)
\dottedline{1.6}(110,130)(90,120)(110,90)(90,80)(110,50)
(90,40)(110,20)
\dottedline{1.6}(90,40)(110,10)
\put(84,120){\makebox(0,0){\small{$\asp 1$}}}
\put(84,80){\makebox(0,0){\small{$\asp 3$}}}
\put(84,40){\makebox(0,0){\small{$\asp 5$}}}
\put(136,120){\makebox(0,0){\small{$\asm 1$}}}
\put(136,80){\makebox(0,0){\small{$\asm 3$}}}
\put(136,40){\makebox(0,0){\small{$\asm 5$}}}
\put(110,156){\makebox(0,0){\small{$\as 0$}}}
\put(110,135){\makebox(0,0){\small{$[\epsilon]$}}}
\put(110,115){\makebox(0,0){\small{$\as 2$}}}
\put(110,95){\makebox(0,0){\small{$[\beta_2]$}}}
\put(110,75){\makebox(0,0){\small{$\as 4$}}}
\put(110,55){\makebox(0,0){\small{$[\beta_4]$}}}
\put(110,45){\makebox(0,0){\small{$\asx 61$}}}
\put(110,35){\makebox(0,0){\small{$\asx 62$}}}
\put(110,25){\makebox(0,0){\small{$[\eta]$}}}
\put(116,10){\makebox(0,0){\small{$[\eta']$}}}
\put(110,0){\makebox(0,0){\ddD 8}}
\end{picture}
\caption{\ddD 6 and \ddD 8: Fusion graphs of $\asp 1$ and $\asm 1$}
\label{Devenpm}
\end{center}
\end{figure}
For \ddD {2\varrho+2} we denote
$[\beta_{2j}]=[\epsilon\circ\alpha_{2j}]$, $1\le j\le\varrho-1$,
and $[\eta]=[\epsilon\circ\alpha^{(1)}_{2\varrho}]$,
$[\eta']=[\epsilon\circ\alpha^{(2)}_{2\varrho}]$.

\subsection{Non-degenerate braidings on orbifold graphs}

Let $\cW_0\subset\cW$ (recall that $\cW$ is the sector
basis in $\LTSN$ corresponding to --- and as a set 
identified with --- $\ASU nk$) be a sector sub-basis,
i.e.\ a subset of $\cW$ which is itself a sector basis.
The following lemma is from \cite{evka3}
(cf.\ also \cite[Eq.\ (5.17)]{reh0}), but we think it is
instructive to give an algebraic instead of a graphical
proof here.

\begin{lemma}
For any sector sub-basis $\cW_0\subset\cW$ we have that if
$\Omega\equiv[\lambda_\Omega]\in\cW_0$ is degenerate in
$\cW_0$ then
$S_{\Lambda,\Omega}S_{0,0}=S_{\Lambda,0}S_{\Omega,0}$
for all $\Lambda\equiv[\lambda_\Lambda]\in\cW_0$.
\lablth{degS}
\end{lemma}

\bproof
As a consequence of the Verlinde formula and
the modular relation $(ST)^3=S^2$ for the modular
matrices $S$ and $T$ we obtain with
$T_{\Lambda,\Lambda'}=\del \Lambda{\Lambda'}
\E^{2\pi\I h_\Lambda} T_{0,0}$ the following formula
(see e.g.\ \cite[Eq.\ (2.35)]{fu1}; compare also \cite{reh0}
for a derivation in the DHR context),
\[ S_{\Lambda,\Omega} = S_{0,0} \sum_{\Phi\in\ASU nk}
\N \Lambda\Omega\Phi \, d_\Phi \,
\E^{2\pi\I (h_\Lambda+h_\Omega-h_\Phi)} \,, \qquad
\Lambda,\Omega\in\ASU nk \,. \]
Now assume that $\Lambda,\Omega$ belong to $\cW_0$.
Since $\cW_0$ is a sector sub-basis it follows
$\N \Lambda\Omega\Phi$ can only be non-zero
if $\Phi$ belongs to $\cW_0$. Moreover, if
$[\lambda_\Omega]$ is degenerate in $\cW_0$ then we
find for the eigenvalues of the monodromy
$\E^{2\pi\I (h_\Lambda+h_\Omega-h_\Phi)}=1$
whenever $\N \Lambda\Omega\Phi\neq 0$ by
\erf{mondiag}. Hence
\[ S_{\Lambda,\Omega} = S_{0,0} \sum_{\Phi\in\ASU nk}
\N \Lambda\Omega\Phi \, d_\Phi = S_{0,0} \, d_\Lambda d_\Phi
=  \frac{S_{\Lambda,0}S_{\Omega,0}}{S_{0,0}} \]
whenever $\Lambda$ belongs to $\cW_0$.
\eproof

Now we return to the $\bbZ_m$ orbifold situation, i.e.\
we write $n=mq$ and consider levels $k$ with $kq\in 2m\bbZ$
if $n$ is even and $kq\in m\bbZ$ if $n$ is odd.
The following lemma is a slight generalization of
\cite[Lemma 3.4]{evka3}, but since there is a mistake in
the proof we give a corrected and generalized proof here.

\begin{lemma}
Let $\cW_0$ be the sector basis of sectors
$[\lambda_\Lambda]$, $\Lambda\in\ASU nk$, with
$t(\Lambda)=0\,\mod\,m$, and $n,m,k,q$ as above.
Then the degenerate elements within $\cW_0$ are
given by $[\sigma_q^j]$, $j=0,1,2,\ldots,m-1$.
\lablth{degW0}
\end{lemma}

\bproof
For any $\Omega\in\ASU nk$ let $s^\Omega$ be the
column vector of the S-matrix of $\SUn_k$
corresponding to the weight $\Omega$, i.e.\ in components
$(s^\Omega)_\Lambda=S_{\Lambda,\Omega}$, $\Lambda\in\ASU nk$.
By the Verlinde formula these vectors are eigenvectors of
the fusion matrices $N_\Lambda$ (defined by
$(N_\Lambda)_{\Lambda',\Lambda''}=\N {\Lambda'}\Lambda{\Lambda''}$),
\[ N_\Lambda \, s^\Omega = \gamma_\Omega (\Lambda) \, s^\Omega \,,
\qquad \gamma_\Omega (\Lambda) =
\frac{S_{\Lambda,\Omega}}{S_{0,\Omega}} \,,
\quad \Lambda\in\ASU nk \,.\]
We may split $s^\Omega$ into $m$ pieces,
$s^\Omega = {}^\mathrm{t}(s^\Omega_0,s^\Omega_1,...,s^\Omega_{m-1})$,
where each vector $s^\Omega_j$ consists of
components $S_{\Lambda,\Omega}$ with
$t(\Lambda)=j\,\mod\,m$. Since we have
$\N {\Lambda'}\Lambda{\Lambda''}=0$ whenever
$t(\Lambda)+t(\Lambda')\neq t(\Lambda'')\,\mod\,n$, hence in
particular if
$t(\Lambda)+t(\Lambda')\neq t(\Lambda'')\,\mod\,m$, and
$t(\Lambda_{(1)})=1$, $t(\Lambda_{(n-1)})=n-1$, we can write
\[ N_{\Lambda_{(1)}} \, s^\Omega_j = \gamma_\Omega (\Lambda_{(1)})
\, s^\Omega_{j+1} \,, \qquad
N_{\Lambda_{(n-1)}} \, s^\Omega_{j+1} = \gamma_\Omega (\Lambda_{(n-1)})
\, s^\Omega_j \,, \]
and the index $j$ can be read $\mod\,m$. Therefore
we find with $N_{\Lambda_{(n-1)}}= {}^\mathrm{t} N_{\Lambda_{(1)}}$,
\[ \gamma_\Omega (\Lambda_{(1)}) \| s^\Omega_{j+1} \|^2 =
\la N_{\Lambda_{(1)}} s^\Omega_j, s^\Omega_{j+1} \ra =
\la s^\Omega_j , N_{\Lambda_{(n-1)}} s^\Omega_j \ra =
\gamma_\Omega (\Lambda_{(1)}) \| s^\Omega_j \|^2 \,, \]
where we used $\overline{\gamma_\Omega (\Lambda_{(n-1)})} =
\overline{S_{\Lambda_{(n-1)},\Omega}/S_{0,\Omega}} =
S_{\Lambda_{(1)},\Omega}/S_{0,\Omega}=\gamma_\Omega (\Lambda_{(1)})$.
Since $\|s^\Omega\|=1$ by unitarity of the S-matrix we first conclude
that $\gamma_\Omega (\Lambda_{(1)})\neq 0$ and hence
$\|s^\Omega_j\|=m^{-1/2}$ for all $j=0,1,2,...,m-1$ and all
$\Omega\in\ASU nk$. Now assume that $[\lambda_\Omega]$ is
degenerate in $\cW_0$. Then
$S_{\Lambda,\Omega}=S_{\Omega,0}S_{\Lambda,0}/S_{0,0}$
by Lemma \ref{degS} for all
$\Lambda\in\ASU nk$ with $t(\Lambda)=0\,\mod\,m$,
and this is $s^\Omega_0=\gamma_0(\Omega) s^0_0$.
By $\|s^\Omega_0\|=m^{-1/2}=\|s^0_0\|$ it follows
$\gamma_0(\Omega)=1$, i.e.\ $S_{\Omega,0}=S_{0,0}$.
This means that $\Omega$ is the weight of a simple current,
i.e.\ either 0 or one of the weights $k\Lambda_{(\ell)}$,
$\ell=1,2,3,...,n-1$. Now take, for example, the weight
$m\Lambda_{(1)}$ which has $t(m\Lambda_{(1)})=m\equiv n/q$.
Therefore we find by Corollary \ref{siglmono},
\[ \eps {\lambda_{m\Lambda_{(1)}}}{\lambda_{k\Lambda_{(\ell)}}}
\eps {\lambda_{k\Lambda_{(\ell)}}}{\lambda_{m\Lambda_{(1)}}}
= \E^{2\pi\I \ell/q} \bfe \,, \qquad \ell = 0,1,2,\ldots, n-1, \]
and therefore $[\lambda_{k\Lambda_{(\ell)}}]$ can be
degenerate in $\cW_0$ only if $\ell$ is a multiple of $q$.
On the other hand it follows similarly from Corollary
\ref{siglmono} that $[\lambda_{k\Lambda_{(\ell)}}]$
is in fact degenerate in $\cW_0$ for $\ell=jq$,
$j=0,1,2,...,m-1$.
Finally we note that $t(k\Lambda_{(jq)})=kjq=0\,\mod\,m$
at the relevant levels, so in fact
$[\sigma_q^j]=[\lambda_{k\Lambda_{(jq)}}]$ belongs to $\cW_0$,
the proof is complete.
\eproof

Defining $\cT$ to be the set of all irreducible
subsectors of $[\aLap]$, $\Lambda\in\ASU nk$ with
$t(\Lambda)=0\,\mod\,m$, gives a sector basis with
$\cT=\cV^+\cap\cV^-$ by Lemma \ref{eqordis},
and it plays exactly the role of the set of
marked vertices in the conformal inclusion case.
One also checks easily that putting
\[ b_{t,\Lambda}=\la \lambda_\Lambda, \sigma_{\beta_t} \ra_N=
\la \aLapm, \beta_t \ra_M \,,\quad t\in\cT\,,\quad
\Lambda\in\ASU nk \,, \]
yields the expression
$Z_{\Lambda,\Lambda'}=\sum_{t\in\cT} b_{t,\Lambda}b_{t,\Lambda'}$.
By Corollary \ref{Tbraid} we find that $\cT$ has braiding,
given by the relative braiding operators. The following
theorem nicely reflects Rehren's conjecture \cite{reh0}
which was proven by M\"uger \cite{mug}.

\begin{theorem}
For any $\bbZ_m$ orbifold inclusion of $\SUn$ the sector basis
$\cT$ as above has a non-degenerate braiding.
\lablth{Tnd}
\end{theorem}

\bproof
First note that $\cT$ is the image of $\cW_0$ by
$\alpha$-induction ($+$ or $-$). Now let
$\beta_t\in\End(M)$ such that $[\beta_t]\equiv t$
is an irreducible subsector of $[\aLap]$ for some
$\Lambda\equiv[\lambda_\Lambda]\in\cW_0$, i.e.\
$t\in\cT$, and assume
$\epsr {\beta_t}{\beta_{t'}}=\epsr {\beta_{t'}}{\beta_t} ^*$
for all $t'\in\cT$, where, as usual, $\beta_{t'}$ denotes
a representative endomorphism for each $t'$ . Since
$[\aLaps]=[\aLams]$ for all $\Lambda'\in\cW_0$ we obtain
$\eps{\lambda_\Lambda}{\lambda_{\Lambda'}}
\eps{\lambda_{\Lambda'}}{\lambda_\Lambda} = \bfe$
for all $\Lambda'\in\cW_0$ by Lemma \ref{monlm=1}.
By Lemma \ref{degW0} we conclude that
$[\lambda_\Lambda]=[\sigma_q^j]\equiv[\lambda_{k\Lambda_{(jq)}}]$
for some $j=0,1,2,...,m-1$. But
$[\alpha^\pm_{k\Lambda_{(jq)}}]=[\id_M]$, hence
$[\beta_t]=[\id_M]$,
showing that the braiding is non-degenerate.
\eproof

For the $\SUz$ orbifold inclusions, the set $\cT$ corresponds
to the even vertices of the D-graphs, constructed as fusion
graphs of either $\asp 1$ or $\asm 1$. So here
Theorem \ref{Tnd} can be rephrased, roughly speaking,
as ``there is a non-degenerate braiding on the even vertices
of the graphs \Deven'', which is a known result, see
\cite{ocng,evka3,tuwe}.
For $\SUd$ the corresponding statement is that
there is a non-degenerate braiding associated to the triality
zero vertices of Kostov's graphs, see \cite[Fig.\ 3b]{kost} or
\cite[Fig.\ 25]{franc}, \cite[Fig.\ 8.32]{evka}. Analogous
statements hold now for a huge variety of orbifold graphs
of the graph $\ASU nk$. Let us finally remark that the
analogue of Theorem \ref{Tnd} for the conformal
inclusions is not very interesting since,
by Proposition \ref{oreps}, we just rediscover the
(non-degenerate) braiding of the
enveloping WZW level 1 theory (cf.\ $\SOf$ and $\Gtwo$ for the
\ddE 6 and \ddE 8 modular invariants in the $\SUz$ case).

\subsection{More conformal inclusions of $\SUn$}

Let us now continue with the treatment of conformal inclusions of
$\SUn$. We first present a useful lemma. Recall that the fusion
rules of the simple current $\lambda_{k\Lambda_{(1)}}$ correspond
to the $\bbZ_n$-rotation $\sigma$ of the Weyl alcove, i.e.\
$[\lambda_{k\Lambda_{(1)}} \circ \lambda_\Lambda]
= [\lambda_{\sigma(\Lambda)}]$ for $\Lambda\in\ASU nk$.
The map $\tau:\ASU nk\rightarrow\bbZ_n$,
$\Lambda\mapsto\tau(\Lambda)=t(\Lambda)\,\mod\,n$, is
sometimes called colouring or ``$n$-ality'', and recall that the
fusion coefficients vanish, $\N \Lambda{\Lambda'}{\Lambda''}=0$,
unless $\tau(\Lambda)+\tau(\Lambda')=\tau(\Lambda'')$.

\begin{lemma}
For conformal inclusions at levels $k\in 2n\bbN$ if $n$ is even and
$k\in n\bbN$ if $n$ is odd we have the following: If $[\canr]$ is
$\bbZ_n$-rotation invariant, i.e.
$[\lambda_{k\Lambda_{(1)}}\circ\canr]=[\canr]$,
then $\la\aLap,\aLam\ra_M=0$ whenever $\tau(\Lambda)\neq 0$, hence
in particular $\la\aLap\circ\aLamb,\aLaps\circ\aLambs\ra_M=0$
whenever $\tau(\Lambda)\neq\tau(\Lambda')$,
$\Lambda,\Lambda'\in\ASU nk$.
\lablth{cihelp}
\end{lemma}

\bproof
As in the proof of Lemma \ref{eqordis} we find if $[\canr]$ is rotation
invariant then
\[ \la\aLapm,\aLapm\ra_M = \la\aLapm,\alpha^\pm_{\sigma(\Lambda)}\ra_M = 
\la\alpha^\pm_{\sigma(\Lambda)},\alpha^\pm_{\sigma(\Lambda)}\ra_M
= \la\canr\circ\lambda_\Lambda,\lambda_\Lambda\ra_N \,, \]
hence $[\aLapm]=[\alpha^\pm_{\sigma(\Lambda)}]$.
Then by \erf{genkota} with $q=1$
we see that at levels $k\in 2n\bbN$ if $n$ is even and $k\in n\bbN$
if $n$ is odd then $h_{\sigma(\Lambda)}-h_\Lambda\notin\bbZ$
if $t(\Lambda)\notin n\bbZ$, i.e.\ if $\tau(\Lambda)\neq 0$.
Hence $\la\aLap,\aLam\ra_M=\la\aLap,\alpha^-_{\sigma(\Lambda)}\ra_M$
vanishes if $\tau(\Lambda)\neq 0$ by Lemma \ref{disj}.
Then clearly
\[ \la\aLap\circ\aLamb,\aLaps\circ\aLambs\ra_M=
\la\aLap\circ\aLapbs,\aLam\circ\aLambs\ra_M=0 \,, \]
since $[\aLapm\circ\aLapmbs]$ decompose into $[\alpha^\pm_\Omega]$'s,
$\Omega\in\ASU nk$, with $\tau(\Omega)=\tau(\Lambda)-\tau(\Lambda')$
non-zero if $\tau(\Lambda)\neq\tau(\Lambda')$.
\eproof

\ddDgx 6 revisited: $\SUd_3 \subset \mathit{SO}(8)_1$:
This is the first case of the $\cD$-series for $\SUd$ and
it happens to be a conformal embedding at the same time, similar
to the \ddD 4 example for $\SUz$. Although the discussion
is in principle covered by our treatment of the orbifold inclusions
it is instructive to do the calculations for this case.
Recall from \cite{boev2} that
\[ \cV^\pm = \{ \asd 00 , \asdpm 10 , \asdpm 11, \asdx 211 ,
\asdx 212 , \asdx 213 \} \,, \]
where $\asd 00$ and $\asdx 21i$, $i=1,2,3$, are the marked vertices
corresponding to the four level 1 representations of $\mathit{SO}(8)$.
Note that Lemma \ref{cihelp} directly yields $\cV^+\cap\cV^-=\cT$
in this case since $[\canr]$ is rotation invariant and
$\asd 00 = \asd 30 = \asd 33$ and $\asd 21$ are the only 
sectors in $\cV^\pm$ of the form $[\alpha^\pm_\Lambda]$
with $\Lambda$ of colour zero.

\begin{proposition}
For the \ddDgx 6 example we have
$[\can]=[\id_M]\oplus[\eta_1]\oplus[\eta_2]$, where
$[\eta_1]$ is an irreducible subsector of $\asdprod 1011$
and $[\eta_2]$ is an irreducible subsector of $\asdprod 1110$.
Hence the equivalent conditions of Proposition
\ref{4equiv} are fulfilled.
\end{proposition}

\bproof
Note that $\la\can,\can\ra_M=\la\canr,\canr\ra_N=3$ and
$d_\can=d_\canr=3$ so that $[\can]$ must decompose into three
sectors of statistical dimension one. As $[\canr]$ is rotation
invariant we learn from Lemma \ref{cihelp} that
\[ \asd 00 \equiv [\id_M] \,, \,\,\,
\asdprod 1011 \,,\,\,\, \asdprod 1110 \,, \]
are disjoint sectors and as $\asdpm 10$ and $\asdpm 11$ are
irreducible they have all one subsector in common with
$[\can]$ by Corollary \ref{congam}.
\eproof 

\ddEgx 8 revisited: $\SUd_5 \subset \mathit{SU}(6)_1$: First
note that $[\canr]$ is not rotation invariant here. However,
the treatment of this example is not particularly difficult.
Recall from \cite{boev2} that $\cV^\pm$ consists of six
marked vertices $\asd 00$, $\asdx 201$, $\asdx 221$, $\asd 50$,
$\asd 55$ (forming a $\bbZ_6$ fusion subalgebra) and six
further sectors $\asdpm 10$, $\asdpm 11$, $\asdpm 40$, $\asdpm 44$,
$\asdpm 51$ and $\asdpm 55$.

\begin{proposition}
For the \ddEgx 8 example we have
$[\can]=[\id_M]\oplus\asdprod 1011$,
hence the equivalent conditions of Proposition
\ref{4equiv} are fulfilled.
\end{proposition}

\bproof
We have $\Nd 421031 =1$ but $\cdd 31 - \cdd 10 = 2/3-1/6=1/2\notin\bbZ$,
hence $\asdp 10 \neq \asdm 10$ by Corollary \ref{channel} and thus
disjoint since $\asdpm 10$ is irreducible. Therefore
\[ \la \aedprod 1011 , \id_M \ra_M = \la \aedp 10, \aedm 10 \ra_M=0\,,\]
but
\[ \la \aedprod 1011 , \can \ra_M = \la \aedp 10, \aedp 10 \ra_M=1 \]
by Corollary \ref{congam}. Now note that
$d_\can=d_\canr=1+d_{(4,2)}=4+2\sqrt 2$, and
$\la\can,\can\ra_M=\la\canr,\canr\ra_N=2$, therefore
$[\can]=[\id_M]\oplus[\delta]$ with $[\delta]$ irreducible
and $d_\delta=3+2\sqrt 2$. But we have
$d_{\aedprod 1011}=d_{(1,0)}^2=(1+\sqrt 2)^2 = 3+2\sqrt 2$.
Hence we must have $[\delta]=\asdprod 1011$.
\eproof

With the results of \cite{boev2} it is easy to write down
the principal graph of $N\subset M$, presented in
Figure \ref{P-cE8}. This graph first appeared in \cite{xu2}
and was, as a principal graph, associated to the conformal
inclusion $\SUd_5 \subset \mathit{SU}(6)_1$ in \cite{xu4}.
\begin{figure}[tb]
\unitlength 0.35mm
\thicklines
\begin{center}
\begin{picture}(320,120)
\multiput(10,100)(50,0){7}{\circle*{3}}
\multiput(40,20)(80,0){4}{\circle*{3}}
\path(10,100)(40,20)(60,100)(200,20)(160,100)(120,20)(60,100)
\path(110,100)(120,20)(260,100)(280,20)(310,100)(200,20)(210,100)
\put(10,111){\makebox(0,0){$\lasd 00$}}
\put(60,111){\makebox(0,0){$\lasd 42$}}
\put(110,111){\makebox(0,0){$\lasd 51$}}
\put(160,111){\makebox(0,0){$\lasd 21$}}
\put(210,111){\makebox(0,0){$\lasd 54$}}
\put(260,111){\makebox(0,0){$\lasd 33$}}
\put(310,111){\makebox(0,0){$\lasd 30$}}
\put(40,8){\makebox(0,0){$[\iotab]$}}
\put(120,8){\makebox(0,0){$[\iotab\circ\aedpm 51]$}}
\put(200,8){\makebox(0,0){$[\iotab\circ\aedpm 54]$}}
\put(280,8){\makebox(0,0){$[\iotab\circ\aedx 301]$}}
\end{picture}
\end{center}
\caption{\ddEgx 8: Principal graph for the conformal inclusion
$\SUd_5\subset\mathit{SU}(6)_1$}
\label{P-cE8}
\end{figure}
With our machinery, we can now easily determine the dual
principal graph of $N\subset M$. Let us first determine
$\cV$. First we check that for
$[\beta_a^\pm],[\beta_b^\pm]\in\cV^\pm$ we have
\[ \la \beta_a^\pm\circ \aedmp 10,\beta_b^\pm\circ \aedmp 10\ra_M
=\la \co{\beta_b^\pm}\circ\beta_a^\pm,
\aedmp 10 \circ \aedmp 11\ra_M = \del ab \,,\]
since
$[\aedmp 10\circ\aedmp 11]=\asd 00 \oplus\asdmp 51 \oplus\asdmp 54$,
and the identity is the only marked vertex on the right hand side.
Hence, besides $[\beta_a^\pm]\in\cV^\pm$ we have the irreducible
sectors $[\beta_a^\pm\circ \aedmp 10]$ in $\cV$. But since
$[[\cV]]=d_\can[[\cV^\pm]]\equiv(1+d_{(1,0)}^2)[[\cV^\pm]]$
by Lemmata \ref{W=dV} and \ref{W=tV}, it follows that these
sectors are already all sectors in $\cV$. Just by looking at
the fusion graph of $\asdp 10$ given in \cite{boev2} (and clearly
the fusion graph of $\asdm 10$ looks the same way) we find
that $[\aedx 201\circ\aedmp 10]=\asdmp 54$ and similarly
$[\aed 50\circ\aedmp 10]=\asdmp 51$. We denote
$[\delta']=\asdprod 4410$. Now using $[\can]=[\id_M]\oplus[\delta]$
with $[\delta]=\asdprod 1011 = \asdprod 1110$ it is straightforward
to compute the $\can$-multiplication and therefore to obtain
(by Theorem \ref{pdpg}) the dual principal graph,
displayed in Figure \ref{DP-cE8}.
\begin{figure}[tb]
\unitlength 0.35mm
\thicklines
\begin{center}
\begin{picture}(300,120)
\multiput(10,100)(40,0){8}{\circle*{3}}
\multiput(45,20)(70,0){4}{\circle*{3}}
\path(10,100)(45,20)(50,100)(115,20)(250,100)(185,20)(50,100)
\path(90,100)(115,20)(130,100)
\path(170,100)(185,20)(210,100)
\path(250,100)(255,20)(290,100)
\put(10,112){\makebox(0,0){$\asd 00$}}
\put(50,112){\makebox(0,0){$[\delta]$}}
\put(90,112){\makebox(0,0){$\asdp 54$}}
\put(130,112){\makebox(0,0){$\asdm 54$}}
\put(170,112){\makebox(0,0){$\asdp 51$}}
\put(210,112){\makebox(0,0){$\asdm 51$}}
\put(250,112){\makebox(0,0){$[\delta']$}}
\put(290,112){\makebox(0,0){$\asdx 301$}}
\put(45,8){\makebox(0,0){$[\iota]$}}
\put(115,8){\makebox(0,0){$[\aedpm 54\circ\iota]$}}
\put(185,8){\makebox(0,0){$[\aedpm 51\circ\iota]$}}
\put(255,8){\makebox(0,0){$[\aedx 301\circ\iota]$}}
\end{picture}
\end{center}
\caption{\ddEgx 8: Dual principal graph for the conformal inclusion
$\SUd_5\subset\mathit{SU}(6)_1$}
\label{DP-cE8}
\end{figure}
The subsystems $\cV^\pm_0$ of $M$-$M$ sectors of the dual principal
graph as in Corollary \ref{subdual} consist obviously of the sectors
$\asd 00$, $\asdpm 51$, $\asdpm 54$ and $\asdx 301$.

Example \ddEgx {12}: $\SUd_9\subset(\mathrm{E}_6)_1$:
The corresponding modular invariant reads
\[ Z_{\cE^{(12)}}= |\chid 00  + \chid 90  + \chid 99 
+ \chid 51 + \chid 84   + \chid 54 |^2
+ 2  | \chid 42 + \chid 72 + \chid 77 |^2 \]
and therefore
\[ [\canr] = \lasd 00 \oplus \lasd 90 \oplus \lasd 99 \oplus \lasd 51
 \oplus \lasd 84 \oplus \lasd 54 \,. \]
With this we can determine the sector bases $\cV^\pm$.
We find
\[ \bearl
 \cV^\pm = \{ \asd 00 , \asdpm 10 , \asdpm 11 , \asdpm 20 ,
\asdpm 21 , \asdpm 22 , \\[.4em]
\qquad\qquad \qquad \qquad \asdxpm 311 ,  \asdxpm 312 , \asdxpm 321 ,
\asdxpm 322 , [\eta_1] , [\eta_2] \} \,,
\eear \]
and some useful identities
\[ \begin{array}{rlrl}
\asdpm 31 &= \asdpm 22 \oplus \asdxpm 311 \oplus \asdxpm 312 \,,\,\,\,
& \asdpm 42 &= 2\, \asdpm 21 \oplus [\eta_1] \oplus [\eta_2] \,,\\[.4em]
\asdpm 32 &= \asdpm 20 \oplus \asdxpm 321 \oplus \asdxpm 322 \,,
&\asdpm 43 &= \asdpm 10 \oplus \asdpm 31 \,, \\[.4em]
\asdpm 52 &=  \asdpm 10 \oplus \asdpm 22 \oplus \asdpm 31 \,,
&\asdpm 21 &= \asdpm 30 = \asdpm 33 \,.
\eear \]
Here $\asd 00$, $[\eta_1]$ and $[\eta_2]$ are the marked vertices
corresponding to the three level 1 representations of \ddE 6. One
checks by matching statistical dimensions that they are simple
sectors and hence are forced to satisfy the $\bbZ_3$ fusion rules
of $(\mathrm{E}_6)_1$. From
\[ \asdpm 10 \times \asdpm 42 = \asdpm 31 \oplus
\asdpm 43 \oplus \asdpm 52 \]
and its conjugation we obtain
\[ \bearll
([\eta_1] \oplus [\eta_2]) \times \asdpm 10
&= \asdxpm 311 \oplus \asdxpm 312 \,,\\[.4em]
([\eta_1] \oplus [\eta_2]) \times \asdpm 11
&= \asdxpm 321 \oplus \asdxpm 322 \,.
\eear \]
We have the freedom to choose the notation such that
\[ [\eta_i] \times \asdpm 10 = \asdxpm 31i \,, \qquad i=1,2, \]
and this will actually provide a nice $\bbZ_3$ symmetry of the fusion
graphs of $\asdpm 10$. We remark that the homomorphisms
$[\alpha^\pm]$ of sector algebras are not surjective as we
cannot isolate $[\eta_1]$ and $[\eta_2]$ separately.
However, the sector algebras associated to $\cV^\pm$ are uniquely
determined, and the fusion graph of either $\asdp 10$ in
$\cV^+$ or $\asdm 10$ in $\cV^-$ is given in Figure \ref{cE121}.
%
%%%%%%%%%%%% cE^12_1 %%%%%%%%%%%%%
\begin{figure}[tb]
\unitlength 0.55mm %0.65mm
\begin{center}
\begin{picture}(110,180)
%%%%
\thinlines
\put(50,170){\circle*{2}}
\put(30,50){\circle*{2}}
\put(90,10){\circle*{2}}
\put(50,170){\circle{4}}
\put(30,50){\circle{4}}
\put(90,10){\circle{4}}
\multiput(30,130)(40,0){2}{\circle*{2}}
\multiput(10,90)(40,0){3}{\circle*{2}}
\multiput(20,70)(40,0){2}{\circle*{2}}
\multiput(50,50)(40,0){2}{\circle*{2}}
\path(50,170)(10,90)(30,50)(90,90)(50,170)
\path(90,90)(90,10)(10,90)
\path(30,130)(50,90)(70,130)(30,130)
\path(20,70)(50,90)(60,70)(20,70)
\path(50,50)(50,90)(90,50)(50,50)
\path(10,90.4)(90,90.4)
\path(10,89.6)(90,89.6)
\spline(10,90)(50,105)(90,90)
\thicklines
\put(40,150){\vector(-1,-2){0}}
\put(20,110){\vector(-1,-2){0}}
\put(60,150){\vector(-1,2){0}}
\put(80,110){\vector(-1,2){0}}
\put(60,110){\vector(-1,-2){0}}
\put(40,110){\vector(-1,2){0}}
\put(50,130){\vector(1,0){0}}
\put(30,90){\vector(1,0){0}}
\put(70,90){\vector(1,0){0}}
\put(15,80){\vector(-1,2){0}}
\put(35,80){\vector(-3,-2){0}}
\put(50,80){\vector(0,-1){0}}
\put(55,80){\vector(-1,2){0}}
\put(75,80){\vector(-3,-2){0}}
\put(40,70){\vector(1,0){0}}
\put(70,70){\vector(-1,1){0}}
\put(90,70){\vector(0,-1){0}}
\put(35,65){\vector(-1,1){0}}
\put(25,60){\vector(-1,2){0}}
\put(45,60){\vector(-3,-2){0}}
\put(70,50){\vector(1,0){0}}
\put(70,30){\vector(-1,1){0}}
\put(90,30){\vector(0,-1){0}}
\put(48.9,101.1){\vector(-1,0){0}}
\put(50,177){\makebox(0,0){\scriptsize{$\asd 00$}}}
\put(84,130){\makebox(0,0){\scriptsize{$\asdpm 11$}}}
\put(16,130){\makebox(0,0){\scriptsize{$\asdpm 10$}}}
\put(1,83){\makebox(0,0){\scriptsize{$\asdpm 20$}}}
\put(11,63){\makebox(0,0){\scriptsize{$\asdxpm 312$}}}
\put(22,44){\makebox(0,0){\scriptsize{$[\eta_2]$}}}
\put(45,41){\makebox(0,0){\scriptsize{$\asdxpm 311$}}}
\put(98,10){\makebox(0,0){\scriptsize{$[\eta_1]$}}}
\put(102,50){\makebox(0,0){\scriptsize{$\asdxpm 321$}}}
\put(102,90){\makebox(0,0){\scriptsize{$\asdpm 22$}}}
\put(62,94){\makebox(0,0){\scriptsize{$\asdpm 21$}}}
\put(65,63){\makebox(0,0){\scriptsize{$\asdxpm 322$}}}
\end{picture}
\caption{$\cE^{(12)}_1$: Fusion graph of either
$\asdp 10$ in $\cV^+$ or $\asdm 10$ in $\cV^-$}
\label{cE121}
\end{center}
\end{figure}
To determine the full induced sector basis $\cV$ is much more involved,
and we do not present the calculations here. However, we briefly
show that we also have $Z=\tilde{Z}$ in this case.

\begin{proposition}
For the \ddEgx {12} example the equivalent conditions
of Proposition \ref{4equiv} are fulfilled.
\end{proposition}

\bproof
We show that each subsector of $[\can]$ is in $\cV$.
Consider the following sectors:
\[ \bearlll \asd 00 \,,\qquad\qquad & [\aedv 10+ \circ \aedv 11-] \,,
\quad & [\aed11+ \circ \aedv 10-] \,, \\[.4em]
[\aedv 20+ \circ \aedv 22-] \,, & [\aedv 22+ \circ \aedv 20-] \,,
&[\aedv 21+ \circ \aedv 21-] \,.
\eear \]
They all have a subsector in common with $[\can]$ by Corollary
\ref{congam}, and we now show that they are all disjoint.
By Lemma \ref{cihelp} we only need to show
\[ \begin{array}{rl}
\la \alpha_{(0,0)} , \aedv 21+\circ\aedv 21- \ra_M &=0 \,,\\[.4em]
\la \aedv 10+ \circ \aedv 11- , \aedv 22+\circ\aedv 20- \ra_M &=0 \,,\\[.4em]
\la \aedv 11+ \circ \aedv 10- , \aedv 20+\circ\aedv 22- \ra_M &=0 \,.
\eear \]
First we find $\asdp 21 \neq \asdm 21$ using Corollary \ref{channel}
since $\Nd 21 51 30 =1$ but
$\cdd 30 - \cdd 21 = 1/2-1/4=1/4\notin\bbZ$. We even have
$\la\aedv 21+,\aedv 21-\ra_M=0$ as $\asdpm 21$ is irreducible,
and this is the first relation. The other relations follow
by use of the sector products
\[ \asdpm 10 \times \asdpm 20 = \asdpm 11 \times \asdpm 22 =
2\, \asdpm 21 \,. \]
Since $\la\can,\can\ra_M=\la\canr,\canr\ra_N=6$ we have already
identified all subsectors of $[\can]$ as subsectors of some
products $\asdp pq \times \asdm rs$.
\eproof

Example \ddEgx {24}: $\SUd_{21}\subset(\mathrm{E}_7)_1$:
The corresponding modular invariant reads
\[ \begin{array}{ll}
Z_{\mathcal{E}^{(24)}}
=& | \chid 00 + \chid {21}0 + \chid {21}{21} +
\chid 84 + \chid {17}4 + \chid {17}{13} \\[.4em]
& \qquad + \chid {11}1 + \chid {11}{10} + \chid {20}{10} +
\chid {12}6 + \chid {15}6 + \chid {15}9 |^2 \\[.4em]
& + | \chid 60 + \chid {21}6 + \chid {15}{15} +
\chid {15}0 + \chid {21}{15} + \chid 66 \\[.4em]
& \qquad + \chid {11}4 + \chid {17}7 + \chid {14}{10} +
\chid {11}7 + \chid {14}4 + \chid {17}{10} |^2 \,,
\end{array} \]
therefore
\[ \begin{array}{ll}
[\canr] =& \lasd 00 \oplus \lasd {21}0 \oplus \lasd {21}{21}
 \oplus \lasd 84  \oplus \lasd {17}4 \oplus \lasd {17}{13} \\[.4em]
& \qquad \lasd {11}1 \oplus \lasd {11}{10} \oplus \lasd {20}{10}
 \oplus \lasd {12}6 \oplus \lasd {15}6 \oplus \lasd {15}9  \,.
\end{array} \]
With this we can determine the sector bases $\cV^\pm$.
We find
\[ \bearll
 \cV^\pm =& \{ \asd 00 , \asdpm 10 , \asdpm 11 , \asdpm 20 ,
\asdpm 21 , \asdpm 22 , \asdpm 30 , \asdpm 31 , \\[.4em]
& \quad \asdpm 32 ,  \asdpm 33 , \asdxpm 401 ,
\asdxpm 402 , \asdxpm 411 , \asdxpm 412 , \asdxpm 421 ,
\asdxpm 422 , \\[.4em] 
& \quad  \asdxpm 431 , \asdxpm 432 , \asdxpm 441 ,
\asdxpm 442 , \asdxpm 501 , \asdxpm 511 , \asdxpm 551 ,
[\epsilon] \} \,.
\eear \]
We also give some irreducible decompositions,
\[ \asdpm 4q = \asdxpm 4q1 \oplus \asdxpm 4q2  \,, \qquad
q= 0,1,2,3,4, \]
and
\[ \begin{array}{rlrl}
\asdpm 50 \!\!\!&= \asdxpm 411 \oplus \asdxpm 501 , 
&\asdpm 51 \!\!\!\!&= 
\asdpm 30 \oplus \asdxpm 421 \oplus \asdxpm 511 , \\[.4em]
\asdpm 52 \!\!\!&= \asdpm 31 \oplus \asdxpm 401 \oplus \asdxpm 511 ,
&\asdpm 53 \!\!\!\!&= \asdpm 32 \oplus
\asdxpm 411 \oplus \asdxpm 441 , \\[.4em]
\asdpm 54 \!\!\!&=  \asdpm 33 \oplus \asdxpm 422 \oplus \asdxpm 511 ,
&\asdpm 55 \!\!\!\!&= \asdxpm 431 \oplus \asdxpm 551 ,
\eear \]
and
\[ \bearl
\asdpm 60 = \asdxpm 422 \oplus \asdxpm 511 \oplus [\epsilon] \,,\\[.4em]
\asdpm 61 = \asdpm 31 \oplus \asdxpm 402 \oplus \asdxpm 431
\oplus \asdxpm 551 \,,\\[.4em]
\asdpm 62 = \asdpm 20 \oplus \asdpm 32 \oplus \asdxpm 411 \oplus
\asdxpm 412 \oplus \asdxpm 442 \,,\\[.4em]
\asdpm 63 = \asdpm 21 \oplus \asdpm 30 \oplus \asdpm 33 \oplus
\asdxpm 421 \oplus \asdxpm 422 \,,\\[.4em]
\asdpm 64 = \asdpm 22 \oplus \asdpm 31 \oplus \asdxpm 402 \oplus
\asdxpm 431 \oplus \asdxpm 432 \,,\\[.4em]
\asdpm 65 = \asdpm 32 \oplus \asdxpm 411 \oplus \asdxpm 442
\oplus \asdxpm 501 \,,\\[.4em]
\asdpm 66 = \asdxpm 421 \oplus \asdxpm 511 \oplus [\epsilon] \,,
\eear \]
and also
\[ \asdpm 70 = \asdxpm 431 \oplus \asdxpm 432 \oplus \asdxpm 551 \,,\,\,\,
\asdpm 77 = \asdxpm 411 \oplus \asdxpm 412 \oplus \asdxpm 501 \,. \]
Here $\asd 00$ and $[\epsilon]$ are the marked vertices
corresponding to the two level 1 representations of \ddE 7.
These formulae are indeed enough to isolate each irreducible sector,
i.e.\ can be inverted; in fact the homomorphisms $[\alpha^\pm]$ are
surjective in this case. The fusion graph of either $\asdp 10$ in
$\cV^+$ or $\asdm 10$ in $\cV^-$ is given in Figure \ref{cE24}.
%
%%%%%%%%%%%%% cE^24 %%%%%%%%%%%%%%
\begin{figure}[tb]
\unitlength 0.65mm %0.80mm
\begin{center}
\begin{picture}(110,180)
%%%%
\thinlines
\put(55,170){\circle*{2}}
\multiput(40,150)(30,0){2}{\circle*{2}}
\multiput(25,130)(30,0){3}{\circle*{2}}
\multiput(10,110)(30,0){4}{\circle*{2}}
\put(55,100){\circle*{2}}
\multiput(25,90)(60,0){2}{\circle*{2}}
\put(55,80){\circle*{2}}
\multiput(10,70)(30,0){4}{\circle*{2}}
\multiput(25,50)(30,0){3}{\circle*{2}}
\multiput(40,30)(30,0){2}{\circle*{2}}
\put(55,10){\circle*{2}}
\put(55,170){\circle{4}}
\put(55,10){\circle{4}}
\path(55,170)(25,130)(25,50)(55,10)(85,50)(85,130)(55,170)
\path(10,110)(10,70)(100,70)(100,110)(10,110)
\path(25,130)(85,130)(70,110)(10,70)(70,150)(40,150)(100,70)
(55,80)(40,110)(25,130)
\path(85,50)(25,50)(40,70)(100,110)(40,30)(70,30)(10,110)
(55,100)(70,70)(85,50)
\path(40,110)(40,70)
\path(70,110)(70,70)
\thicklines
\put(47.5,160){\vector(-3,-4){0}}
\put(55,150){\vector(1,0){0}}
\put(62.5,160){\vector(-3,4){0}}
\multiput(32.5,140)(30,0){2}{\vector(-3,-4){0}}
\multiput(47.5,140)(30,0){2}{\vector(-3,4){0}}
\multiput(40,130)(30,0){2}{\vector(1,0){0}}
\multiput(32.5,120)(30,0){2}{\vector(-3,4){0}}
\multiput(47.5,120)(30,0){2}{\vector(-3,-4){0}}
\put(25,120){\vector(0,-1){0}}
\put(85,120){\vector(0,1){0}}
\put(32.5,110){\vector(-1,0){0}}
\put(55,110){\vector(1,0){0}}
\put(77.5,110){\vector(-1,0){0}}
\put(32.5,105){\vector(4,-1){0}}
\put(62.5,105){\vector(-3,-2){0}}
\multiput(17.5,100)(60,0){2}{\vector(3,-4){0}}
\multiput(32.5,100)(60,0){2}{\vector(3,4){0}}
\put(70,100){\vector(0,1){0}}
\put(76,94){\vector(3,2){0}}
\put(10,90){\vector(0,1){0}}
\put(50,90){\vector(-1,2){0}}
\put(60,90){\vector(1,-2){0}}
\put(100,90){\vector(0,-1){0}}
\put(34,86){\vector(-3,-2){0}}
\multiput(17.5,80)(60,0){2}{\vector(-3,-4){0}}
\multiput(32.5,80)(60,0){2}{\vector(-3,4){0}}
\put(40,80){\vector(0,-1){0}}
\put(47.5,75){\vector(3,2){0}}
\put(77.5,75){\vector(-4,1){0}}
\put(32.5,70){\vector(1,0){0}}
\put(55,70){\vector(-1,0){0}}
\put(77.5,70){\vector(1,0){0}}
\multiput(32.5,60)(30,0){2}{\vector(3,4){0}}
\multiput(47.5,60)(30,0){2}{\vector(3,-4){0}}
\put(25,60){\vector(0,-1){0}}
\put(85,60){\vector(0,1){0}}
\multiput(40,50)(30,0){2}{\vector(-1,0){0}}
\multiput(32.5,40)(30,0){2}{\vector(3,-4){0}}
\multiput(47.5,40)(30,0){2}{\vector(3,4){0}}
\put(47.5,20){\vector(3,-4){0}}
\put(55,30){\vector(-1,0){0}}
\put(62.5,20){\vector(3,4){0}}
\put(55,177){\makebox(0,0){\scriptsize{$\asd 00$}}}
\put(30,150){\makebox(0,0){\scriptsize{$\asdpm 10$}}}
\put(80,150){\makebox(0,0){\scriptsize{$\asdpm 11$}}}
\put(15,130){\makebox(0,0){\scriptsize{$\asdpm 20$}}}
\put(95,130){\makebox(0,0){\scriptsize{$\asdpm 22$}}}
\put(43,134){\makebox(0,0){\scriptsize{$\asdpm 21$}}}
\put(0,110){\makebox(0,0){\scriptsize{$\asdxpm 412$}}}
\put(48,113){\makebox(0,0){\scriptsize{$\asdpm 31$}}}
\put(62,113){\makebox(0,0){\scriptsize{$\asdpm 32$}}}
\put(110,110){\makebox(0,0){\scriptsize{$\asdxpm 432$}}}
\put(63,99){\makebox(0,0){\scriptsize{$\asdxpm 421$}}}
\put(17,90){\makebox(0,0){\scriptsize{$\asdpm 30$}}}
\put(93,90){\makebox(0,0){\scriptsize{$\asdpm 33$}}}
\put(47,81){\makebox(0,0){\scriptsize{$\asdxpm 422$}}}
\put(0,70){\makebox(0,0){\scriptsize{$\asdxpm 401$}}}
\put(48,66){\makebox(0,0){\scriptsize{$\asdxpm 411$}}}
\put(62,66){\makebox(0,0){\scriptsize{$\asdxpm 431$}}}
\put(110,70){\makebox(0,0){\scriptsize{$\asdxpm 441$}}}
\put(15,50){\makebox(0,0){\scriptsize{$\asdxpm 402$}}}
\put(95,50){\makebox(0,0){\scriptsize{$\asdxpm 442$}}}
\put(67,46){\makebox(0,0){\scriptsize{$\asdxpm 511$}}}
\put(30,30){\makebox(0,0){\scriptsize{$\asdxpm 501$}}}
\put(80,30){\makebox(0,0){\scriptsize{$\asdxpm 551$}}}
\put(55,3){\makebox(0,0){\scriptsize{$[\epsilon]$}}}
\end{picture}
\caption{\ddEgx {24}: Fusion graph of either
$\asdp 10$ in $\cV^+$ or $\asdm 10$ in $\cV^-$}
\label{cE24}
\end{center}
\end{figure}

To determine the full induced sector basis $\cV$ is much more involved,
and we do not present the calculations here, but we just
show the following

\begin{proposition}
For the \ddEgx {24} example the equivalent conditions of
Proposition \ref{4equiv} are fulfilled.
\end{proposition}

\bproof
As all elements of $\cV^\pm$ except the marked vertex $[\epsilon]$
appear as subsectors of some $\asdpm pq$, $0\le q\le p \le 5$,
it suffices to show $\la\aedp pq,\aedm pq\ra_M=0$ for all
$0\le q \le p \le 5$, except $(p,q)=(0,0)$, in order to prove
$\cV^+\cap\cV^-=\cT\equiv\{\asd 00,[\epsilon]\}$. 
Since $[\canr]$ is rotation invariant we only need to show
$\la\aedp pq,\aedm pq\ra_M=0$ for the colour zero
cases, i.e.\ for $(p,q)=(2,1),(3,0),(3,3),(4,2),(5,1),(5,4)$,
by Lemma \ref{cihelp}. As $\asdpm 30$ is irreducible it
suffices to show $\asdp 21 \neq \asdm 21$. It follows from
$\Nd 218463=1$ and $\cdd 63 - \cdd 21 = 5/8-1/8=1/2\notin\bbZ$.
Similarly $\la\aedp 30,\aedm 30\ra_M=0$, since $\Nd 308451=1$
but $\cdd 52 - \cdd 30=1/2-1/4=1/4\notin\bbZ$ and then we obtain
$\la\aedp 33,\aedm 33\ra_M=0$ by conjugation. Now note that
$\asdpm 42$ is a (reducible) subsector of $\asdpm 63$ but
$\cdd 63-\cdd 42=5/8-1/3=7/24\notin\bbZ$, it follows
$\la\aedp 42,\aedm42\ra_M=0$ by Lemma \ref{disj}. For
$\la\aedp 51,\aedm 51\ra_M=0$ it only remains to be shown
that $\asdxp 511 \neq \asdxm 511$, and this follows since
$\asdxpm 511$ is a subsector of $\asdpm 60$ but
$\cdd 60 - \cdd 51 = 3/4-1/2=1/4\notin\bbZ$. Finally
$\la\aedp 54,\aedm 54\ra_M=0$ follows by conjugation.
\eproof

In turn one can also show that the following sectors,
\[ \bearl
\asdprod 1011 \,,\quad \asdprod 2220 \,,\quad 
\asdprod 1110 \,,\quad \asdprod 2022 \,,\\[.4em]
\asd 00 \,,\quad \asdprod 2121 \,,\quad \asdprod 3033 \,,\quad
\asdprod 4242 \,,
\eear \]
are all disjoint and have one sector in common with $[\can]$,
and that the two further disjoint sectors $\asdprod 4341$ and
$\asdprod 4143$ have two sectors in common with $[\can]$. This
already yields all the subsectors of $[\can]$ since
$\la\can,\can\ra_M=\la\canr,\canr\ra_N=12$.

$\mathit{SU}(4)_4\subset\mathit{SO}(15)_1$ revisited. We first
remark that $[\canr]$ is not $\bbZ_4$-rotation invariant here.
Recall from \cite{boev2} that
\[ \bearll
\cV^\pm =& \{ \asf 000 , \asfpm 100 , \asfpm 110 , \asfpm 111 ,
\asf 400 , \asfx 3211 , \\[.4em] & \qquad\qquad\qquad
\asfxpm 210i , \asfxpm 220i , \asfxpm 221i , \quad i=1,2 \} \,,
\eear \]
and
\[ \cT = \{ \asf 000 , \asf 400 , \asfx 3211 \} \,. \]
This example is the first one which leads to non-commutative
chiral sector algebras, however, it is not an exception in the
sense that the following holds.

\begin{proposition}
For the conformal embedding $\mathit{SU}(4)_4\subset\mathit{SO}(15)_1$
the equivalent conditions of Proposition \ref{4equiv} are fulfilled.
\end{proposition}

\bproof
We first claim that $\la\aefp 210,\aefm 210\ra_M=0$.
This follows since $\asfpm 210 = \asfpm 322$ but
$\cdf 322 - \cdf 210 = 55/64-39/64=1/4\notin\bbZ$.
By conjugation we obtain $\la\aefp 221,\aefm 221\ra_M=0$.
Note that we have the irreducible decompositions
\[ \bearll
\asfpm 210 =& \asfpm 111 \oplus \asfxpm 2101
\oplus \asfxpm 2102 \,,\\[.4em]
\asfpm 221 =& \asfpm 100 \oplus \asfxpm 2211
\oplus \asfxpm 2212 \,,
\eear \]
therefore we also find $\asfp 100 \neq \asfm 100$ and
$\asfp 111 \neq \asfm 111$. Further we recall that
$\asfpm 220 = \asfxpm 2201 \oplus \asfxpm 2202$ is a
subsector of $\asfpm 211$ but
$\cdf 220 - \cdf 211 = 3/4-1/2=1/4\notin\bbZ$, hence also
$\la\aefp 220,\aefm 220\ra_M=0$. Now $\asfpm 220$ appears
(twice) in the square of $\asfpm 110$, hence also
$\asfp 110 \neq \asfm 110$. We have
established $\cT=\cV^+\cap\cV^-$.
\eproof

In turn one easily checks that
\[ \asf 000 \,,\,\,\, \asfprod 100111 \,,\,\,\,
\asfprod 111100 \,,\,\,\, \asfprod 110110 \,, \]
are disjoint sectors and they all have a subsector in common
with $[\can]$, exhibiting all subsectors of $[\can]$ since
$\la\can,\can\ra_M=\la\canr,\canr\ra_N=4$.

\section{Summary and Outlook}

We have analyzed the structure of the induced sector systems
obtained by mixing the $\pm$-inductions for conformal and
$\bbZ_m$ orbifold embeddings of $\SUn$. We proved the formula
$Z_{\Lambda,\Lambda'}=\la\aLap,\aLams\ra_M$,
$\Lambda,\Lambda'\in\ASU nk$, for the associated modular invariant
mass matrix for all $\bbZ_m$ orbifold inclusions and several
conformal inclusions. As a consequence, all subsectors of
$[\can]$ can be obtained by decomposing suitable sectors
$[\aLap\circ\aLams]$, the ``global index'' of the induced
sector basis is maximal, i.e.\ coincides with the one of
the original $\SUn_k$ fusion algebra, and we have
$Z_{\Lambda,\Lambda'}\neq 0$ if and only if
$\Lambda,\Lambda'\in\Exp$ as well as
$Z_{\Lambda,\Lambda}\neq 0$ if and only if $\Lambda\in\Exp^+$.
Our results cover in particular all type \nolinebreak I
modular invariants of $\SUz$ and $\SUd$.
The proof for the conformal inclusions is,
unfortunately, case by case and therefore covers only a limited
number of examples. However, we believe that it holds
for all of them:

\begin{conjecture}
The equivalent conditions of Proposition \ref{4equiv} hold
for any conformal inclusion of $\SUn$.
\end{conjecture}

Recall from the proof of Lemma \ref{nonv} that the ``regular''
representation of the induced fusion algebra, given in terms
of the sector product matrices $M_x$, $x\in\cV$, decomposes
into representations $B_{\Lambda,\Lambda'}$ labelled by
the set of exponents, namely
$M_x=\bigoplus_{(\Lambda,\Lambda')\in\Exp}B_{\Lambda,\Lambda'}(x)$,
$x\in\cV$. We believe that this decomposition is minimal
in the following sense:

\begin{conjecture}
For any conformal or orbifold inclusion of $\SUn$ we have
$B_{\Lambda,\Lambda'}\simeq \pi_{(\Lambda,\Lambda')}
\otimes\bfe_{Z_{\Lambda,\Lambda'}}$, where the
$\pi_{(\Lambda,\Lambda')}$'s are the irreducible,
pairwise inequivalent representations of the full
induced sector algebra, and the dimension of
$\pi_{(\Lambda,\Lambda')}$ is $Z_{\Lambda,\Lambda'}$,
$(\Lambda,\Lambda')\in\Exp$. In consequence,
$\dim\,\Eig(\Lambda,\Lambda')=Z_{\Lambda,\Lambda'}^2$
for $\Lambda,\Lambda'\in\ASU nk$.
\lablth{regV}
\end{conjecture}

Our results provide powerful methods to compute the induced
sector bases $\cV$ and their algebraic structure,
yet the computations may become more
and more involved with increasing rank and level. (For large
$n$ and $k$ it might not even be possible to determine
$\cV$ completely with our results at hand.) However, if $\cV$
and its sector algebra
is determined one can easily write down the principal and
the dual principal graph of the conformal inclusion subfactors,
(this is certainly less interesting for the orbifold inclusions
since there the subfactors $N\subset M$ are just of $\bbZ_m$
type), and we have illustrated these powerful methods by
several examples, including the computation of the dual
principal graph for the conformal inclusion
$\SUd_5\subset\mathit{SU}(6)_1$, which has,
to the best of our knowledge, not been computed before.
Thus our theory can also be used to determine basic
invariants of new subfactor examples.

Let us finally remark that there are
type \nolinebreak I modular invariants
which come neither from conformal inclusions nor from
simple current extensions as, for instance, the exceptional
$\mathit{SU}(10)$ level $2$ modular invariant found in
\cite{walt} which arises by level-rank duality from the
\ddE 6 modular invariant of $\SUz$. It is natural to presume
that there will still be an associated net of subfactors
such that $\alpha$-induction can be applied. If so, the
next thing to check is whether the equivalent conditions
of Prop.\ \ref{4equiv} even hold for these cases.
However, it does not seem reasonable to expect that there
are associated nets of subfactors for all type \nolinebreak I
modular invariants. In fact there are type \nolinebreak I
modular invariants for which there is no ``fixed point
resolution'' of the S-matrix, see \cite[Sect.\ 4]{fss1}.
For example, there is a type \nolinebreak I modular invariant
of $\mathit{SU}(5)$ at level $5$ of this kind
\cite[Eq.\ (B.3)]{scya3} which has the same vacuum block
as the different (type \nolinebreak I) modular invariant
\cite[Eq.\ (B.6)]{scya3} corresponding to the conformal
embedding $\mathit{SU}(5)_5\subset\mathit{SO}(24)_1$.
So here we expect an associated net of subfactors to
exist only for the conformal embedding invarant.
Type \nolinebreak I invariants without fixed point resolution
appear to be rather rare, however.

\vspace{0.5cm}
\begin{footnotesize}
\noindent{\it Acknowledgement.}
We would like to thank A.\ N.\ Schellekens for sending us the
(huge) output files for the $\SUd_9$ and $\SUd_{21}$ fusion
coefficients. We are grateful to T.\ Gannon for explanations,
in particular concerning the combinatorics relevant for the
$\bbZ_m$ orbifold modular invariants, and to M.\ Izumi for
providing us a preliminary version of \cite{izu4}.
We would like to thank Y.\ Kawahigashi for various discussions
and in particular for the idea of using the sum matrix
$Q=\sum_{\Lambda,\Lambda'} R^{\Lambda,\Lambda'}$
in the proof of Proposition \ref{4equiv}. We are grateful
to K.-H.\ Rehren for useful comments.
 
This project is supported by the EU TMR Network in
Non-Commutative Geometry.
\end{footnotesize}

%%%%%%%%%%%%%%%%%%%%%%%%%%%%%%%%%%%%%%%%%%%%%%%%%%%%%%%%%%%%%%%%%%%%%%%%%%%

%%%%%%%%%%%%%%%%%%%%%%%%% bibliography %%%%%%%%%%%%%%%%%%%%%%%%%%%%%%%%%%%%

\newcommand\biba[7]   {\bibitem{#1} {\sc #2:} {\sl #3.} {\rm #4} {\bf #5}
                      { (#6) } {#7}}% \hspace*{\fill} {\small\tt {#1}}}
\newcommand\bibb[4]   {\bibitem{#1} {\sc #2:} {\it #3.} {\rm #4}}
                      %\hspace*{\fill} {\small\tt {#1}}}
\newcommand\bibx[4]   {\bibitem{#1} {\sc #2:} {\sl #3} {\rm #4}}
                      %\hspace*{\fill} {\small\tt {#1}}}
\def\AAM              {Acta Appl.\ Math.}
\def\AIP              {Ann.\ Inst.\ H.\ Poincar\'e (Phys.\ Th\'eor.)}
\def\CMP              {Com\-mun.\ Math.\ Phys.}
\def\IJM              {Intern.\ J. Math.}
\def\JFA              {J.\ Funct.\ Anal.}
\def\JMP              {J.\ Math.\ Phys.}
\def\LMP              {Lett.\ Math.\ Phys.}
\def\RMP              {Rev.\ Math.\ Phys.}
\def\Inv              {Invent.\ Math.}
\def\npbp             {Nucl.\ Phys.\ {\bf B} (Proc.\ Suppl.)}
\def\nupb             {Nucl.\ Phys.\ {\bf B}}
\def\adma             {Adv.\ Math.}
\def\coma             {Con\-temp.\ Math.}
\def\physa            {Physica {\bf A}}
\def\ijmp             {Int.\ J.\ Mod.\ Phys.\ {\bf A}}
\def\jpa              {J.\ Phys.\ {\bf A}}
\def\FdP              {Fortschr.\ Phys.}
\def\PLB              {Phys.\ Lett.\ {\bf B}}
\def\RIMS             {Publ.\ RIMS, Kyoto Univ.}

%%%%%%%%%%%%%%%%%%%%%%%%%%%%%%%%%%%%%%%%%%%%%%%%%%%%%%%%%%%%%%%%%%%%%%%%%%%

\begin{footnotesize}

\end{footnotesize}
\end{document}